\documentclass[%
    reprint,
    superscriptaddress,
    nofootinbib,
    amsmath,amssymb,
    aps,
]{revtex4-2}

\usepackage{graphicx}
\usepackage{dcolumn}
\usepackage{bm}
\usepackage{booktabs}
\usepackage[hidelinks]{hyperref}

\usepackage{siunitx}

\usepackage{xspace}

\usepackage[normalem]{ulem}

\usepackage[dvipsnames]{xcolor}
\usepackage{verbatim}

\usepackage{float}
\makeatletter
\let\newfloat\newfloat@ltx
\makeatother

\usepackage{algorithm}
\usepackage{algpseudocode}

\usepackage{txfonts}

\usepackage{url}

\newcommand{\geant}{\textsc{Geant4}\xspace}
\newcommand{\ddhep}{\textsc{DD4hep}\xspace}
\newcommand{\pandora}{\textsc{PandoraPFA}\xspace}
\newcommand{\DDML}{\textsc{DDML}\xspace}
\newcommand{\LFlows}{\textsc{ConvL2LFlows}\xspace} 
\newcommand{\CCiii}{\textsc{\mbox{CaloClouds3}}\xspace}
\newcommand{\CCii}{\textsc{\mbox{CaloClouds II}}\xspace}
\newcommand{\CC}{\textsc{\mbox{CaloClouds}}\xspace}
\newcommand{\bibae}{\textsc{\mbox{BiBAE}}\xspace}
\newcommand{\optimumXOne}{\textsc{\mbox{Optimum (x1)}}\xspace}
\newcommand{\optimumXNine}{\textsc{\mbox{Optimum (x9)}}\xspace}
\newcommand{\optimumSteps}{\textsc{\mbox{Optimum (steps)}}\xspace}

\begin{document}

\newcommand{\PJM}[1]{\textcolor{PineGreen}{PM: #1}}
\newcommand{\ThB}[1]{\textcolor{RoyalBlue}{ThB: #1}}
\newcommand{\AK}[1]{\textcolor{Orange}{AK: #1}}
\newcommand{\FG}[1]{\textcolor{Red}{FG: #1}}
\newcommand{\TM}[1]{\textcolor{LimeGreen}{TM: #1}}
\newcommand{\HDH}[1]{\textcolor{Purple}{HDH: #1}}

\preprint{APS/123-QED}

\title{A First Full Physics Benchmark for Highly Granular Calorimeter Surrogates}

\author{Thorsten Buss}
\affiliation{Institut f\"ur Experimentalphysik, Universit\"at Hamburg, Luruper Chaussee 149, 22761 Hamburg, Germany}
\affiliation{Deutsches Elektronen-Synchrotron DESY, Notkestr. 85, 22607 Hamburg, Germany}

\author{Henry Day-Hall}
\affiliation{Deutsches Elektronen-Synchrotron DESY, Notkestr. 85, 22607 Hamburg, Germany}

\author{Frank Gaede}
\affiliation{Deutsches Elektronen-Synchrotron DESY, Notkestr. 85, 22607 Hamburg, Germany}

\author{Gregor Kasieczka}
\affiliation{Institut f\"ur Experimentalphysik, Universit\"at Hamburg, Luruper Chaussee 149, 22761 Hamburg, Germany}

\author{Katja Krüger}
\affiliation{Deutsches Elektronen-Synchrotron DESY, Notkestr. 85, 22607 Hamburg, Germany}

\author{Anatolii Korol}
\email{anatolii.korol@desy.de}
\affiliation{Deutsches Elektronen-Synchrotron DESY, Notkestr. 85, 22607 Hamburg, Germany}

\author{Thomas Madlener}
\affiliation{Deutsches Elektronen-Synchrotron DESY, Notkestr. 85, 22607 Hamburg, Germany}

\author{Peter McKeown}
\email{peter.mckeown@cern.ch}
\affiliation{Deutsches Elektronen-Synchrotron DESY, Notkestr. 85, 22607 Hamburg, Germany}
\affiliation{CERN, 1211 Geneva 23, Switzerland}

\date{November 20, 2025}

\begin{abstract}
    The physics programs of current and future collider experiments necessitate the development of surrogate simulators for calorimeter showers. While much progress has been made in the development of generative models for this task, they have typically been evaluated in simplified scenarios and for single particles. This is particularly true for the challenging task of highly granular calorimeter simulation. 
    For the first time, this work studies the use of highly granular generative calorimeter surrogates in a realistic simulation application. We introduce DDML, a generic library which enables the combination of generative calorimeter surrogates with realistic detectors implemented using the DD4hep toolkit. We compare two different generative models -- one operating on a regular grid representation, and the other using a less common point cloud approach. In order to disentangle methodological details from model performance, we provide comparisons to idealized simulators which directly sample representations of different resolutions from the full simulation ground-truth. We then systematically evaluate model performance on post-reconstruction benchmarks for electromagnetic shower simulation. Beginning with a typical single particle study, we introduce a first multi-particle benchmark based on di-photon separations, before studying a first full-physics benchmark based on hadronic decays of the tau lepton. Our results indicate that models operating on a point cloud can achieve a favorable balance between speed and accuracy for highly granular calorimeter simulation compared to those which operate on a regular grid representation.
\end{abstract}

\maketitle


\section{\label{sec:Introduction} Introduction}
Accurate and efficient simulations of particle detectors are essential for modern collider experiments, where they are used for detector design and physics analysis. Traditionally, simulations rely on Monte Carlo (MC) methods, which, despite their high accuracy, incur substantial computational costs~\cite{HEPSoftwareFoundation:2017ggl,Boehnlein:2022}. Especially high computational demands arise from the simulation of particle showers in calorimeter systems, where a single incident particle can trigger a large cascade of particles and interactions.

In recent years, generative surrogate models have emerged as a promising solution. These models aim to replicate the complex patterns of particle showers while significantly accelerating simulation speed. To this end, various generative paradigms have been applied, including generative adversarial networks (GANs)~\cite{Paganini:2017hrr,Paganini:2017dwg,deOliveira:2017rwa,Erdmann:2018kuh,Erdmann:2018jxd,Carminati:2018khv,Musella:2018rdi,Belayneh:2019vyx,Butter:2020qhk,ATLAS:2020,Ghosh:2020kkt,ATLAS:2021pzo,ATLAS:2022jhk,Hashemi:2023ruu,FaucciGiannelli:2023fow,Dogru:2024gpk}, variational autoencoders (VAEs)~\cite{Buhmann:2020pmy,Buhmann:2021lxj,Buhmann:2021caf,ATLAS:2022jhk,Cresswell:2022tof,Bieringer:2022cbs,Diefenbacher:2023prl,Hoque:2023zjt,Liu:2024kvv}, classical normalizing flows (NFs)~\cite{Krause:2021ilc,Krause:2021wez,Schnake:2022,Krause:2022jna,Diefenbacher:2023vsw,Xu:2023xdc,Buckley:2023daw,Pang:2023wfx,Ernst:2023qvn,Schnake:2024mip,Du:2024gbp,Buss:2024orz}, auto-regressive models~\cite{Birk:2025wai}, and diffusion and continuous flow models~\cite{Mikuni:2022xry,Buhmann:2023bwk,Acosta:2023zik,Mikuni:2023tqg,Amram:2023onf,Buhmann:2023kdg,Jiang:2024ohg,Kobylianskii:2024ijw,Jiang:2024bwr,Favaro:2024rle,Brehmer:2024yqw,Buss:2025cyw,Raikwar:2025fky}. For a recent taxonomy see~\cite{Hashemi:2023rgo}.

These models have been trained and evaluated on a wide range of datasets, making it difficult to draw general conclusions about their performance. To address this, the recent CaloChallenge~2022~\cite{Krause:2024avx} was started. It provided a valuable, large-scale comparison of generative models for calorimeter simulation on common datasets.

However, the evaluation of generative models has typically been limited to simplified scenarios, such as single particle showers with fixed impact angles (usually normal to the calorimeter surface) and fixed positions within the detector volume. While these benchmarks are useful for assessing the basic capabilities of generative models, they do not show how these models perform in a realistic experimental setup. 

A notable exception is the ATLAS experiment, where GANs are already used in production~\cite{ATLAS:2021pzo, ATLAS:2022jhk}, with NFs and diffusion models being explored for use in its next generation of fast simulation~\cite{ATL-SOFT-PUB-2025-003}. However, sufficient performance of generative calorimeter surrogates in realistic experimental setups has yet to be demonstrated for highly granular calorimeters, such as those designed for the CMS HGCAL~\cite{CMS_hgcal} and future collider experiments~\cite{ILDConceptGroup:2020sfq, Bacchetta:2019fmz}. These detectors are able to resolve significantly finer substructure in calorimeter showers, and therefore require generative surrogates to operate on orders of magnitude higher data dimensionality and sparsity.

To address this gap, we introduce \DDML~\cite{ddml_zenodo}, a flexible library designed to integrate generative models into the full simulation chain of various particle physics experiments. It is built on top of the \ddhep~\cite{Frank:2014zya} toolkit which is widely adopted in the HEP community. Using \DDML, we present the first full physics benchmark for highly granular calorimeter surrogates, studying the International Large Detector (ILD)~\cite{ILDConceptGroup:2020sfq} as an example. While ILD is used as a case study of a highly granular detector, the \DDML library already supports several detector concepts for future colliders, and could easily be adapted to other detector designs. We integrate two state-of-the-art generative models, \LFlows~\cite{Buss:2024orz} and \CCiii~\cite{buss2025caloclouds3ultrafastgeometryindependenthighlygranular}. 
 These models are trained solely on Geant4 photon showers in an idealized detector geometry.
 We use these surrogate models and the traditional Geant4 simulation and compare the simulation and reconstruction results with three physics benchmark datasets: single-photon showers, di-photon separation, and $\tau$-pair events.
In addition, we provide comparisons to three idealized models that directly sample from the ground-truth full simulation to analyze the effects that the choice of data representation and modeling assumptions have.

We systematically evaluate these models on three post-reconstruction benchmarks, specifically chosen to test highly granular calorimeter surrogates for electromagnetic showers. Beginning with a typical evaluation of single particle performance, we then introduce a first multi-particle benchmark in this context by studying di-photon separations, before performing a full physics benchmark based on hadronic decays of the tau lepton.

The remainder of the paper is structured as follows. Section~\ref{sec:Train_Valid_data} describes the datasets and benchmarks used in this work. Section~\ref{sec:Models} provides descriptions of the \LFlows and \CCiii models which are the subjects of this study. Section~\ref{sec:Results} presents the results of our benchmarks for single particles, di-photon events, and tau decay events. Finally, Section~\ref{sec:Discussion} provides a discussion.

\section{\label{sec:Train_Valid_data} Datasets and Benchmarks}

This section describes all datasets used in this study. Section~\ref{sec:ILD} introduces the ILD detector concept, which relies on highly granular sampling calorimeters. The different calorimeter shower representations studied are introduced in Sections~\ref{sec:Representations}, together with \geant reference generators for each representation. The training dataset for the surrogate models is introduced in Section~\ref{sec:TrainingDataset}. Section~\ref{sec:BenchmarkDatasets} introduces the approach to benchmarking adopted in this study, followed by the datasets used for single-particle validation, the di-photon benchmark, and the tau physics benchmark.

\subsection{\label{sec:ILD} The International Large Detector}
This work focuses on the International Large Detector (ILD) \cite{ILDConceptGroup:2020sfq}, a next generation detector for a future $e^{+}e^{-}$ Higgs factory, originally proposed for the International Linear Collider (ILC), and currently also under investigation for use at the FCC-ee. ILD is optimized for the Particle Flow approach to reconstruction, and as such features highly hermetic detector systems, highly granular calorimeters and a minimal material budget in front of the calorimeters. The ILD detector model studied in this work consists of a polyhedral barrel geometry with a total radius of $7.8$ \si{\meter} and a total length of $13$ \si{\meter}. The tracking system consists of a number of silicon pixel detectors, which are encapsulated in a time projection chamber (TPC). Outside of the tracking system are placed highly granular sampling calorimeters, separated into an electromagnetic calorimeter (ECAL) system and a hadronic calorimeter (HCAL) system. Both of these calorimeter systems consist of an octagonal barrel region with the ends closed by flat endcap disks. The primary focus of this study is the Si-W option for the ECAL \cite{CALICE:2008gxs}. This calorimeter consists of $30$ layers of passive tungsten absorbers and active silicon sensors. The first $20$ tungsten absorbers have a thickness of $2.1$ \si{\mm}, while the last $10$ layers have a thickness of $4.2$ \si{\mm}. The total thickness of the calorimeter therefore corresponds to approximately $24$ radiation lengths. The silicon layers feature cells of size $5 \times 5$ \si{\mm}$^2$, and have a thickness of $0.525$ \si{\mm}.
Behind the ECAL is placed the analogue hadronic calorimeter (AHCAL) \cite{CALICE:2010fpb}. It consists of $48$ layers with stainless steel absorbers, each with a thickness of $17.2$ \si{\mm}, and $3$ \si{\mm} thick active layers. The active layers feature $3 \times 3$ \si{\cm}$^2$ scintillator tiles, each individually read out by a silicon photomultiplier.
These detector systems sit inside a superconducting solenoid coil, which produces a magnetic field of strength $3.5$ \si{\tesla} orientated along the beam axis. An iron return yoke with integrated muon system and tail catcher calorimeter is placed outside of the coil.

\begin{figure*}[tbhh]
    \centering
    \includegraphics[width=0.43\textwidth]{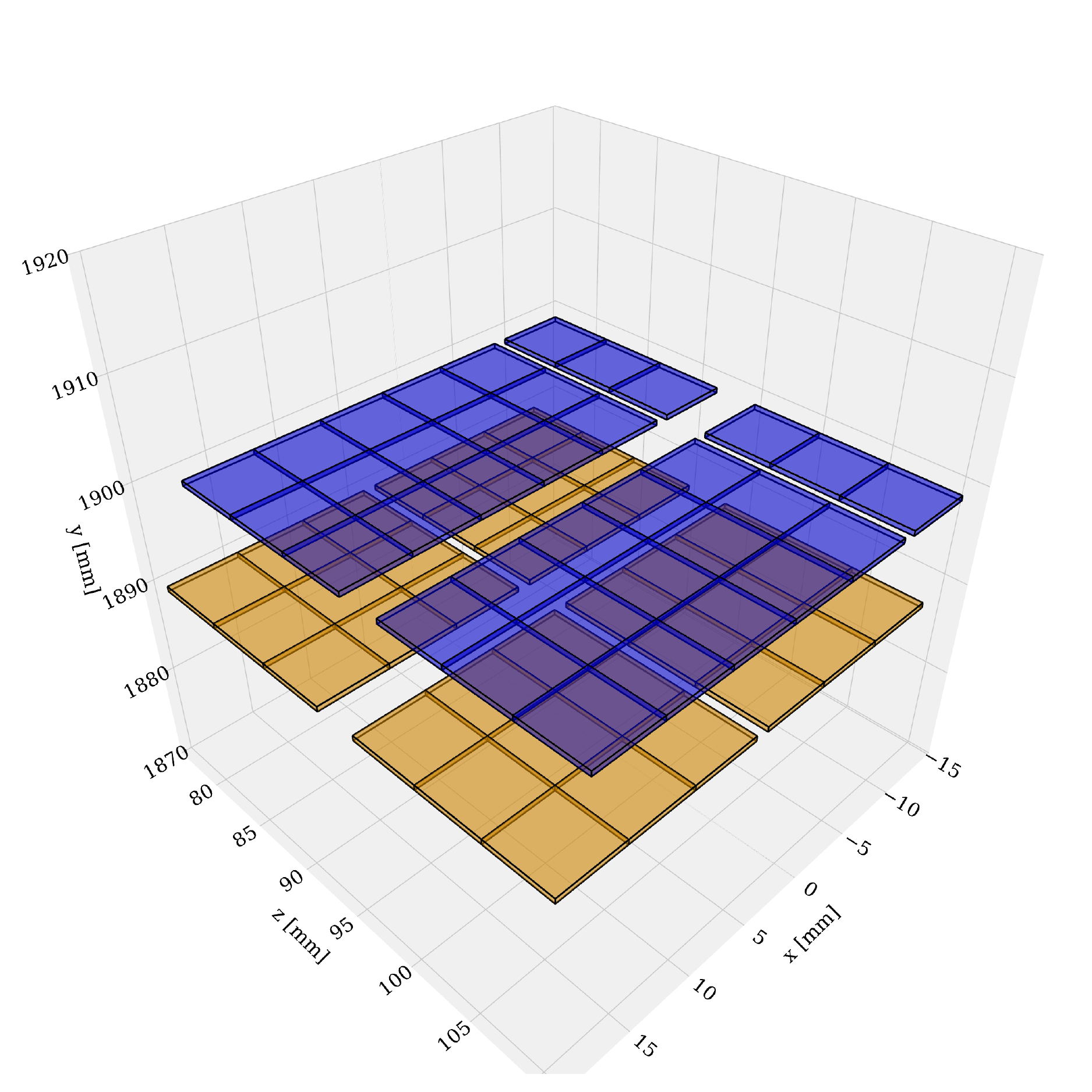}
    \includegraphics[width=0.43\textwidth]{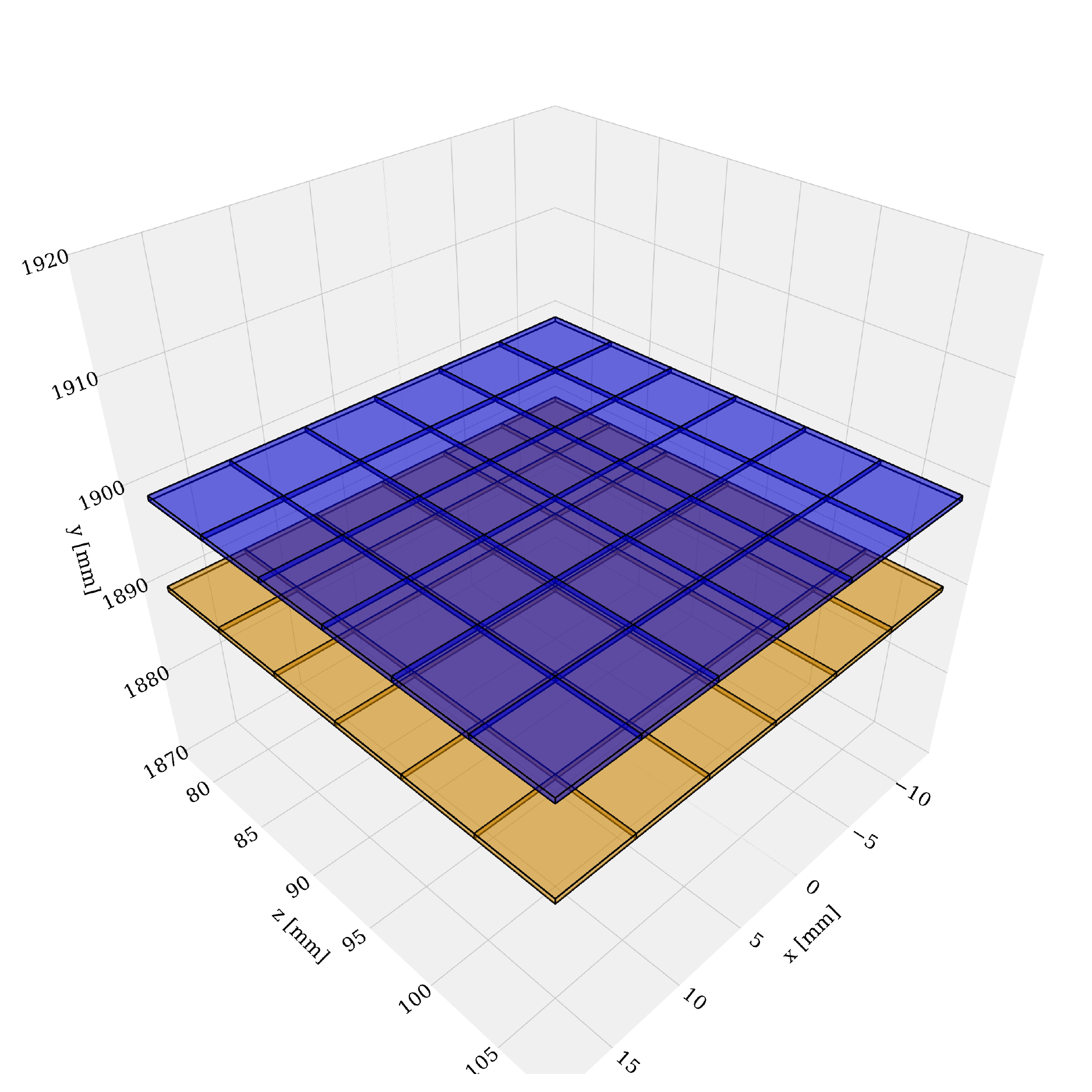}
    \caption[Visualization of geometry maps for (left) the physical geometry and (right) the regularized geometry for a section of two sensitive layers in the calorimeter. The physical geometry includes gaps between the cells arising from insensitive volumes such as structural supports and readout electronics, as well as a staggering effect between layers. The regularized geometry consists purely of sensitive material, with the cells being perfectly aligned from one layer to the next]{Visualization of geometry maps for (left) the physical geometry and (right) the regularized geometry for a section of two sensitive layers in the calorimeter. The physical geometry includes gaps between the cells arising from insensitive volumes such as structural supports and readout electronics, as well as a staggering effect between layers. The regularized geometry consists purely of sensitive material, with the cells being perfectly aligned from one layer to the next. Figure from \cite{McKeown:2024}.}
    \label{fig:Geometry_Map}
\end{figure*}

For this study, we employ the \textsc{Key4hep}~\cite{Key4hep:2022jnk} software ecosystem, using \geant \cite{GEANT4:2002zbu} version $11.2.2$ with the \textsc{QGSP\_BERT} physics list, and \ddhep~\cite{Frank:2014zya} version $1.30$. A realistic and detailed model of the ILD detector geometry\footnote{The version of the ILD detector geometry used in this study is \texttt{ILD\_l5\_o1\_v02}}, described using the \ddhep toolkit, is used.
Of particular relevance, is the geometrical structure of the sensitive layers of the ECAL, a visualization of which is shown for two active layers in Figure \ref{fig:Geometry_Map} (left). This illustrates that the readout geometry of the detector is irregular, featuring two types of insensitive volumes. The smaller insensitive volumes are present only in the active layers, and lie at the edge of the silicon wafers. A staggering effect is present in their positioning between layers. The larger insensitive volumes arise from the presence of gaps between sensors or structural supports in the calorimeter. These are therefore aligned between layers, and present in both the absorber and the active layers.
Such an irregular readout geometry, which will be present in every realistic calorimeter, creates a number of difficulties when conceiving a scheme to allow a model to be used to simulate showers at varying positions on and angles to the calorimeter face. For models relying on a regular grid representation of calorimeter layers, it would require a means of removing projection artifacts~\cite{Buhmann:2020pmy} for all possible incident positions and angles, which would be infeasible. An additional problem that would affect models generally, is the variation in the fraction of the active layer that is sensitive depending on the local geometry near the incidence position. Since \geant discards energy depositions that do not land in a region of the detector that is not assigned to be sensitive, this would result in the potential loss of information when trying to simulate a shower at a different position than the one the model was originally trained at.

To combat this challenge, a modified \ddhep description of the ILD ECAL with a regularized readout geometry was created, as shown in Figure \ref{fig:Geometry_Map} (right). In this geometry, the readout segmentation of the sensitive layers was altered such that the layer contained no insensitive volumes, meaning that all energy depositions in these layers are recorded. This alteration leaves the structure of the detector, both in terms of material composition and longitudinal layer placement (i.e. into the depth of the calorimeter), unaffected. By using this regularized version of the calorimeter to create a training dataset, maximum information about energy depositions in the sensitive layers can be retained, allowing them to be dropped selectively during simulation (see Section \ref{sec:Integration}), depending on the local readout geometry and thereby avoiding artifacts from sensitive gaps in the training showers.

\subsection{\label{sec:Representations} Shower Representations}
\geant produces individual energy deposits, so called \textit{\geant steps}, with a much higher spatial resolution than the physical detector readout. If a generative model is to be used to simulate showers across different incident positions, intuitively having a mechanism for generating showers at a higher resolution than the detector readout should reduce artifacts and edge effects. To better understand the consequences and potential limitations of projecting showers into a realistic detector geometry, we consider three different shower representations, 
each accompanied by a corresponding truth-based reference, which we denote as \textit{optimal shower generators}: \textsc{Optimum (x1)}, \textsc{Optimum (x9)}, and \textsc{Optimum (steps)}.

Each of these optimal shower generators is derived from simulations run with \geant on the regularized 
ILD ECAL introduced in section \ref{sec:Train_Valid_data}, from which all individual \geant steps within the sensitive layers are extracted. This enables us to quantify the effect of projecting a regular grid with a given granularity onto the actual detector geometry, and allows us to isolate the effects of the data representation from the performance of a given generative model.

\begin{figure*}[tbhh]
    \centering
    \includegraphics[width=0.9\textwidth]{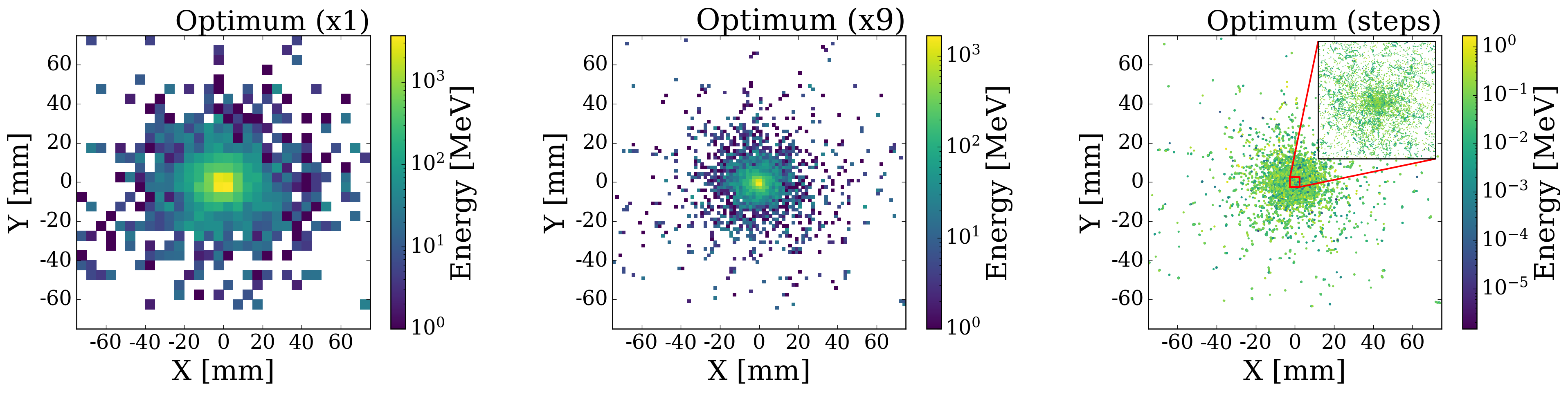}
    \caption{Visualization of the same 90 GeV electromagnetic shower in lateral projection of the ILD ECAL, represented using the three optimal shower generators: \textsc{Optimum (x1)} (left), \textsc{Optimum (x9)} (center), and \textsc{Optimum (steps)} (right).}
    \label{fig:single_shower}
\end{figure*}

The first representation $\mathcal{R}_{\times1}$ features a granularity identical to that of the ILD ECAL ($5 \times 5$ \si{\mm}$^2$), and has a corresponding optimal shower generator \textsc{Optimum (x1)} shown in Figure \ref{fig:single_shower} (left). Here, each simulated step from \geant is projected onto a virtual regular grid, exactly matching the realistic detector cell sizes. Thus, \textsc{Optimum (x1)}, provides the ideal reference for the performance achievable with any model trained on the readout geometry of the physical detector.

The second representation $\mathcal{R}_{\times9}$ increases the lateral granularity by a factor of three in each dimension, resulting in nine times more cells ($1.67 \times 1.67$ \si{\mm}$^2$) per layer compared to $\mathcal{R}_{\times1}$. This representation has an optimal shower generator \textsc{Optimum (x9)}, shown in Figure \ref{fig:single_shower} (center). An increased granularity such as this allows for a finer spatial resolution of showers, reducing projection-related effects. It serves as a reference for understanding the impact of increasing the granularity of the data representation on the fidelity of the projected showers.

Finally, representation $\mathcal{R}_{\text{(steps)}}$, with corresponding generator \textsc{Optimum} (steps) shown in Figure \ref{fig:single_shower} (right), provides the most detailed reference scenario by avoiding any spatial projection onto a predefined grid altogether. Instead, it directly utilizes the ultimate resolution -- \geant simulation steps -- as they occur within the sensitive material. This results in the highest achievable spatial resolution, completely free from projection artifacts, and represents the ultimate benchmark for evaluating the accuracy and potential information loss associated with any spatial discretization scheme in the regularized detector geometry.

For all optimal shower generators, a wide bounding box is used to select shower hits, with a side length of $800$ \si{mm}. This box cut is necessary to exclude rare low energy hits resulting from backscatter that typically occur at the opposite end of the detector to which the showers occur. This ensures that all relevant hits in the shower are contained.

By leveraging these optimal shower generators -- \textsc{Optimum (x1)}, \textsc{Optimum (x9)}, and \textsc{Optimum} (steps), we gain insight into the intrinsic limits imposed by projection artifacts, independent of generative model performance. Moreover, they provide invaluable baselines against which generative models can be assessed.

\subsection{Training Dataset}\label{sec:TrainingDataset}

The training dataset used in the study consisted of $\sim3$ million samples, created by photons with incident energies uniformly distributed in the range of  1-126 GeV. The photons were created at a position of $[x=0,~y = 1804.7~\text{\si{\mm}},~z = - 50~\text{\si{\mm}}]$ at the front face of the ECAL. Here the ILD global coordinate system is defined with the $z$ axis along the beam direction, the $x$ axis pointing horizontally, and the $y$ axis pointing vertically upwards. The incident angles were varied within a cone of up to $60^\circ$ in $\theta'$ (relative to the normal to the calorimeter layers), with an azimuthal angle $\varphi'$ uniformly distributed in $[0^\circ, 360^\circ)$. These angles are defined in the local frame of reference, where the $z'$ axis is aligned with the normal to the calorimeter layers. 

This configuration enabled uniform sampling over the space of possible incident directions on the calorimeter face, and conforms with the coordinate system convention used in the DDML library, which is described in Appendix~\ref{sec:Integration:DDML}\footnote{Throughout this paper, we use primed coordinates to represent the local calorimeter coordinate system, otherwise coordinates are assumed to be in the global ILD coordinate system.}.

\subsection{Benchmark datasets}\label{sec:BenchmarkDatasets}

We introduce the generic library DDML~\cite{ddml_zenodo}, which allows the inclusion of different generative models designed for fast calorimeter shower simulation in full simulation applications using the \ddhep toolkit~\cite{Frank:2014zya}. This library and the details of the implementation for the ILD detector used in this study are described in Appendix~\ref{sec:Integration}. Crucially, this allows the generation of particle showers in the standard software chain of ILD. This makes it possible to run the standard reconstruction of ILD, in particular particle flow reconstruction with \pandora~\cite{Marshall:2015rfa}, enabling realistic physics benchmarking of generative models. All benchmarking samples used in this paper have gone through the complete software chain including event reconstruction.

\subsubsection{Benchmarking Methodology}
While studying single-shower observables, as is now standard in the literature, is important to gauge the performance of a model, in real physics events showers from multiple particles may overlap. This significantly increases the complexity of evaluating the performance of a model, as the breadth of the phase space makes disentangling the interplay of overlapping showers with reconstruction algorithms a challenging task. For this reason we take a step-by-step approach, ultimately building towards benchmarking the model in a full physics setting. 

We begin by studying the performance of the model in terms of single particle observables. Next, we study the simplest scenario for a multi-particle test -- two photons fired into the face of the ECAL. This is a standard benchmark which has also been used for the development of reconstruction algorithms, such as \pandora \cite{Xu:2016rcz}. This approach provides an isolated and controllable means of probing generative model performance, as well as allowing connections to be drawn to the performance on single particle observables.

Finally, we study the performance of generative models for the simulation of photon showers in a full physics process. As the performance required of a fast simulation tool will depend heavily on the physics process for which the tool is used, we desire a physics process that provides a stringent test of a fast simulation tool for electromagnetic showers. To this end, we choose hadronic decay modes of the tau lepton in the process $e^{+}e^{-}\rightarrow \tau^{+} \tau^{-}$.

The tau lepton is of interest for many precision studies planned for future $e^{+}e^{-}$ collider experiments \cite{Dam:2021ibi, Calibbi:2021pyh, Jahedi:2024kvi, Gutierrez-Rodriguez:2022mtt, deBlas:2019rxi}. As a result of its high mass, it has the strongest coupling to the Higgs of any lepton, and is the only lepton in the Standard Model which decays to hadrons. This, combined with the ability to precisely reconstruct the spin state of the tau from its decay products \cite{Tran:2015nxa, Jeans:2015vaa}, makes it a prime candidate for probing the Higgs sector, including its CP structure \cite{Berge:2013jra, Jeans:2018anq}.

Approximately $65$\% of tau decays involve hadrons, with the reduced number of neutrinos involved compared to purely leptonic decays meaning that hadronic decay modes are preferred for precision measurements. Hadronic decay modes of the tau frequently involve one or more neutral pions, often occurring via the $\rho(770)$ or $a1(1200)$ intermediate resonances \cite{ParticleDataGroup:2024cfk}. Reconstructing the correct decay mode of the tau, which is essential when determining its spin state, therefore often involves correctly reconstructing the number of $\pi^{0}$s produced, each of which almost always decay via $\pi^{0}\rightarrow \gamma \gamma$~\cite{ParticleDataGroup:2024cfk}. This is challenging, due to the high boost and collimation of the decay products, making the reconstruction of hadronic tau decays a standard benchmark of the performance of an electromagnetic calorimeter \cite{Tran:2015nxa,Jeans:2015vaa}. The presence of numerous overlapping photon showers in such events therefore makes them ideal for exposing any flaws in the performance of a fast simulation tool for electromagnetic calorimeters. Correctly reconstructing these $\pi^{0}s$ involves not only distinguishing the number of photons from overlapping showers (the performance of which can be linked to the aforementioned isolated di-photon benchmark), but also correctly inferring the kinematics from shower-level observables (which can be linked to the aforementioned single particle benchmark). 

\subsubsection{Single Particle Dataset}\label{SingleParticle:Dataset}
In order to validate the generalization capability of the trained models, independent single shower test samples were generated at multiple positions, uniformly distributed over a single ECAL barrel segment. Test samples were produced at fixed photon energies between 10 and 100 GeV in 10 GeV increments. For each test energy, a sample consisted of 3,000 photon showers with positions selected randomly on the front surface of the ECAL. The incident directions were chosen such that they mimicked particles originating from the interaction point (IP), across the angular ranges $43^\circ < \theta < 137^\circ$ and $79^\circ < \varphi < 109^\circ$ in the ILD global coordinate system, thereby effectively covering one complete stave of the barrel.
This configuration ensures full coverage of the region where models are expected to operate, exposing them to a wide variety of incident directions and local geometries of the sensitive layers -- thereby enabling the evaluation of model performance under realistic conditions.

\subsubsection{Di-photon Dataset}\label{DiPhoton:Dataset}
The di-photon benchmark dataset consists of three sets of 15,000 samples containing two photons. The three sets of samples have different incident photon energies of 5 GeV, 20 GeV and 100 GeV. Within each sample, the energies of each of the two photons are identical.

The photons are produced directly at the face of the ILD ECAL barrel, with their direction of flight being orientated such that they appear to have been produced at the IP\footnote{This mimics photons coming from the IP, e.g. from $\pi^0$ decays, without having to address material interactions, like pair-creation, in the tracking detector on the way to the calorimeter.}.

The photon positions are randomly sampled to expose the di-photon system to different local geometries of the sensitive layers, while ensuring that the separation between the two photons varies uniformly from 0 to 90 mm.
The photons impact the upper barrel module, within a narrow angular range of $\theta \in [83^\circ, 87^\circ]$ and $\phi \in [88^\circ, 92^\circ]$, corresponding to a small localized patch of the detector. To ensure that all photon pairs remain fully contained within the fast simulation trigger region (see Section \ref{sec:Integration:DDML}) and avoid contamination from showers simulated with \geant, the center of this patch is placed well inside the boundaries of the trigger region. These samples are created for \geant,  and each of the shower generation approaches under study. In each sample, both showers are produced with the respective generator.

\subsubsection{Tau Physics Dataset}\label{TauBench:Dataset}

The dataset used in this study consists of samples of the process $e^{+}e^{-} \rightarrow \tau^{+} \tau^{-}$ in an ILC running scenario at a center-of-mass energy of $250$ GeV. MC Generator samples, provided by the ILD Software Working Group in the MC-2020 production \cite{MC2020}, were created with WHIZARD \cite{Kilian:2007gr} version $2.8.5$. A realistic ILC beam energy spectrum and crossing angle, as well as the effects of bremsstrahlung and initial state radiation were included. All samples contained beams of $100$\% polarized left-handed electrons and right-handed positrons ($e_{L}^{-}e_{R}^{+}$). The decay of the tau leptons in the samples was simulated with the TAUOLA library \cite{Jadach:1990mz}.

In order to enable a direct comparison between the various shower representations and models described in Sections \ref{sec:Representations} and \ref{sec:Models}, the same set of MC generator inputs were used in all cases. This means that for each case, only the detector simulation differs, removing any differences in underlying event topologies or physics processes. In addition, no background is overlaid onto events. These two choices reduce event dilution that would only serve to obscure the performance of the calorimeter shower simulators. However, it should be noted that the detector simulation itself can significantly alter the signature of the event depending on what interactions occur prior to the calorimeter. A particularly pertinent example for this study is the case in which a photon converts into an electron-positron pair. This case provides a potentially easier reconstruction scenario, as the charged electron and positron have an associated track and larger separation at the calorimeter face.

Several selection criteria were put on the events, to further enhance the sensitivity of the analysis to the performance of the calorimeter shower simulators. It was required that all events contained at least one $\pi^{0}$ produced in a tau decay, with the $\pi^{0}$ then decaying into two photons. Both of the photons were required to have an energy above $5$ GeV and to satisfy the geometry region trigger described in Appendix~\ref{sec:Integration:ILD_DDML}. Detector simulation was then performed for each of the various shower representations and models described in Sections~\ref{sec:Representations} and \ref{sec:Models}. The software configuration described in Section~\ref{sec:ILD} was employed, using the \DDML implementation described in Appendix~\ref{sec:Integration:ILD_DDML}. All photons with an energy above $5$ GeV incident on the electromagnetic calorimeter which passed the geometry region trigger were then simulated with the appropriate simulator. 

For each calorimeter shower simulator, samples were created with three different random seeds for the detector simulation. This provides a means of estimating the uncertainties on post-reconstruction physics observables that would otherwise be difficult to estimate, given the high level of correlation arising from the use of identical generator level input. In total this meant that samples, consisting of $3$ random seeds each containing $6791$ events, were generated for \geant, each of the optimal shower generators \optimumXOne, \optimumXNine, \optimumSteps described in Section~\ref{sec:Representations} and for both the \CCiii and \LFlows models, which will be described in Section~\ref{sec:Models}. The standard ILD reconstruction chain, as described was then applied to each sample.

\section{\label{sec:Models} Models}

We compare two state-of-the-art generative models operating on different data representations -- \LFlows, a regular grid-based architecture, and \CCiii, a point cloud model. Each model is trained on different shower representations described in Section \ref{sec:Representations}.

\subsection{\label{sec:Models:L2LFlows} Convolutional L2LFlows}
\LFlows~\cite{Buss:2024orz} is a generative surrogate based on normalizing flows~\cite{Dinh:2014mzt,Durkan:2019nsq}, designed for fast and accurate simulation of electromagnetic showers in calorimeters. It operates on a fixed-grid representation, where each shower is discretized into a three-dimensional grid. This voxelized representation allows for convolution-based architectures, such as U-Nets~\cite{Ronneberger:2015}, to effectively model the complex spatial dependencies in calorimeter showers.

\LFlows generates calorimeter responses sequentially, layer by layer, where the generation of each layer is conditioned on the previous ones. The underlying flow-based architecture allows for single-shot sampling. Further details of the model architecture can be found in~\cite{Buss:2024orz}.

As demonstrated in the CaloChallenge 2022~\cite{Krause:2024avx}, \LFlows achieves one of the best trade-offs between accuracy and generation speed among the submitted models on a fixed-grid dataset, making it well-suited for fair comparison between fixed-grid and point-cloud-based models.

While the original \LFlows model was restricted to a fixed incident point and angle, we have extended its capabilities with respect to these conditions to enable its application in this study. Firstly, the model is trained on the $\mathcal{R}_{\times9}$ representation of showers described in Section~\ref{sec:Representations}. Given that the granularity present in this representation is significantly finer than the detector readout, it was necessary to impose a bounding box of side length $150$ \si{mm}. While this produces a noticeable cut in the tails of the shower, it is necessary to handle the sparsity present in the shower and constrain the model size within reasonable limits. The resulting grid has $90\times90\times30$ voxels. Secondly, conditioning on the incident angle enables the model to generate
showers for a range of impact angles, rather than being limited
to a $90$-degree impact angle. Technically, the angle is given as
a unit vector to preserve the topological structure. To prevent
showers from developing outside the bounding box, all layers
are shifted to center the shower core. This effectively results
in a tilted bounding box. These improvements broaden the
applicability of \LFlows, enabling it to be used for a
large fraction of highly energetic photon showers in the ILD
ECAL.

\subsection{\label{sec:Models:CaloClouds} CaloClouds}
The second model is \CC~\cite{Buhmann:2023bwk}, a point cloud generative model that was developed to address the challenges induced by the irregular structure of the detector readout layers described in Section \ref{sec:Train_Valid_data}, and to more efficiently handle the very high sparsity present in highly granular calorimeter showers.
In this study we employ a third iteration of the model, denoted as \CCiii~\cite{buss2025caloclouds3ultrafastgeometryindependenthighlygranular}, which constitutes an extension of the \CCii~\cite{Buhmann:2023kdg} architecture.

\CCiii utilizes a grid with a 25x higher granularity than the original cells, which we refer to as the $\mathcal{R}_{\times25}$ representation. Additional preprocessing is applied to dequantize the hit positions, together with cuts which enforce a bounding box of side width $500$ \si{mm} for shower hit selection. This box cut only removes the outer tails of the shower very far from the shower core. For more details on preprocessing, see~\cite{buss2025caloclouds3ultrafastgeometryindependenthighlygranular}.

Like the improved \LFlows model, this version of \CC also incorporates angular conditioning to generate showers for a range of incident directions. This is achieved through a similar data preprocessing and conditioning methodology as used for \LFlows. The angular conditioning is achieved by explicitly providing the unit vector of the particle's momentum direction as additional conditioning parameter, enabling the model to account for the direction of the incident particle during generation. Such a capability is essential for realistic applications within the simulation chain, where different directions of incoming particles are expected.

In addition to this functional advancement, \CCiii features a simplified architecture. Both the \textsc{PointWiseNet} of the diffusion model and the \textsc{ShowerFlow} components have been optimized, resulting in a reduction in the inference time by a factor of $\sim2$ compared to the previous iteration, \CCii, while preserving the generative fidelity. \CCiii is described in more detail in a dedicated paper~\cite{buss2025caloclouds3ultrafastgeometryindependenthighlygranular}.

\section{Results}\label{sec:Results}

\begin{figure*}[tbhh]
    \centering
    \includegraphics[width=0.4\textwidth]{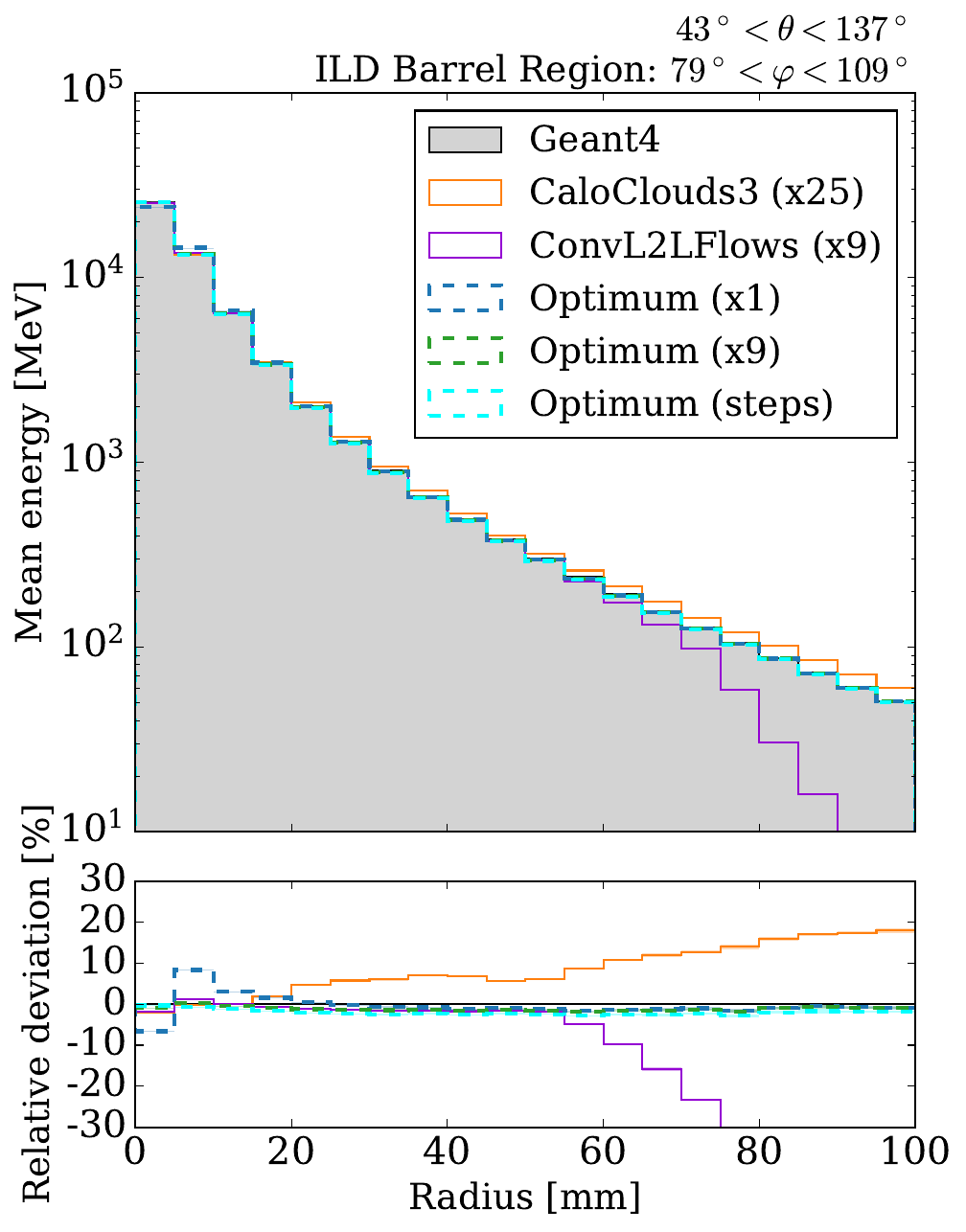}
    \includegraphics[width=0.398\textwidth]{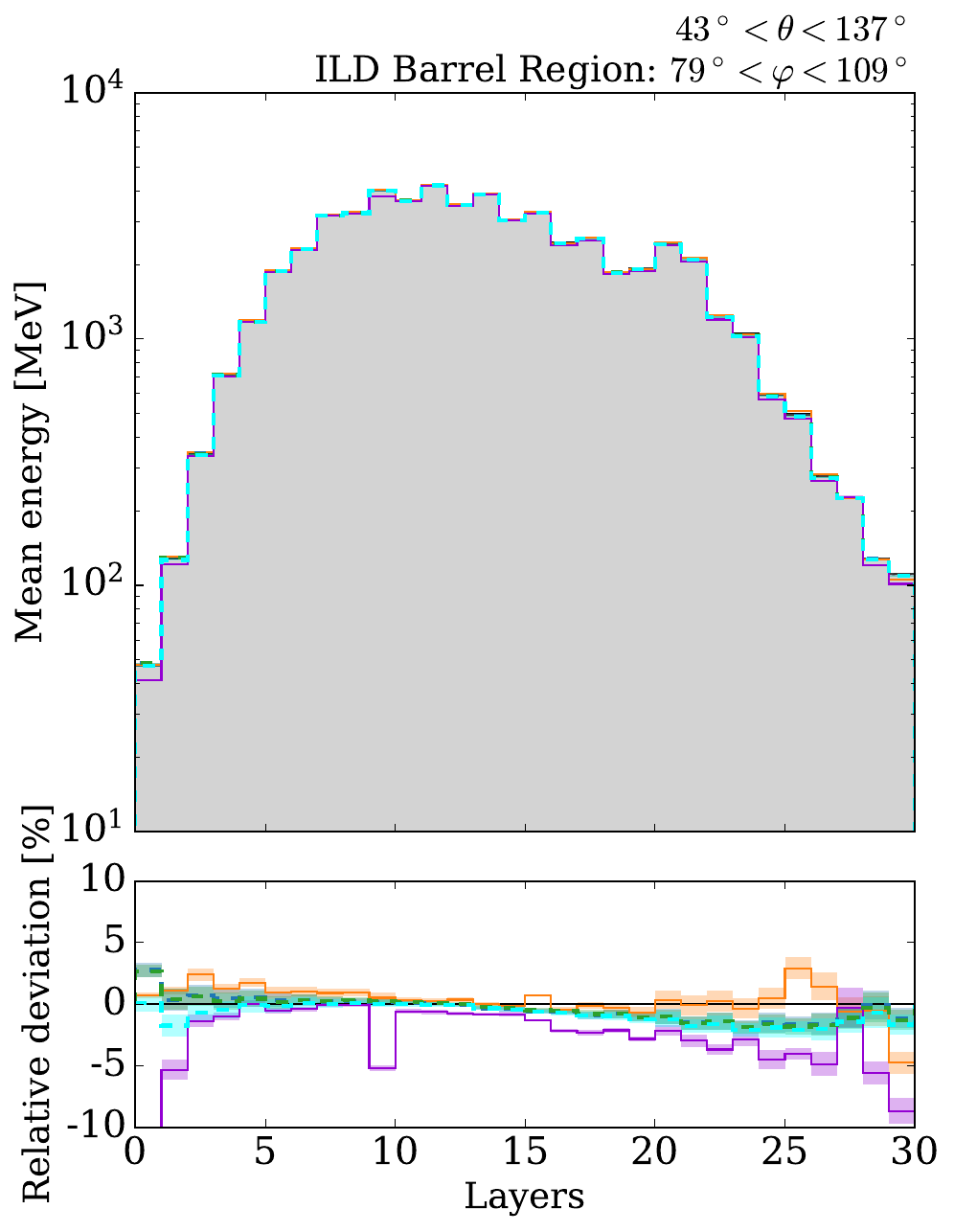}
    \caption{
        Radial (left) and longitudinal (right) energy profiles of electromagnetic showers, computed at the reconstruction level after integration of the generative models.
        The radial profile shows the mean reconstructed energy as a function of distance from the shower axis, while the longitudinal profile shows the mean reconstructed energy per calorimeter layer.
        These observables provide a detailed characterization of the transverse and longitudinal shower structure and are critical benchmarks for assessing how well generative models replicate \geant showers in realistic detector geometry settings.
        The color coding corresponds to the different generative models: \CCiii (orange), \LFlows (violet), \optimumXOne (blue), \optimumXNine (green), and \optimumSteps (cyan).
        The \geant reference is shown in the light grey filled histogram.
        The color coding is consistent across all figures in this section.
        Shaded bands indicate statistical uncertainties; lower panels show relative deviations with respect to the \geant baseline.
    }
    \label{fig:radial_longitudinal_energy}
\end{figure*}

We present the results of simulator performance for each benchmark, beginning with the single particle performance in Section~\ref{sec:SingleParticle}, followed by the di-photon performance in Section~\ref{sec:DiPhoton:Results}, and finally the results of the full physics benchmark using tau decays in Section~\ref{sec:PhysicsBenchmark:Results}. We conclude the section by presenting the results for single shower generation times in Section~\ref{sec:Timing}.

\subsection{\label{sec:SingleParticle} Single Particle Performance}

A detailed validation of individual particle showers is crucial to ensure that these models correctly capture essential physics characteristics before being used in more complex, multi-particle scenarios or physics analyses.

We present an evaluation of key calorimetric observables for electromagnetic showers generated by single photons using the generative models described previously. We study radial and longitudinal shower profiles, energy resolution, linearity, and the intrinsic shower angle reconstruction. These observables are usually chosen in the literature because they encapsulate the essential features of electromagnetic shower development, and significantly influence the performance of particle identification and reconstruction algorithms.

 We further benchmark our generative models against the optimal shower generators introduced in Section~\ref{sec:Representations}. These optimal shower generators represent idealized performance scenarios that isolate and quantify the intrinsic limitations arising from spatial discretization effects and detector irregularities, independent of the generative model itself. By comparing the performance of the generative models against both these optimal shower generators and the standard \geant simulation, we are able to clearly distinguish between artifacts arising from the data representation and the intrinsic capabilities of the approaches to generative modeling explored. 

This structured approach enables a detailed assessment of generative model fidelity and highlights areas requiring further improvement.

\subsubsection{Shower Profiles}

We begin the performance evaluation of the generative models by examining their ability to reproduce the characteristic EM shower shapes, assessed through comparisons of radial and longitudinal energy profiles, which are key observables reflecting the spatial energy deposition pattern within the calorimeter.

The following observables are computed at the reconstruction level, after the generative models have been fully integrated into the simulation pipeline.
This ensures that any potential geometric or systematic effects introduced by the integration framework are accounted for in the performance evaluation.

The radial energy profile, shown in Figure~\ref{fig:radial_longitudinal_energy} (left), illustrates the mean deposited energy as a function of orthogonal distance from the axis aligned with the direction of the incident particle to the center of the cell.
The longitudinal energy profile, displayed in Figure~\ref{fig:radial_longitudinal_energy} (right), represents the average energy deposited per calorimeter layer along the depth of the detector.

Both models reproduce the longitudinal profile with good accuracy.
The \CCiii model achieves the closest agreement with \geant, with deviations typically within a few percent.
The \LFlows model performs similarly well, although it exhibits deviations of up to $15\%$ near the start and end of the calorimeter. Most of these deviations are due to the finite simulation volume required for fixed grid models. 

At first inspection, the radial energy profile appears to be well reproduced only by the optimal shower generators -- \optimumXOne, \optimumXNine, and \optimumSteps, while the \CCiii and \LFlows models have notable discrepancies at larger radial distances.
However, it is important to note that the radial profile of electromagnetic showers is inherently steep and narrowly peaked, with the energy density rapidly decreasing with distance from the shower axis.
In fact, more than 90$\%$ of the total energy of the shower is contained within a radius of 30~\si{\milli\meter}, indicating that deviations at large radii have a limited impact on the overall shower description.

\begin{figure}[tbhh]
    \centering
    \includegraphics[width=0.4\textwidth]{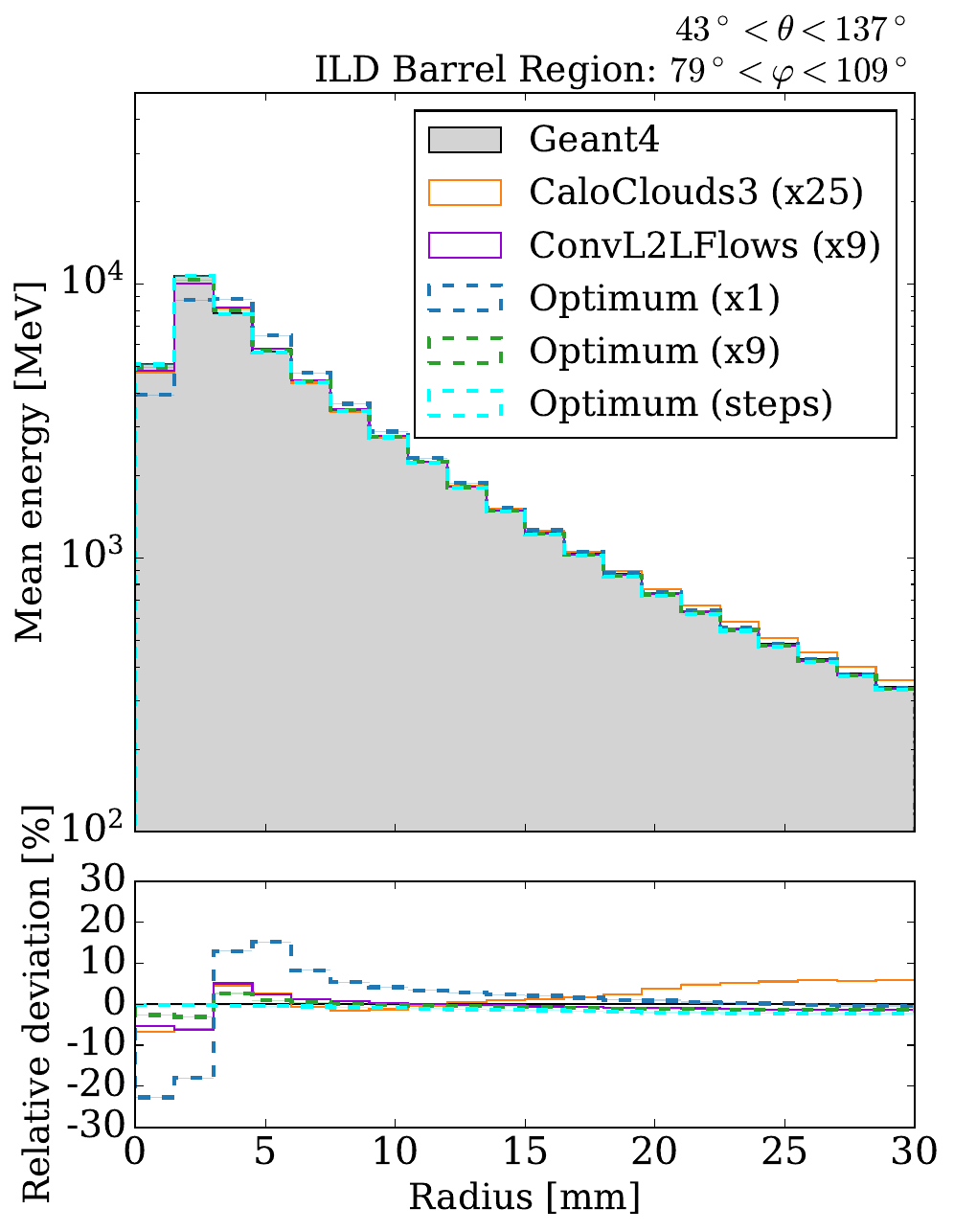}
    \caption{
        Radial energy profile of the showers, zoomed in to the first 30 mm from the shower axis.
        The shaded error bands correspond to the statistical uncertainty in each bin.
        The lower subplot shows the relative deviation of the radial energy profile with respect to the \geant reference.
        The color coding is consistent with Figure~\ref{fig:radial_longitudinal_energy}.
    }
    \label{fig:radial_energy:30mm}
\end{figure}

To better assess the fidelity of the models in the region that dominates the shower energy density, Figure~\ref{fig:radial_energy:30mm} zooms into the first 30 mm from the shower axis. Note that the dip in the first bin arises from the fact that the binning here is applied at a distance less than the width of a cell, whereas the hits in the shower are necessarily at the center of a cell after reconstruction.
This is the most relevant part of the shower which plays a crucial role in the separation of overlapping showers during particle reconstruction.

In this region, the \CCiii and \LFlows models show good agreement with \geant, with relative deviations generally remaining below $10\%$. By contrast, \optimumXOne underestimates the energy density by up to $20\%$ in the innermost bins.
This demonstrates the intrinsic limitation imposed by the $\mathcal{R}_{\times1}$ representation, where the coarser lateral granularity fails to resolve the sharply peaked energy profile near the shower axis.
The lack of detail in this region can significantly impact the reconstruction of particle showers with a large degree of overlap. This highlights a significant disadvantage of generative models trained on fixed grids using the true detector granularity.

\subsubsection{Resolution and Linearity}\label{sec:EnergyResolution}

\begin{figure*}[tbhh]
    \centering
     \includegraphics[width=0.4\textwidth]{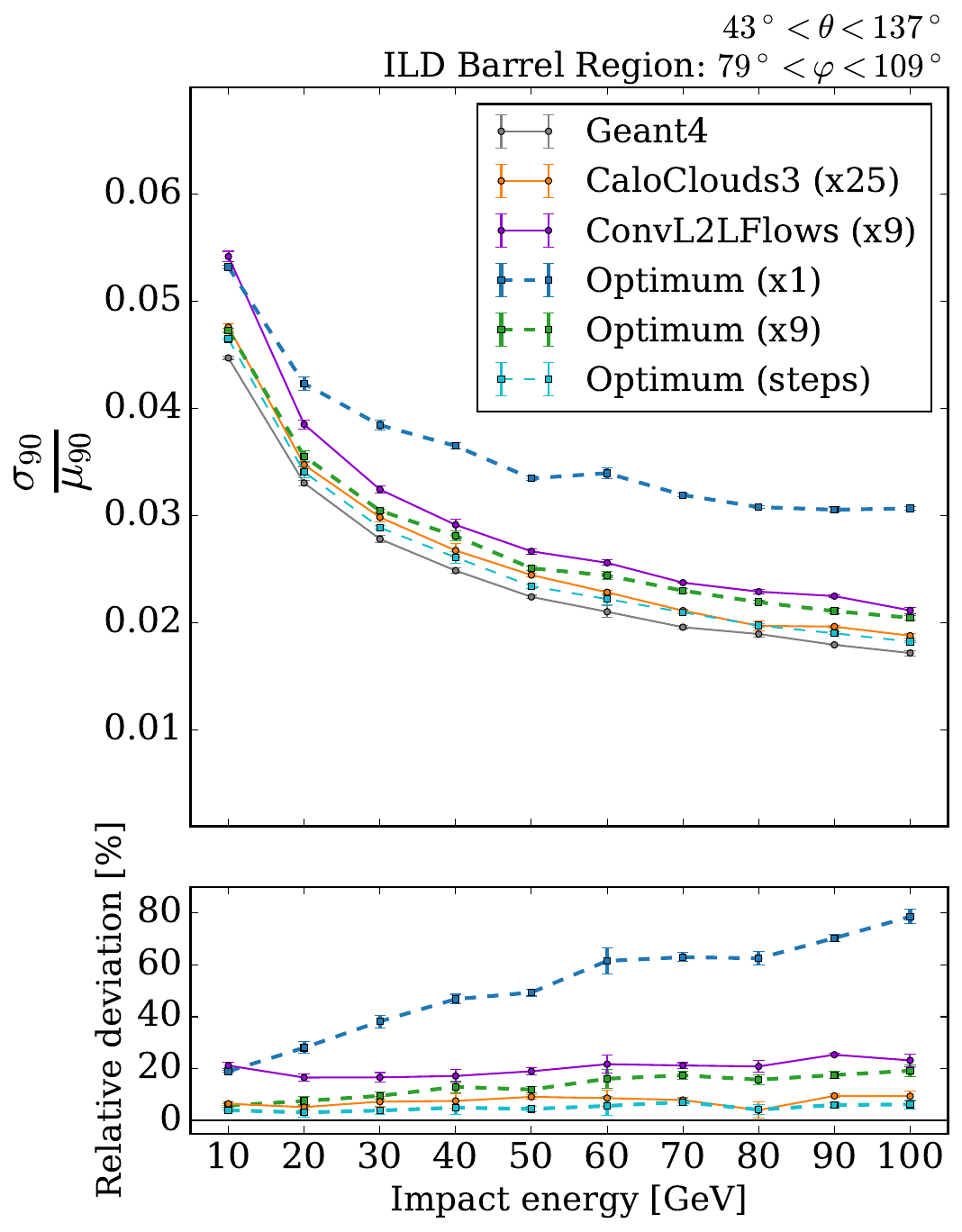}
    \includegraphics[width=0.4\textwidth]{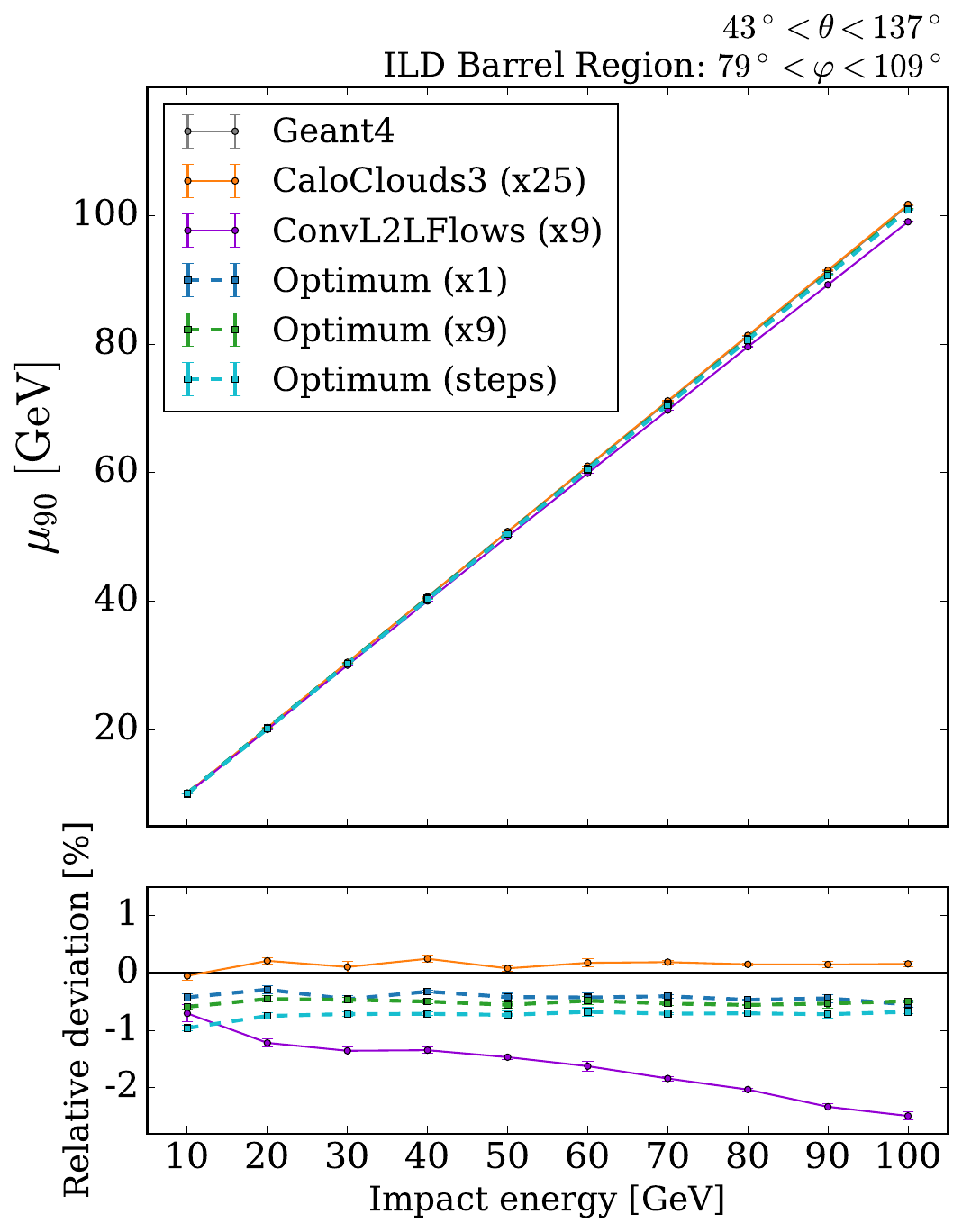}
    \caption{Energy resolution (left) and linearity (right) of reconstructed photon showers in the ILD ECAL. The resolution is defined as the relative width $\sigma_{90}/\mu_{90}$ of the central 90\% interval of the reconstructed energy distribution. The linearity is given by the mean $\mu_{90}$ of this central interval as a function of the incident photon energy.}
    \label{fig:resolution}
\end{figure*}

We now investigate the energy resolution and linearity of the generative models, two critical performance metrics for calorimeter simulation. The energy resolution quantifies the model's ability to accurately reproduce the fluctuations in the deposited energy, directly affecting the precision with which particle energies can be measured. Linearity assesses how accurately the reconstructed energy scales with the true particle energy, essential for ensuring unbiased energy measurements across a large range of incident particle energies.

The energy resolution is evaluated by measuring the relative width of the reconstructed energy distribution for photons at various fixed energies, defined as $\frac{\sigma_{90}}{\mu_{90}}$, where $\sigma_{90}$ and $\mu_{90}$ are standard deviation and mean of the central 90$\%$ of the distribution, shown as a function of the incident photon energy in Figure~\ref{fig:resolution} (left). Figure~\ref{fig:resolution} (right) displays the linearity, expressed as the mean reconstructed energy divided by the true incident energy.

Similar to the radial profile results, the resolution plot shows the same trend. Among the optimal shower generators, \optimumXOne performs the worst, showing significant deviation from the \geant baseline. This demonstrates that the coarse $\mathcal{R}_{\times1}$ representation lacks sufficient granularity to accurately capture the intrinsic energy fluctuations of electromagnetic showers. Despite being derived from full \geant simulation, this representation inherently limits the achievable fidelity due to the loss of fine spatial information.

With increased granularity, \optimumXNine shows improvement, more closely tracking the \geant resolution. Finally, \optimumSteps, which directly uses the individual \geant steps without any spatial discretization, comes closest to reproducing the full simulation, representing the maximum achievable performance.

Importantly, the two generative models follow the trends of their respective representations. The \CCiii model, trained on de-quantized $\mathcal{R}_{\times25}$, almost reaches the performance of \optimumSteps. \LFlows trained on the regular $\mathcal{R}_{\times9}$ data representation, performs comparably to \optimumXNine, although the deviations from the optimal representation are larger than for \CCiii.

These findings demonstrate that both models have successfully learned from their respective training data representations, achieving performance that closely matches the optimal shower generators derived from the same representation.

The linearity is reproduced well by all models being within a $\sim 3\%$ relative deviation from \geant, over the entire energy range tested -- from 10 to 100 GeV with 10 GeV steps. Notably, the \CCiii model exhibits the best agreement, showing negligible deviation from \geant. This is a result of a simple scaling applied during integration, where each generated shower is rescaled by a constant factor determined from the difference in average visible energy at 50 GeV between the model and \geant.

In principle, similar scaling could be applied to all models. However, for \LFlows, and any grid-based model in general, such calibration is more challenging due to its restricted generation region, where an increasing fraction the shower's energy leaks out with increasing energy of the incident particle. As a result, the discrepancy between generated and true visible shower energy becomes more pronounced at higher energies, limiting the effectiveness of global rescaling.

This highlights a fundamental trade-off faced by grid-based models -- they must balance performance against computational efficiency. Increasing the size of the generation volume can improve accuracy by reducing energy leakage, but it also significantly increases memory usage and inference time. As a result, achieving high fidelity with grid-based models requires careful tuning of the generation volume to remain computationally feasible while minimizing physical artifacts.

\begin{figure*}[!tbp]
    \centering
    \includegraphics[width=0.4\textwidth]{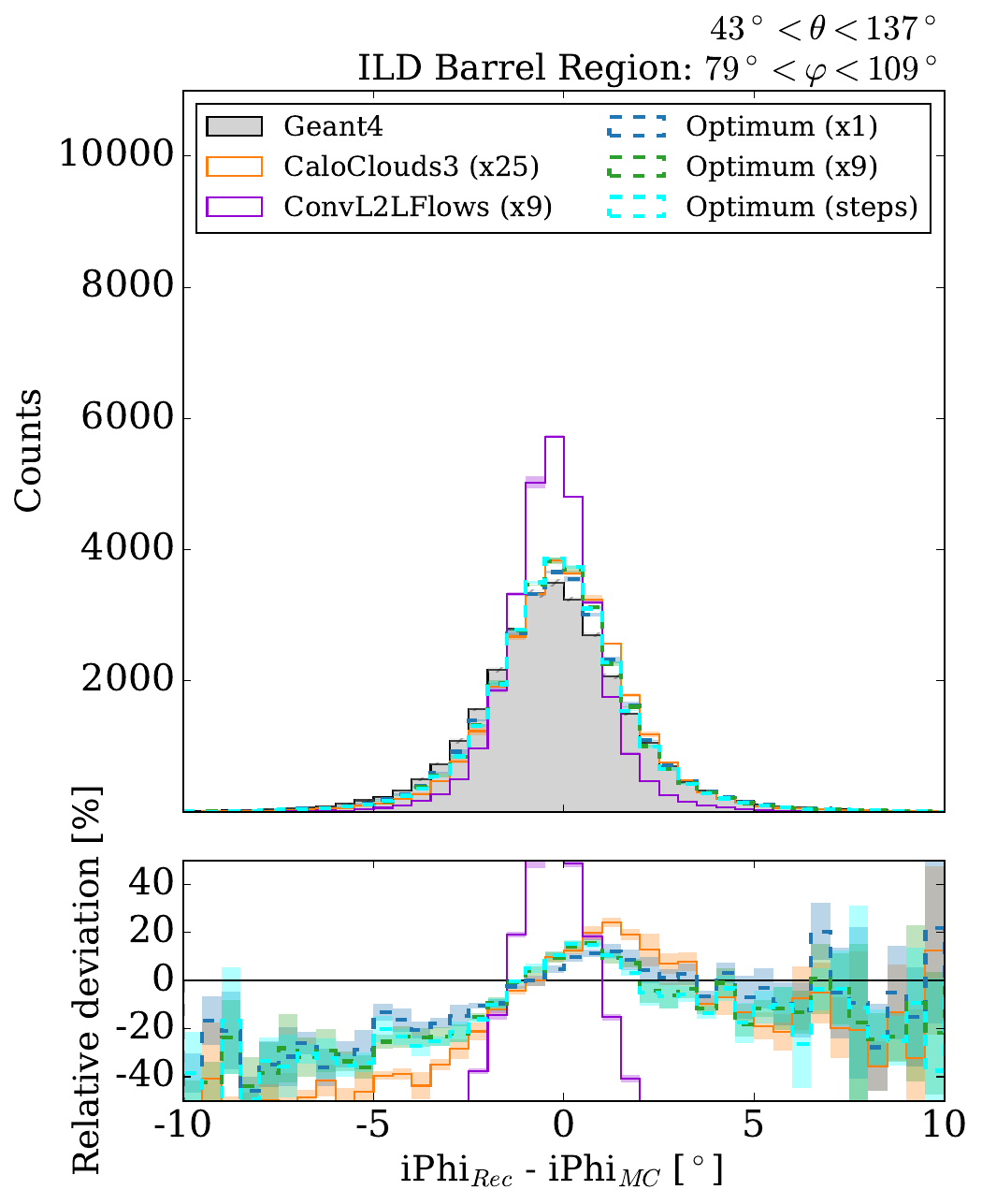}
    \includegraphics[width=0.4\textwidth]{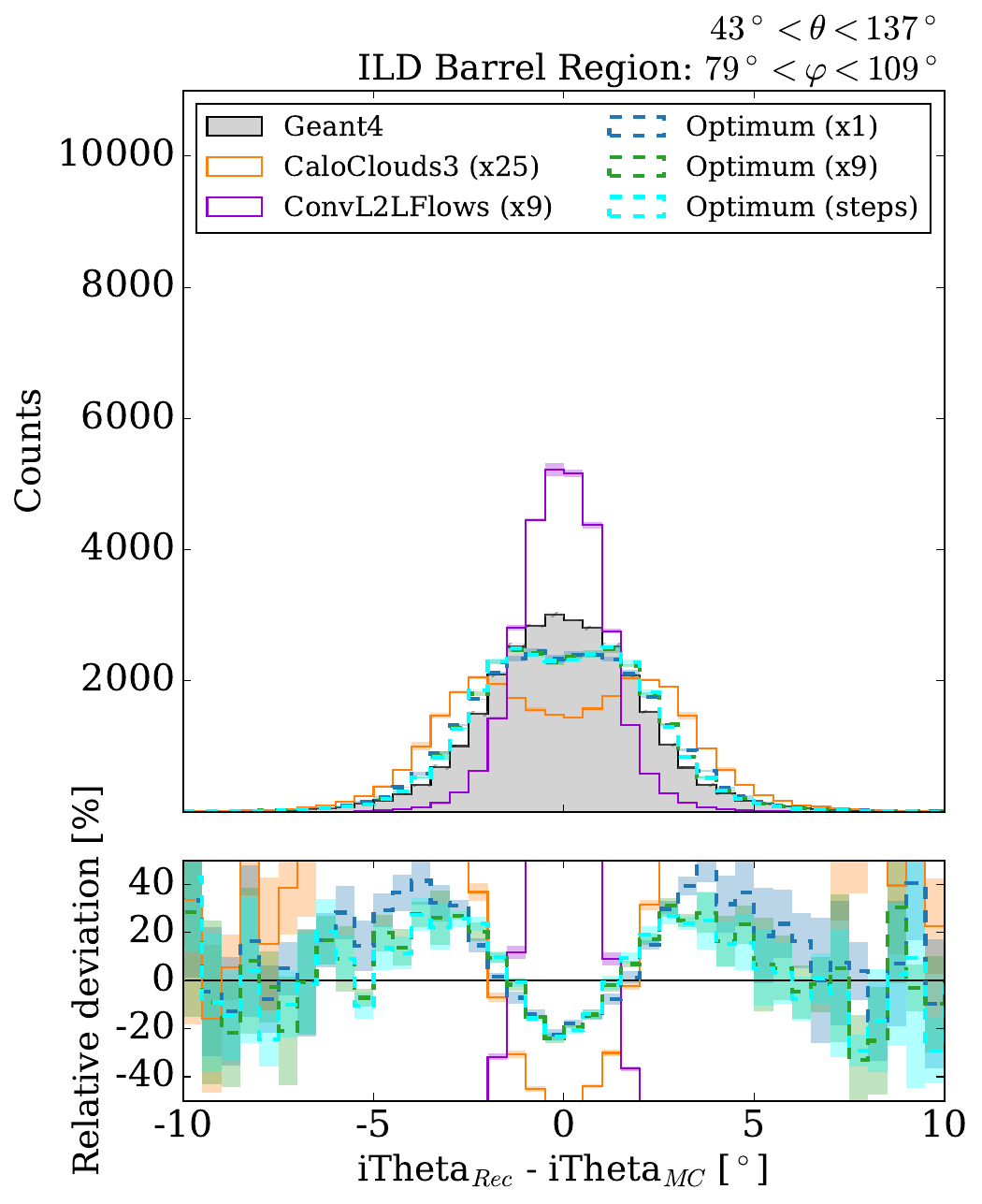}
    \caption{
        Distributions of the differences between reconstructed and true intrinsic angles for showers in the ILD barrel region ($43^\circ < \theta < 137^\circ$, $79^\circ < \varphi < 109^\circ$), comparing \geant to various generative models. Left: azimuthal angle ($iPhi$). Right: polar angle ($iTheta$).
        The top panel shows the angle residual distributions, while the bottom panel presents the relative deviation with respect to \geant. PCA is applied to all reconstructed hits to extract the principal axis of the shower.
    }
    \label{fig:angles_no_cut}
\end{figure*}

\begin{figure*}[!tbp]
    \centering
    \includegraphics[width=0.4\textwidth]{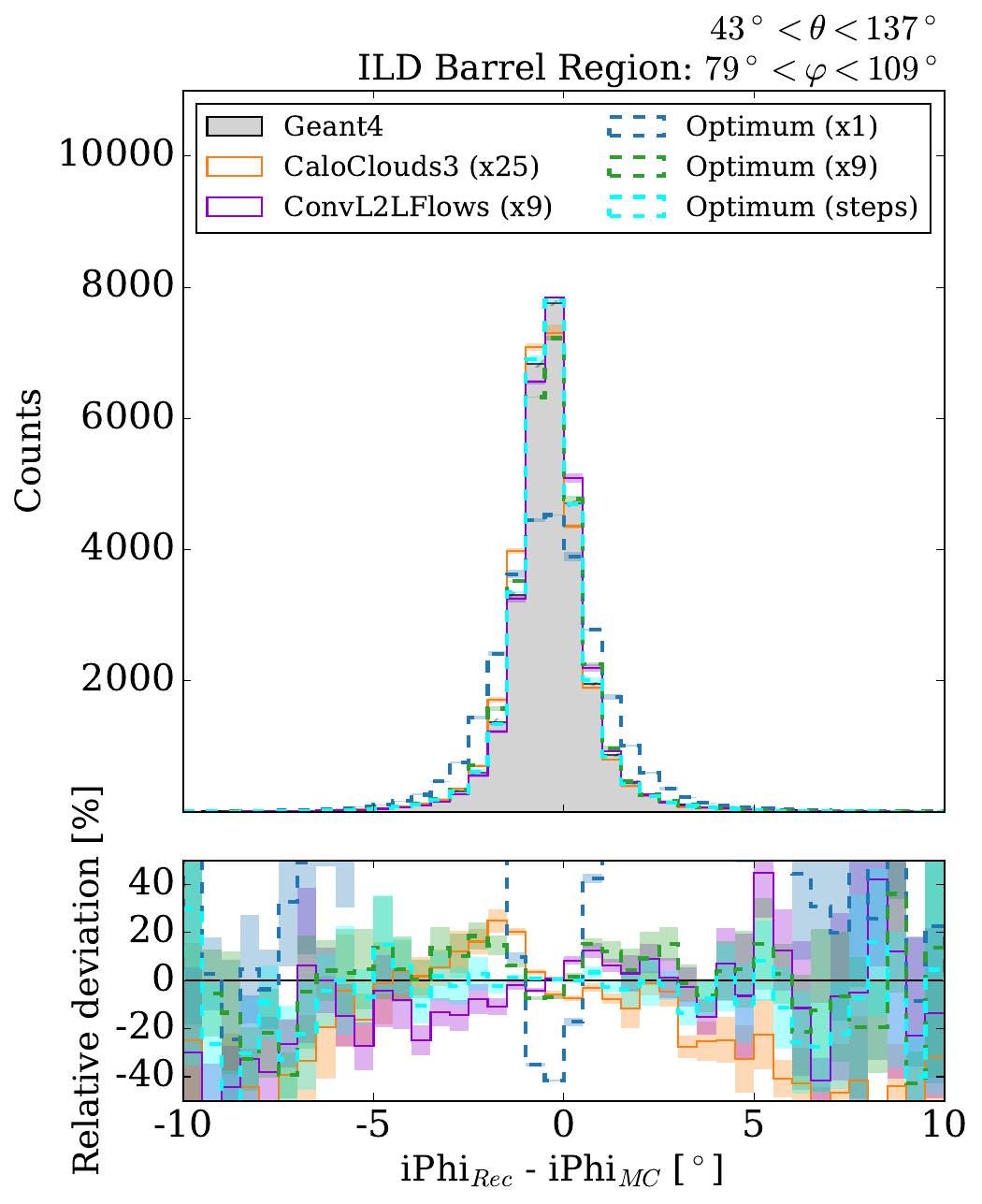}
    \includegraphics[width=0.4\textwidth]{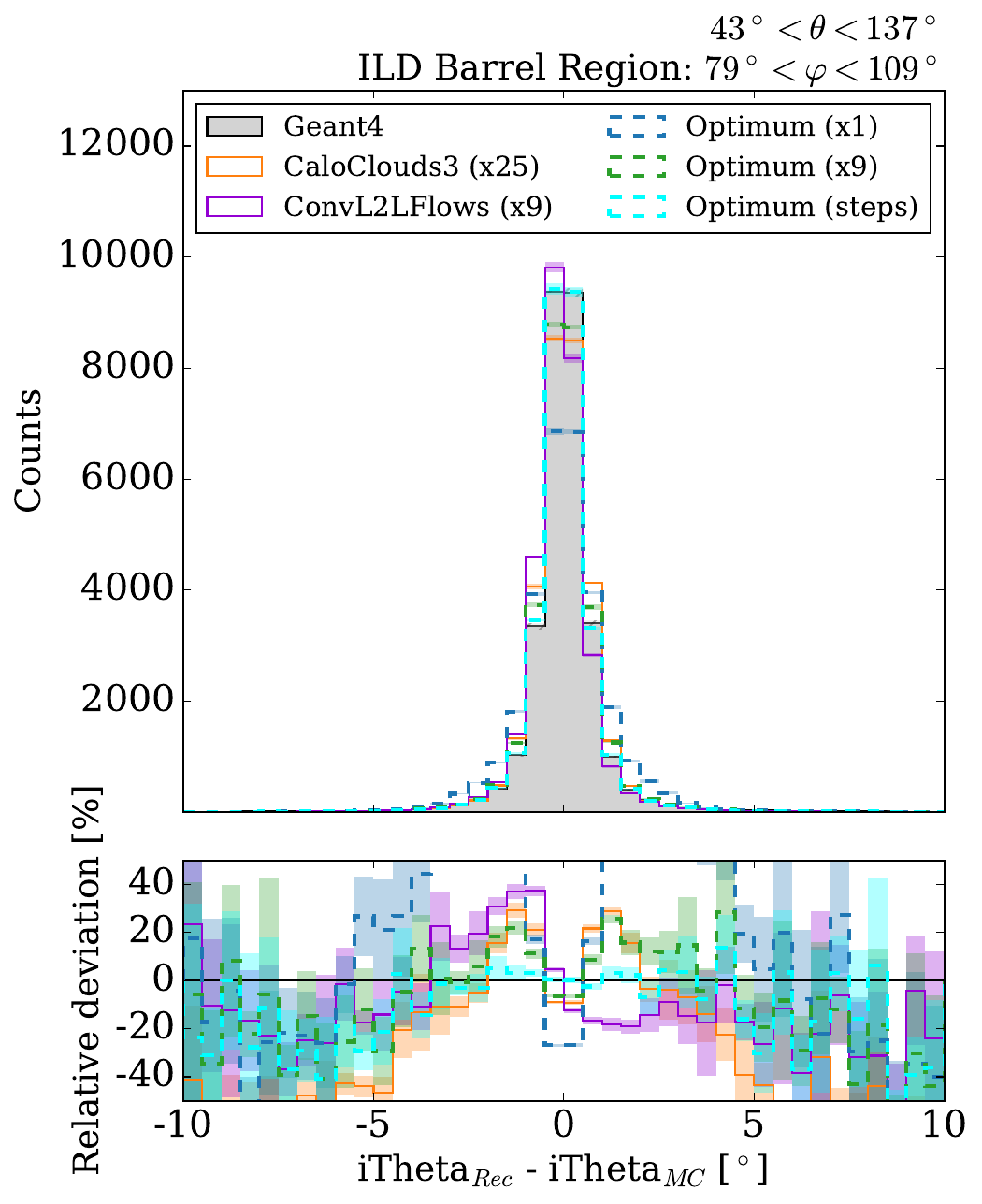}
    \caption{
        Distributions of the differences between reconstructed and true intrinsic angles for showers in the ILD barrel region ($43^\circ < \theta < 137^\circ$, $79^\circ < \varphi < 109^\circ$), comparing \geant to various generative models. Left: azimuthal angle ($iPhi$). Right: polar angle ($iTheta$).
        The top panel shows the angle residual distributions, while the bottom panel presents the relative deviation with respect to \geant. Here, only the top 4$\%$ most energetic hits are used for PCA-based extraction of the shower axis, resulting in significantly improved angular resolution compared to using all hits.
    }
    \label{fig:angles_with_cut}
\end{figure*}

\subsubsection{Intrinsic Angle Reconstruction}\label{sec:Angles}

In a similar fashion to previous studies using the \bibae{} model~\cite{Diefenbacher:2023prl, McKeown:2024}, the previous state-of-the-art generative model applied to the ILD detector, we evaluate the angular response of the generative models by comparing the reconstructed intrinsic angles of showers simulated with \geant to those generated by the models. A principal component analysis (PCA) is applied to all reconstructed hits of each shower to determine its principal axis. The resulting angular distributions of the azimuthal and polar angles, denoted as $iPhi$ and $iTheta$, respectively, are presented as the difference between the reconstructed and true angles of the incoming particle direction. These distributions are shown in Figure~\ref{fig:angles_no_cut}.

As shown in Figure~\ref{fig:angles_no_cut}, both \geant and the generative models produce angular distributions that are centered around zero with comparable widths for both $iPhi$ and $iTheta$, indicating that the models qualitatively reproduce the angular response observed in detailed simulation.

The \CCiii and optimal shower generators, however, show a tendency to overestimate the polar angle, resulting in a double-peaked structure in the $iTheta$ distribution. This indicates that this double-peak effect is likely related to the methodology of placing showers generated in the regularized detector into the real detector readout (see Appendix~\ref{Append:Systematics}). This effect is more pronounced for \CCiii, and is likely compounded by the fact that \CCiii slightly overestimates the energy for larger radii (i.e. further from the shower axis).

\LFlows on the other hand shows a noticeably sharper peak in both the polar and azimuthal angle distributions. This observation, combined with the fact that \LFlows generates showers within a tightly constrained spatial region, as described in Section~\ref{sec:Models:L2LFlows}, suggests that a simple PCA applied to all hits in the shower may be a suboptimal procedure for reconstructing the intrinsic shower angle. This is likely due to the high sensitivity of PCA to hits located far from the shower core, which can disproportionately influence the estimated principal axis.

To address this, we apply an energy-based hit selection, retaining only the top 4$\%$ most energetic hits before running PCA. This suppresses noise from peripheral hits and improves the stability of the reconstructed direction, meaning that this approach results in an improved angular reconstruction algorithm\footnote{For more details on this reconstruction improvement, see \cite{buss2025caloclouds3ultrafastgeometryindependenthighlygranular}}. The results are shown in Figure~\ref{fig:angles_with_cut}. All optimum shower generators, including \CCiii and \LFlows, now demonstrate angular resolutions that closely match the \geant reference. It is notable that the distribution produced by the \optimumXOne shower generator shows the largest discrepancy, producing a slightly wider distribution than \geant. The improved reconstruction shows better alignment with \geant, and the double-peak structure previously observed in the polar angle distribution disappears entirely. 

\subsection{Timing Benchmark}\label{sec:Timing}

\begin{figure*}[!tbp]
    \centering
    \includegraphics[width=0.39\textwidth]{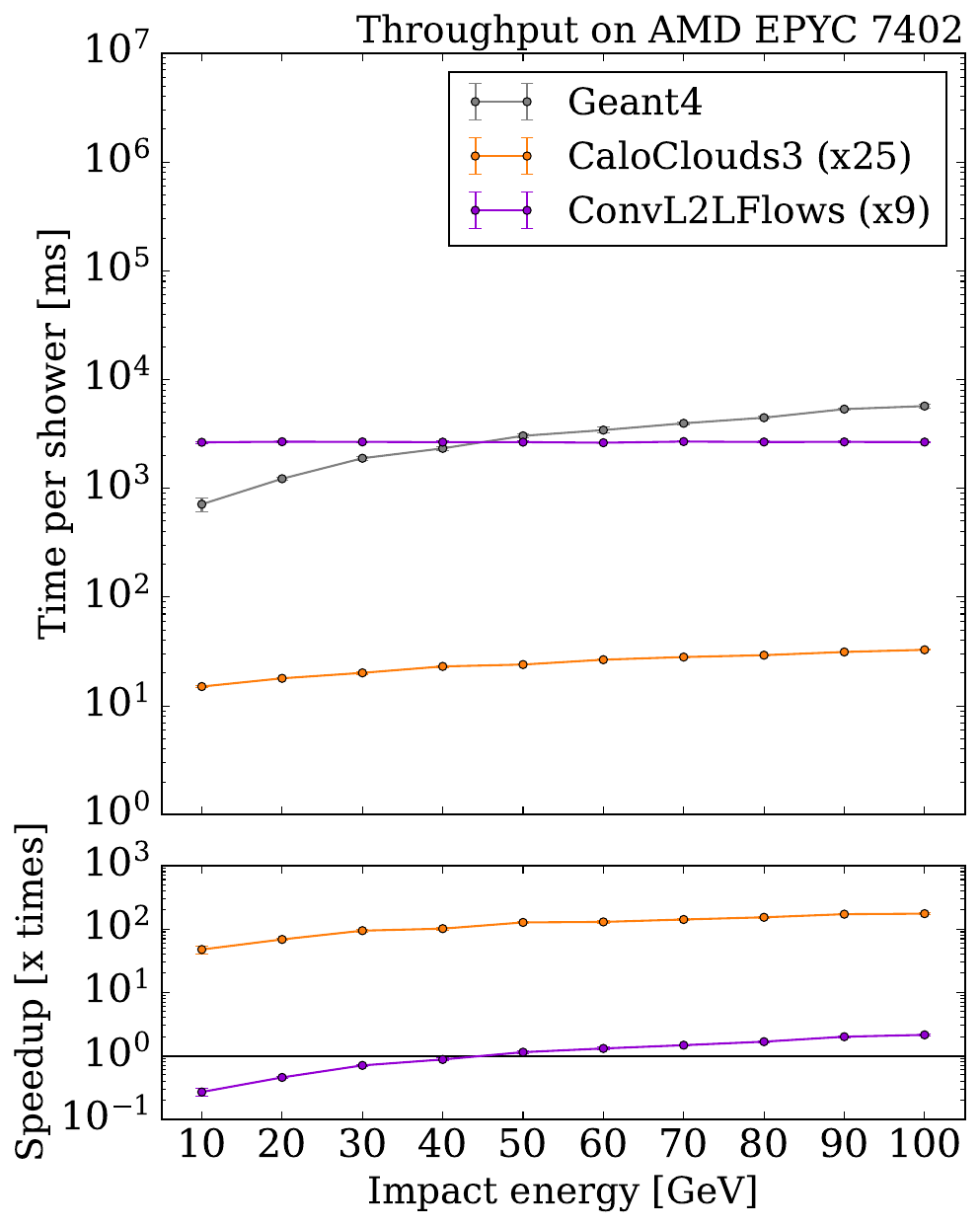}
    \includegraphics[width=0.39\textwidth]{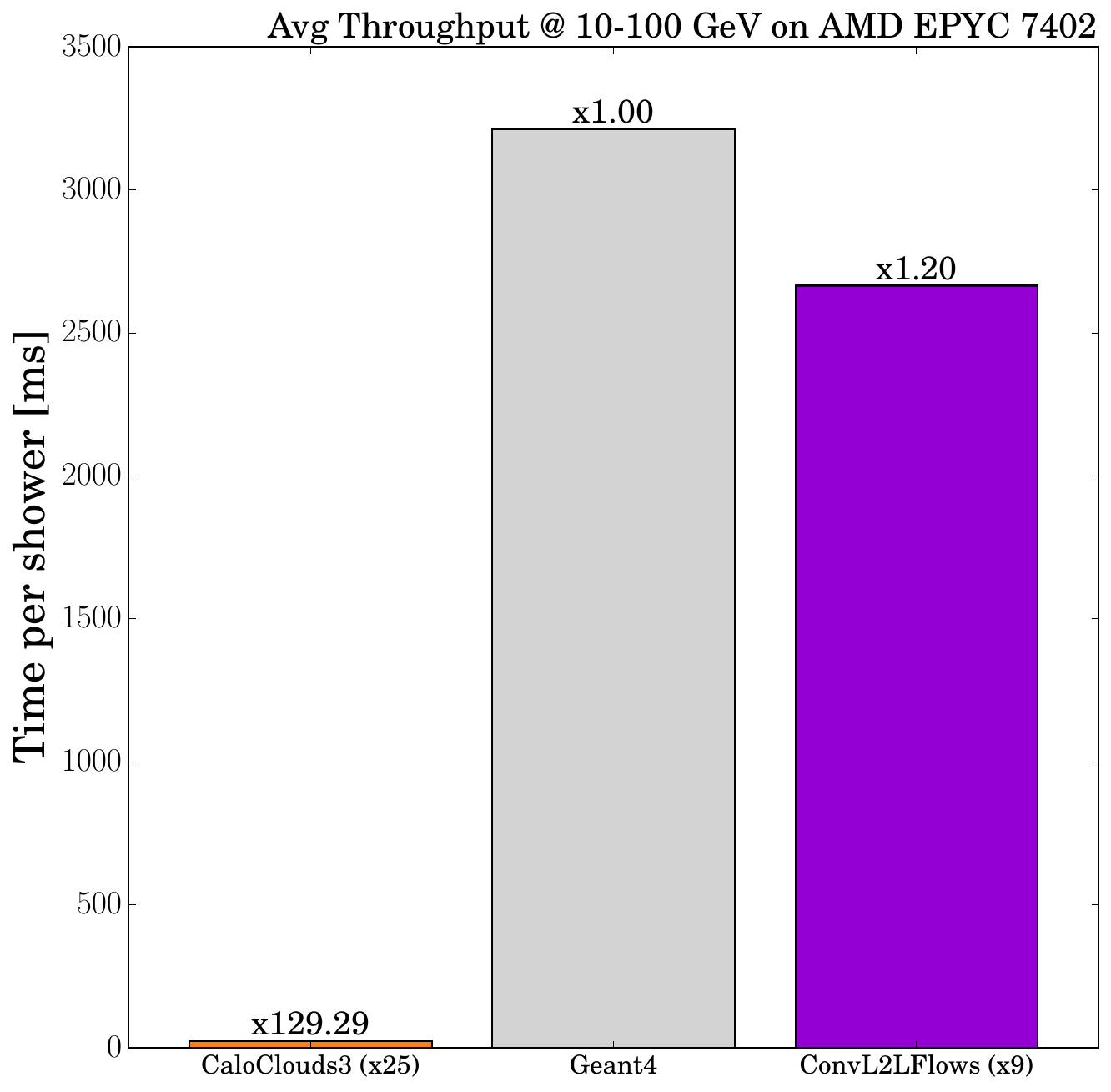}
    \caption{Single particle timing on a single core of an AMD EPYC\textsuperscript{\tiny\texttrademark} 7402 CPU within the full ILD simulation chain. Left: wall-clock time per shower (top) and speed-up vs.\ \geant (bottom) for photons with incident energies in the range of 10 to 100 GeV with 10 GeV steps. Right: average single core throughput, with the speed-up relative to \geant highlighted.}
    \label{fig:timing}
\end{figure*}

To quantify the speed advantage of the fast simulators in the full ILD software chain, we measure the wall-clock time per shower on a single CPU core of an AMD EPYC\textsuperscript{\tiny\texttrademark}~7402. The measurements for \geant and for all generative models were taken on the same machine and software setup. As a result of the integration of the models using the DDML library described in Appendix~\ref{sec:Integration}, all timing measurements are directly comparable to those of \geant, thanks to the use of an identical software configuration, including overheads such as hit placement in the detector geometry. This enables us to perform a fair and realistic timing benchmark, which is not possible without model integration. 

Figure~\ref{fig:timing} (left, top) reports the time per shower as a function of photon energy 10-100~GeV. The \geant baseline lies at the few second level and grows linearly with energy, reflecting the increasing number of interaction steps. \CCiii shows a similar trend, but with a much smaller slope. By contrast, the grid-based model \LFlows, shows no energy dependence and remains flat across the full range.

Complementary to the time per shower as a function of photon energy, Fig.~\ref{fig:timing} (right) shows the average single core throughput over 10-100~GeV, normalized to the \geant baseline (${\times1}$). We observe single-shower simulation speed-up factors of ${\times129.29}$ for \CCiii, and ${\times1.20}$ for \LFlows across this range of incident photon energies. The striking gain for \CCiii reflects its lightweight point-wise inference.

Overall, the scaling behavior is favorable for grid-based models as their inference time is constant with respect to shower energy because the computational cost is fixed by the voxelized volume. As a consequence, the relative speed-up over \geant grows with energy. At the low-energy end of the spectrum, where \geant showers contain fewer steps, the absolute latency gap narrows and the grid models can approach \geant in wall-clock time. Toward higher energies, where \geant scales approximately linearly with the number of interaction steps, the flat cost of \LFlows (and other grid models such as the \bibae~\cite{Buhmann:2020pmy, Buhmann:2021caf, Diefenbacher:2023prl} becomes increasingly advantageous. By contrast, the point-cloud based \CCiii shows a mild energy dependence, with its simulation time rising with energy as the number of generated points grows with incident energy. However, the slope remains well below that of \geant, yielding a more uniform though less asymptotically dramatic speed-up across the full range. For completeness, we note that model initialization (loading weights, caching etc.) is excluded from the timings shown, as including it only affects small sample sizes and does not change the observed scaling trends.

In addition to inference speed, the memory footprint and model size are important practical factors for deployment within large-scale simulation workflows. The compiled \CCiii model has a total weight size of approximately 27~MB, corresponding to a memory footprint of 198~MB during inference. By contrast, the grid-based \LFlows model is substantially larger, with a compiled weight size of 2.5~GB and a peak memory footprint of about 4.4~GB. This reflects the higher parameter count and larger activation maps inherent to convolutional grid-based architectures.

While the scaling behavior is favorable for grid-based models -- their inference time is essentially constant with shower energy because the compute is fixed by the voxelized volume. They also exhibit larger physics performance deviations at higher energies due to leakage from the bounded generation volume. As energy increases, a growing fraction of the shower can reach the box boundary and leak out, degrading containment and biasing observables (see Sec.~\ref{sec:EnergyResolution} and the single particle profiles in Sec.~\ref{sec:SingleParticle}). Mitigations such as expanding the generation volume reduce leakage but increase memory footprint and latency, reducing part of the speed advantage.

\subsection{\label{sec:DiPhoton:Results} Di-photon Benchmark}

\begin{figure*}[tbhh]
    \centering
    \includegraphics[width=0.33\textwidth]{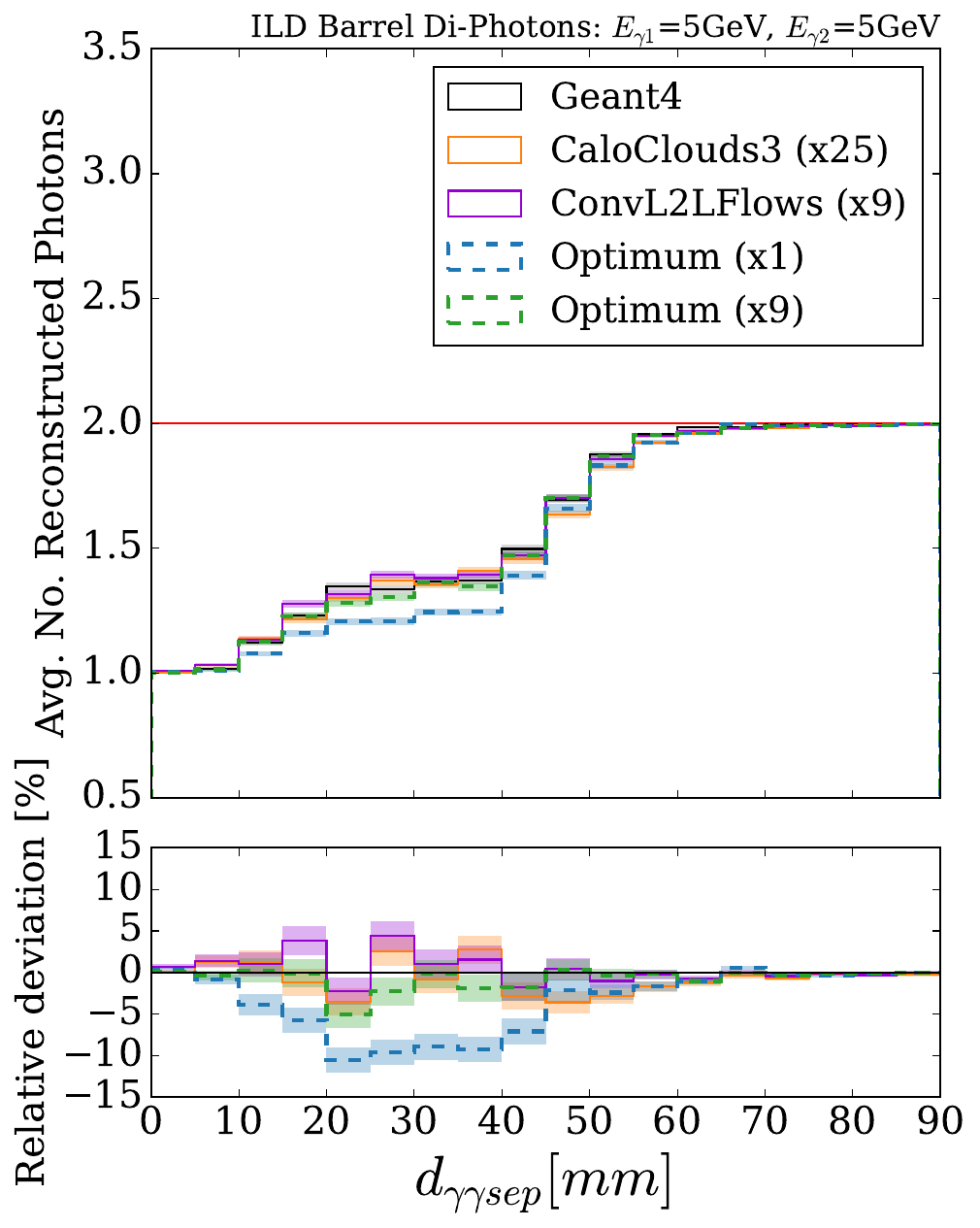}
    \includegraphics[width=0.33\textwidth]{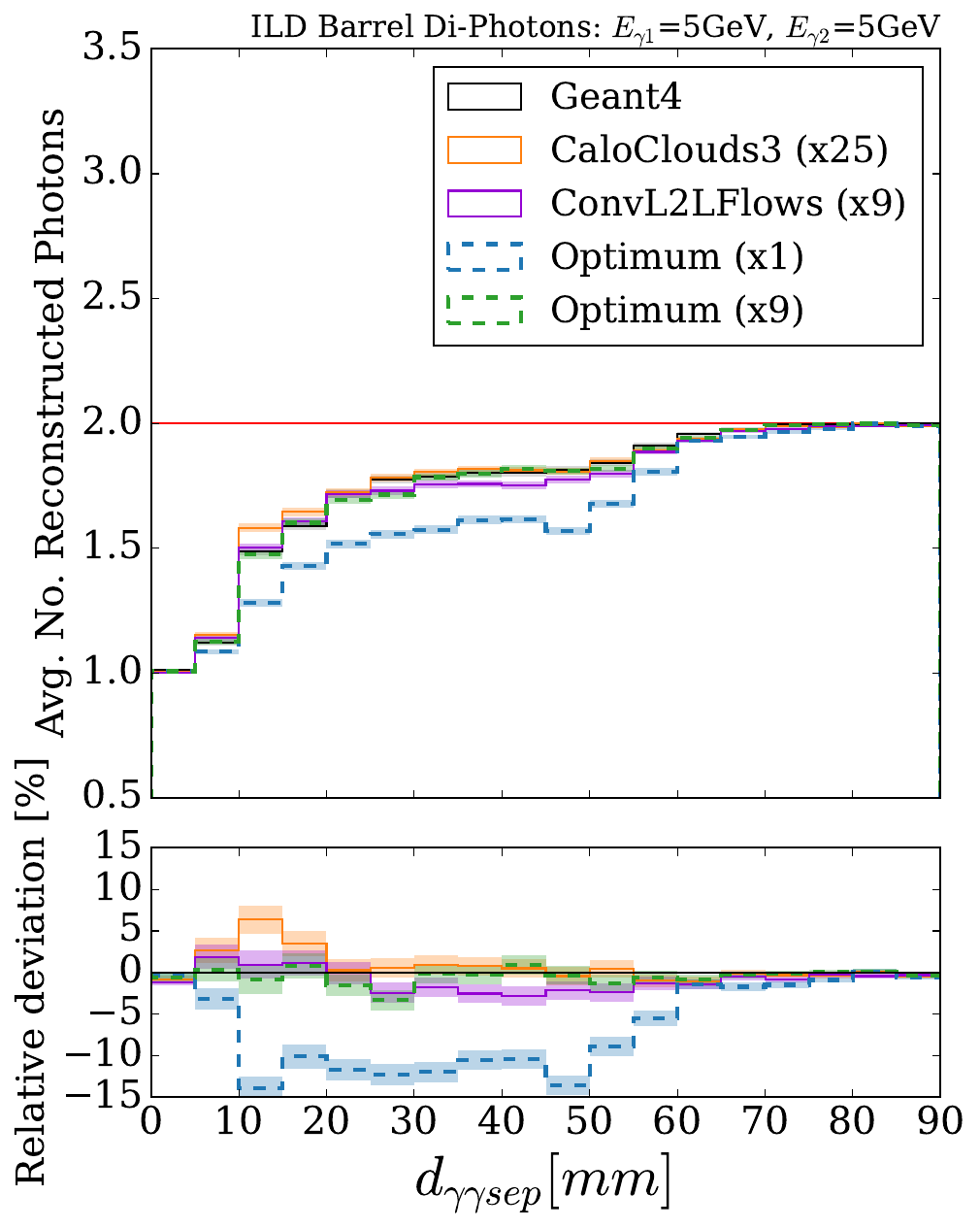}
    \includegraphics[width=0.33\textwidth]{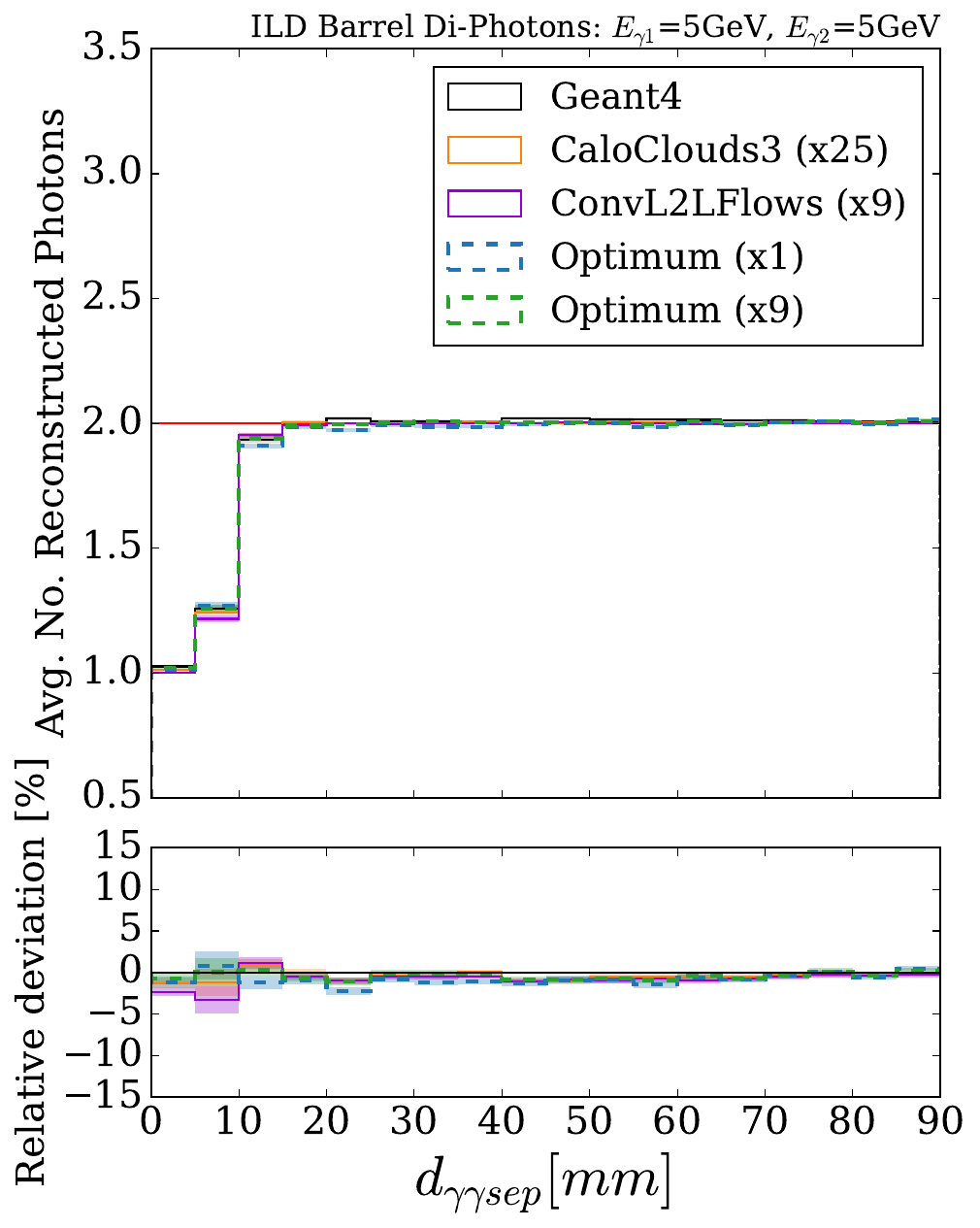}
    \caption{Average number of reconstructed photons against the di-photon separation for identical incident photon pair energies of 5~\si{GeV} (left), 20~\si{GeV} (middle) and 100~\si{GeV} (right). Curves are shown for \geant (light gray), the \CCiii (orange) and \LFlows (violet) models and the \optimumXOne (blue) and \optimumXNine (green) generators. Note that \optimumSteps is not included to aid visibility, as the performance was found to be comparable to that of \geant.}
    \label{fig:DiPhoton_sep}
\end{figure*}

The di-photon benchmark allows the performance of simulators to be studied in a scenario where multiple showers overlap. This emphasizes the relevance of particular shower characteristics that are not directly probed by studying single shower observables, while maintaining a controlled environment that prevents any contamination which may be present in a realistic physics process. To this end, the number of reconstructed photons is plotted against the separation between the two photons in Figure~\ref{fig:DiPhoton_sep}, for symmetric di-photon energies of 5~\si{GeV} (left), 20~\si{GeV} (middle) and 100~\si{GeV} (right). The performance is shown for \geant, \CCiii, \LFlows, \optimumXOne, \optimumXNine. Note that \optimumSteps is not included in order to aid visibility in the plot, as it performed on a similar level to \geant. The error bars represent the binomial error in each case. The red line present in each plot represents two photons being reconstructed on average, which is the optimal case for this reconstruction scenario. 

For all energies, at separations of less than approximately 6~\si{mm}, only a single photon is reconstructed on average. This corresponds to slightly more than one cell's worth of separation, meaning that the two individual shower cores cannot be resolved, with all hits being clustered into a single photon. In addition, it is easier for two photons to be resolved as their incident energy is increased. For photon pairs with incident energies of 100~\si{GeV}, the average number of reconstructed photons quickly rises to two for separations of only 10--20~\si{mm}, while for incident photon pairs with energies of 5~\si{GeV} the average number of reconstructed photons rises much slower, and only reaches an average of two for separations of around 60--65~\si{mm}. This is because incident photons with a higher energy have a higher energy core, which is also more densely populated than for lower energy photon showers, making them easier to distinguish~\cite{Xu:2017lgs}.

Due to the relative simplicity of separating higher energy photons, both models and the optimal shower generators agree well with \geant in terms of the distribution of the average number of reconstructed photons for photon pairs with incident energies of 100~\si{GeV}. Relative deviations in this instance appear around the level of a few percent, with the largest differences arising for the \LFlows model. At lower photon pair energies of 5~\si{GeV} and 20~\si{GeV}, more significant deviations occur. Here, the \optimumXOne shower generator shows significant discrepancies across a large range of separations, reaching relative deviations around the 10--15\% level respectively. Both the \CCiii and the \LFlows models show more contained relative deviations, typically less than 5\%, with the most noticeable exception being for the \CCiii model at separations of around 10-20~\si{mm} where the relative deviation slightly exceeds this level. 

These results indicate that training directly on the detector readout causes major discrepancies in the reconstruction performance. This is likely due to artefacts created when placing hits back into the detector geometry. The radial profile is of particular importance when separating such showers, as the reconstruction is especially sensitive to how the profiles of the two showers interfere. The effects observed around the core of the radial profile for photon showers produced with \optimumXOne in Figure~\ref{fig:radial_energy:30mm} are therefore considered a major factor in the poor performance of this generator. The differences observed for \LFlows and \CCiii are linked to the deviations in the radial profile observed in Figure~\ref{fig:radial_longitudinal_energy} (left), although these are less influential as they occur further out from the shower core. The steep fall in the radial profile of the \LFlows model, predicated by the constrained cut required for a regular grid, model seems to have little influence on the reconstruction performance in this test, as at large separations it is easy to distinguish the showers from their core alone.

\subsection{\label{sec:PhysicsBenchmark:Results} Full Physics Benchmark}

\begin{figure*}[tbhh]
    \centering
    \includegraphics[width=0.33\textwidth]{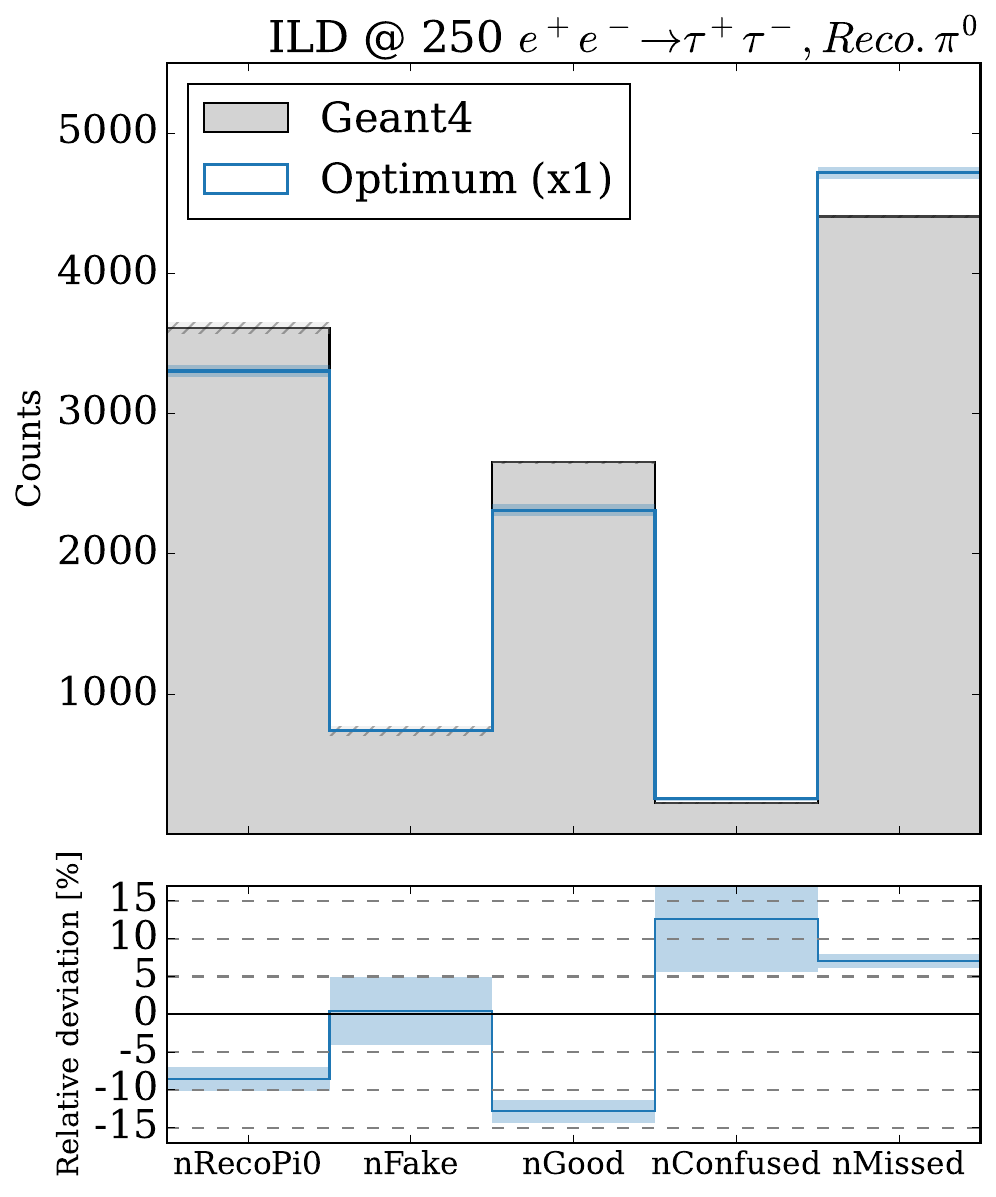}
    \includegraphics[width=0.33\textwidth]{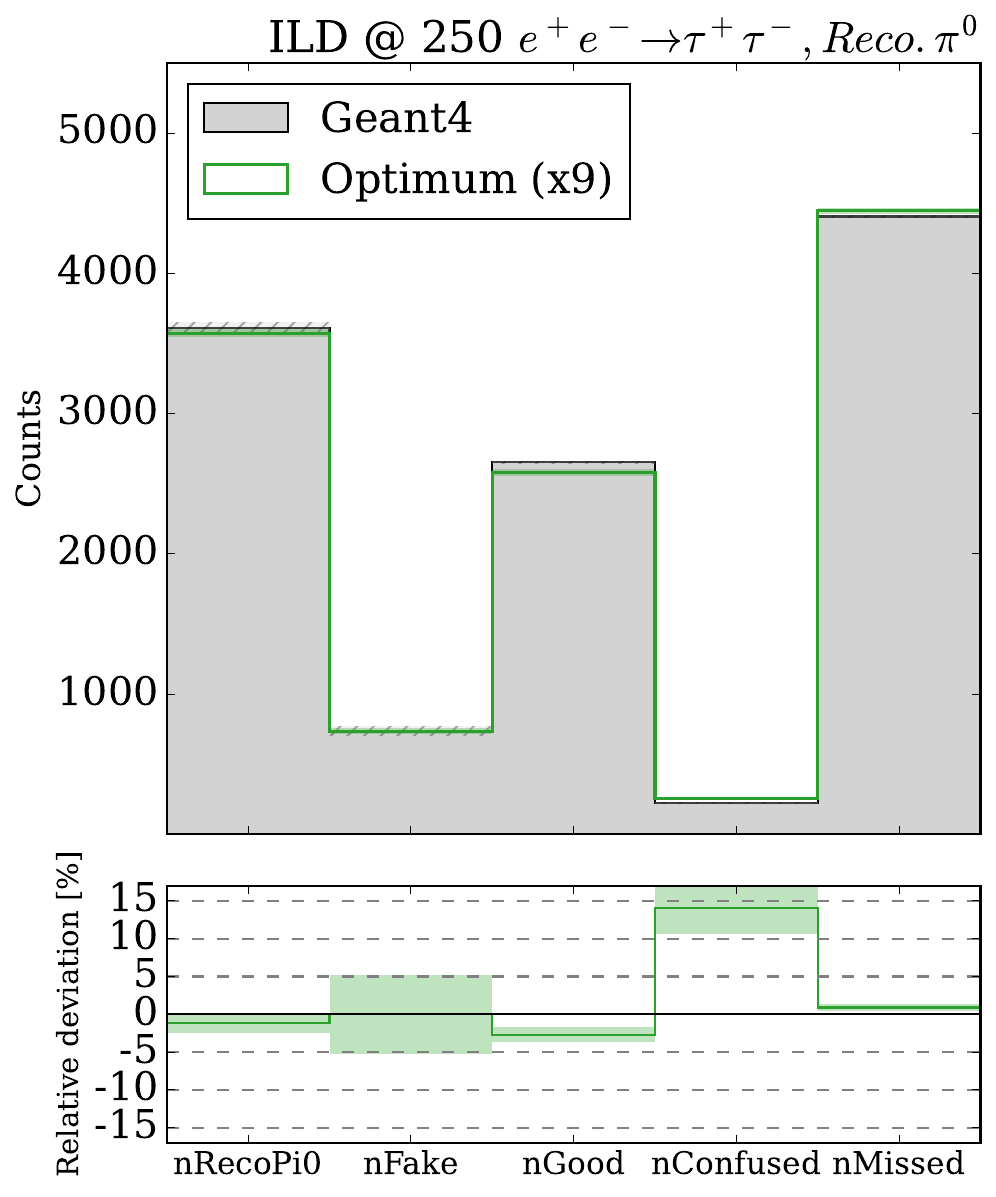}
    \includegraphics[width=0.33\textwidth]{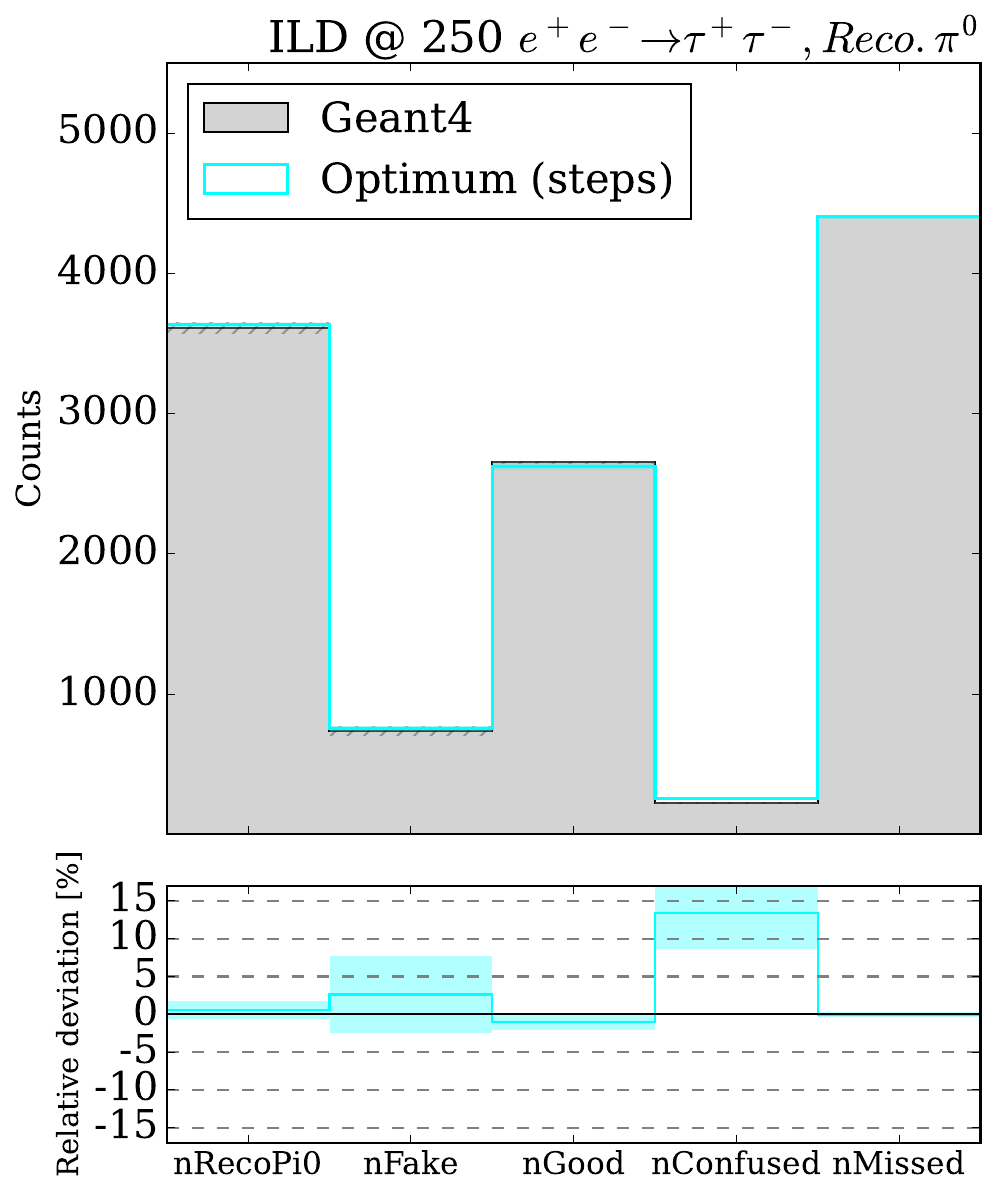}
    \includegraphics[width=0.33\textwidth]{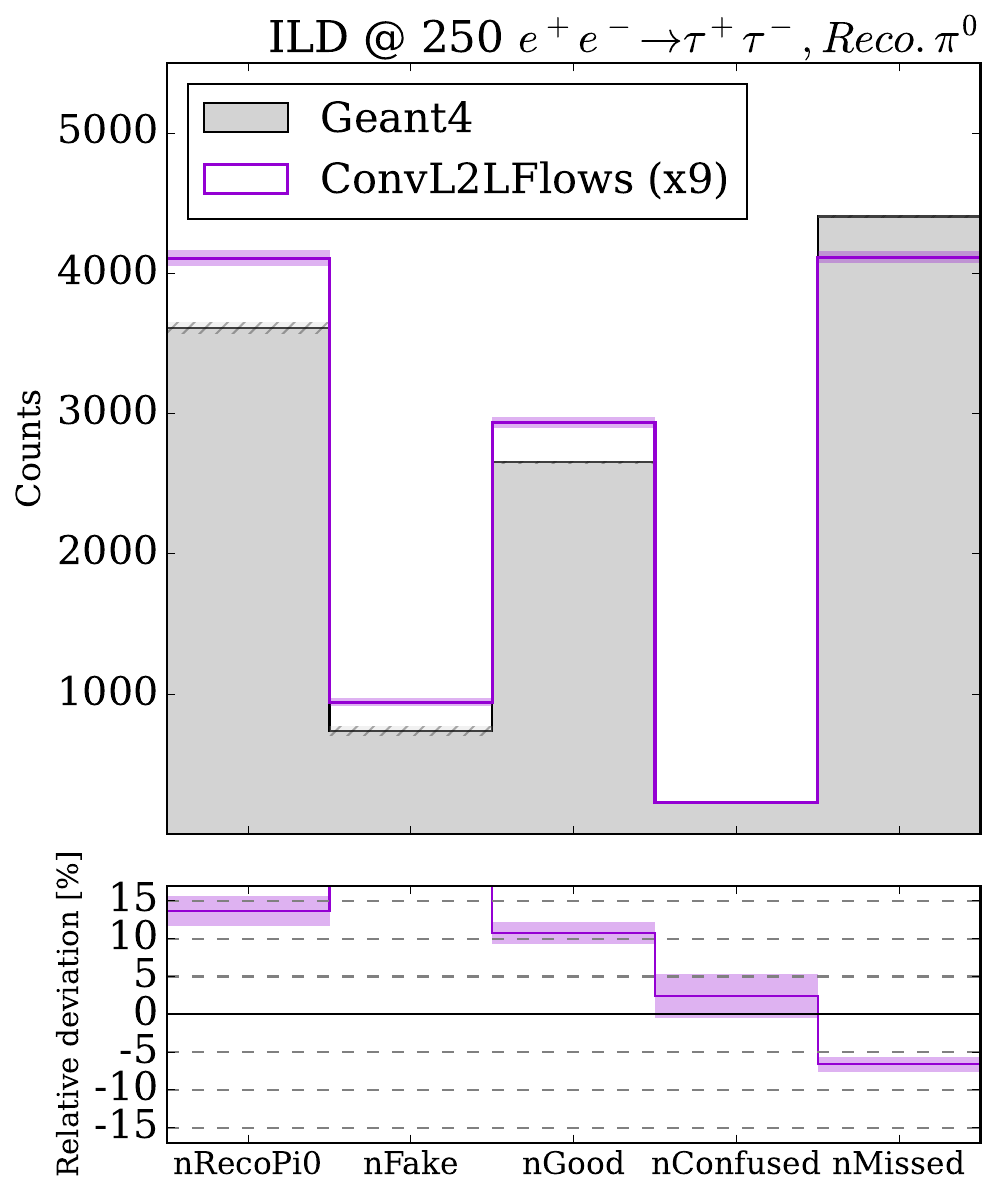}
    \includegraphics[width=0.33\textwidth]{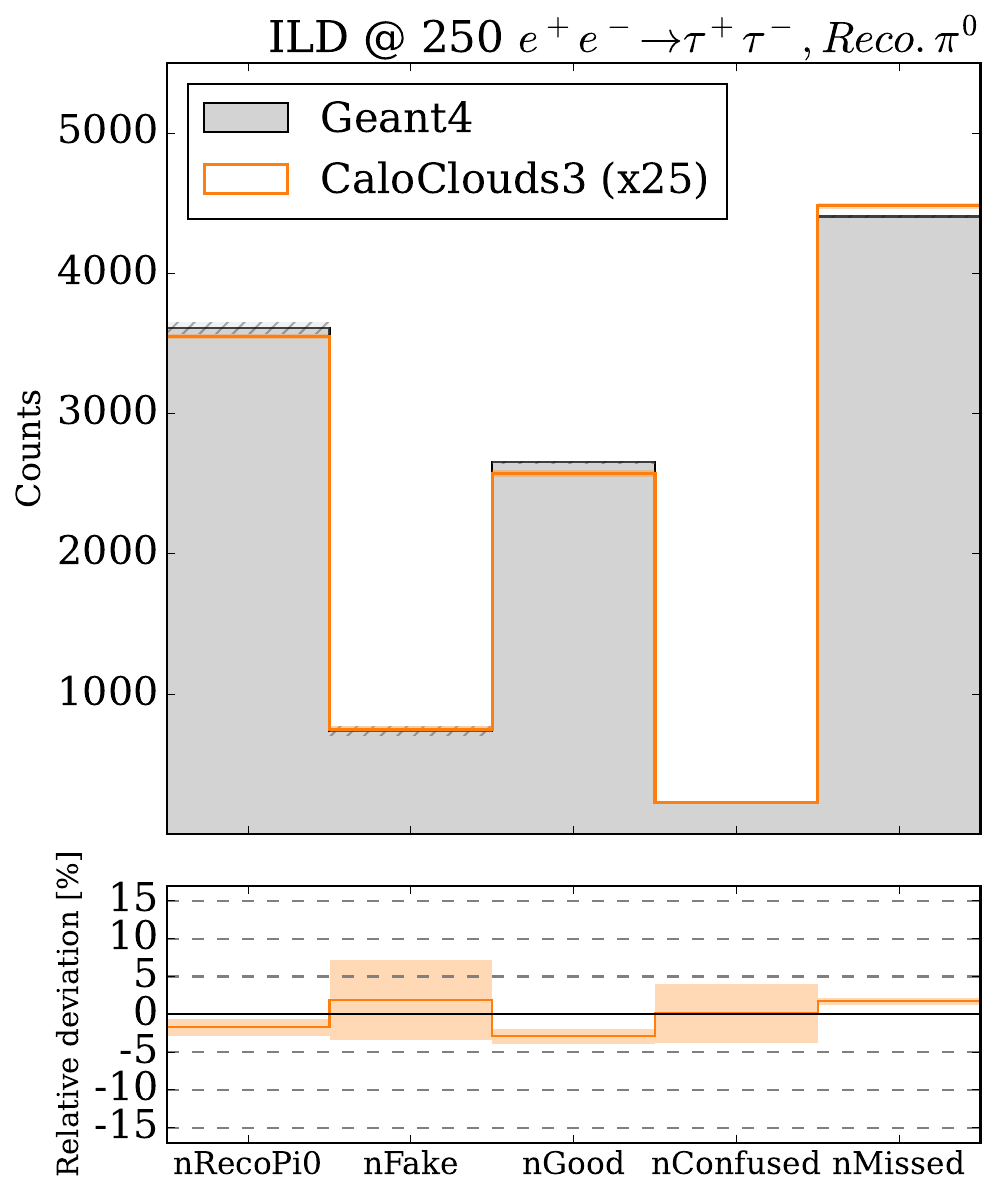}
    \caption{Overall reconstruction quality of $\pi^{0}s$ produced by tau decays in the process $e^{+}e^{-} \rightarrow \tau^{+}\tau^{-}$ for the \optimumXOne (top left, blue), \optimumXNine (top middle, green) and \optimumSteps (top right, cyan) generators, and the \LFlows (bottom left, violet) and \CCiii (bottom right, orange) models in comparison to \geant (grey). The quality of the reconstruction is characterized by five different categories; $\text{nRecoPi0}$, the total number of $\pi^{0}$s reconstructed, $\text{nFake}$, the number of $\pi^{0}$s reconstructed without a corresponding $\pi^{0}$ in the MC Truth, $\text{nGood}$, the number of correctly reconstructed $\pi^{0}$s, $\text{nConfused}$, the number of $\pi^{0}$s reconstructed only partially correctly and $\text{nMissed}$, the number of MC Truth-level $\pi^{0}$s that were not reconstructed. The errors on each category are derived from the standard deviation across three different random seeds used for the detector simulation. In each case, the lower panel represents the relative deviation from \geant in each category.}
    \label{fig:pi0_reco}
\end{figure*}

We now study the post-reconstruction performance of the various simulation approaches for the full physics benchmark based on $\pi^{0}$s produced in hadronic decays of the tau lepton. These results are split into an investigation into the global performance of the $\pi^{0}$ reconstruction in the process $e^{+}e^{-} \rightarrow \tau^{+} \tau^{-}$ in Section~\ref{sec:PhysicsBenchmark:Results:Global}, followed by a study of the modeling of key $\pi^{0}$ physics observables in Section~\ref{sec:PhysicsBenchmark:Results:Observables}.

\subsubsection{Global Reconstruction Performance}\label{sec:PhysicsBenchmark:Results:Global}

We begin by studying the overall quality of the reconstruction of $\pi^{0}$s produced in the tau decays for each shower simulator in comparison to \geant. For this evaluation, it is necessary to create a relation between Monte Carlo (MC) Truth particles and reconstructed particles. The weight of each relation is determined by the energy weighted contribution of each MC Truth particle to a reconstructed particle. If multiple MC Truth particles have a relation to a given reconstructed particle (or vice versa), the one with the highest weight is taken.

We now use these relations to define four different categories of reconstruction quality into which a given reconstructed $\pi^{0}$ may fall. Firstly, we define $\text{nGood}$ as the number of correctly reconstructed $\pi^{0}$s. Secondly, we define $\text{nFake}$ as the number of $\pi^{0}$s reconstructed without being linked to an MC Truth $\pi^{0}$. Thirdly, $\text{nConfused}$ is defined as the number of $\pi^{0}$s reconstructed where only one of the constituent photons was reconstructed correctly. Finally $\text{nMissed}$ represents the number of MC Truth $\pi^{0}$s for which there was no reconstructed $\pi^{0}$. We also calculate the total number of $\pi^{0}$s reconstructed, $\text{nRecoPi0}$. $\text{nFake}$ is constrained on a per-event basis by the equation $\text{nFake} = \text{nRecoPi0} - \text{nGood} - \text{nConfused}$. The results for each of the optimum generators \optimumXOne, \optimumXNine, \optimumSteps, together with the \LFlows and \CCiii models, are shown in Figure \ref{fig:pi0_reco}. In each case, the results are plotted in comparison to the \geant result. For each category, the results show the average over the $3$ different \geant random seeds, with the error calculated from the standard deviation across seeds.

For the optimum generators, the cell-level generator \optimumXOne shows the clearest mismatches. Large deviations from the \geant samples are observed in all categories except $\text{nFake}$. This includes on average almost $10\%$ fewer $\pi^{0}$s being reconstructed in total. On average more than $10\%$ fewer $\pi^{0}$s are correctly reconstructed, while many more $\pi^{0}$s are reconstructed incorrectly ($\text{nConfused}$ is on average more than $10\%$ greater, while $\text{nMissed}$ is on average more than $5\%$ larger) with respect to \geant. It should be noted that while $\text{nConfused}$ exhibits a high relative deviation, $\text{nConfused}$ itself is low meaning that the absolute deviation is small.

Increasing the granularity of the resolution drastically decreases the deviations with respect to the \geant samples, with the \optimumXNine generator showing much closer agreement. However, deviations are still clearly visible, with on average fewer $\text{nGood}$ $\pi^{0}$s relative to \geant and a similar excess in $\text{nConfused}$ as was observed for \optimumXOne.

At the step-level resolution in \optimumSteps, the average for each category is consistent with \geant within error except for $\text{nConfused}$. Again, the deviation of $\text{nConfused}$ relative to \geant is of a similar magnitude to that found for the \optimumXOne and \optimumXNine representations. These results demonstrate that increasing the granularity of the representation past that of the detector readout results in improved physics performance for this use case. 

Turning to the performance of the generative models, \LFlows shows major differences to \geant in a large number of categories, including relative deviations in excess of $10\%$ for $\text{nRecoPi0}$ and $\text{nGood}$, and in excess of $15\%$ for $\text{nFake}$. The deviation with respect to \geant for $\text{nMissed}$ exceeds the $5\%$ level, while $\text{nConfused}$ shows the closest agreement with a relative deviation of only a few percent. These deviations are consistent with the significant discrepancies observed for single particle observables produced with the \LFlows model in Section \ref{sec:SingleParticle}.

By contrast, the \CCiii model shows much closer agreement with the \geant baseline, with all categories showing relative deviations only at the level of a few percent. While at first appearance it may be surprising that \CCiii is able to achieve better performance than \optimumSteps, for example in the $\text{nConfused}$ observable, it should be noted that an extra calibration was applied to the \CCiii model as described in Section \ref{sec:EnergyResolution}.

\subsubsection{Physics Observable Performance}\label{sec:PhysicsBenchmark:Results:Observables}
\begin{figure*}[tbhh]
    \centering
    \includegraphics[width=0.4\textwidth]{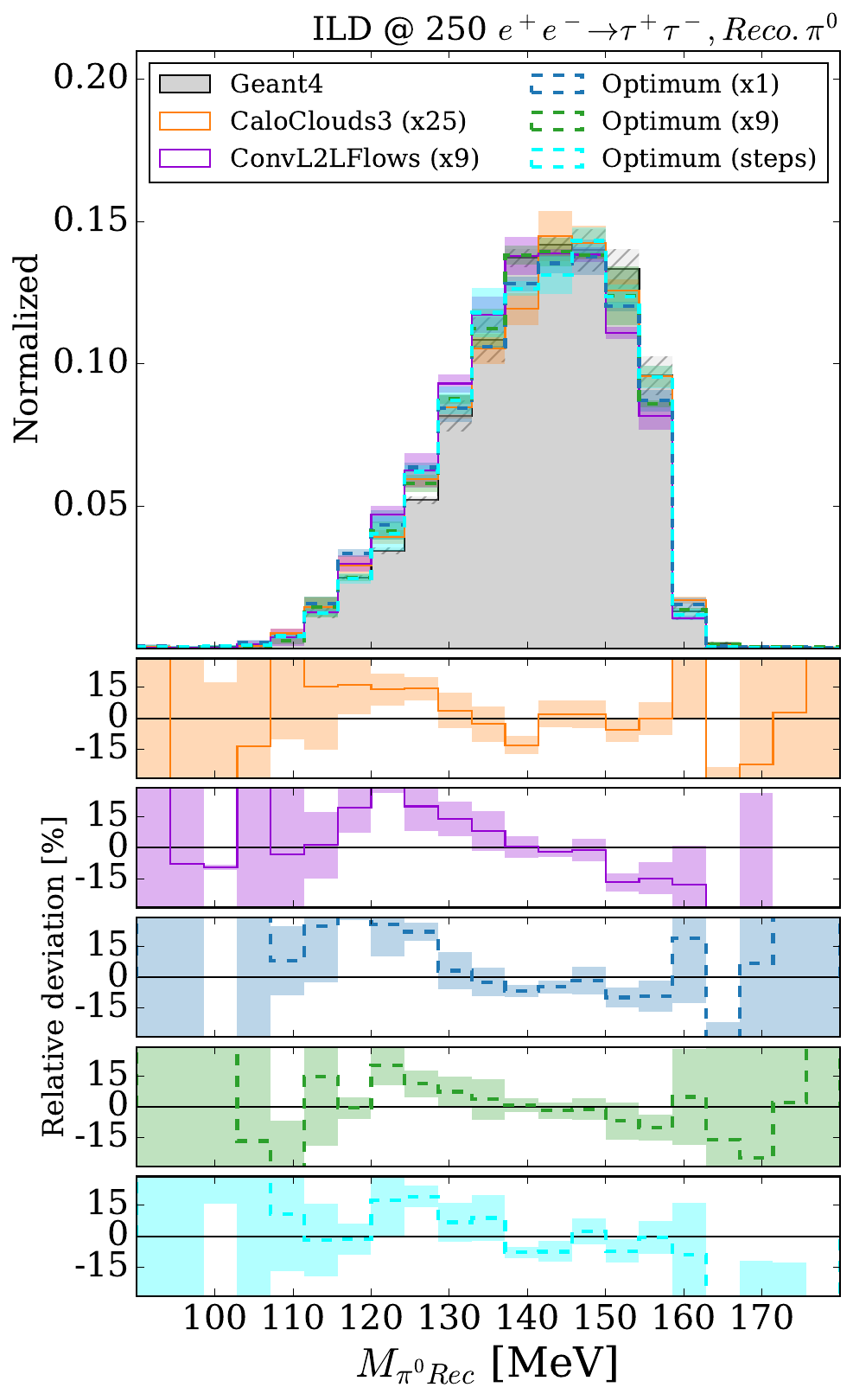}
    \includegraphics[width=0.4\textwidth]{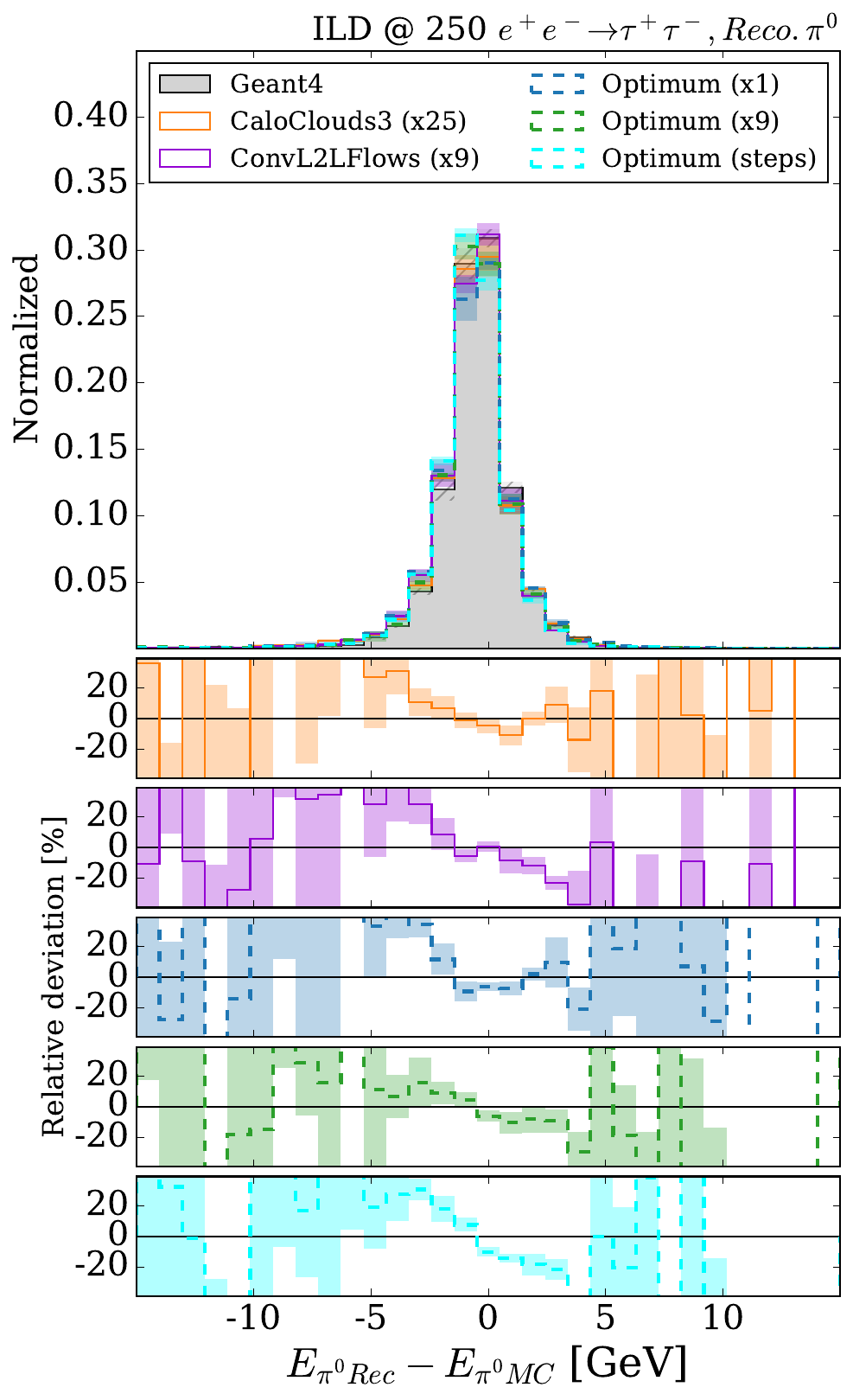}
    \caption{Key physics observables for $\pi^{0}$s produced by tau decays in the process $e^{+}e^{-} \rightarrow \tau^{+}\tau^{-}$ : the reconstructed $\pi^{0}$ invariant mass $M_{\pi^{0}Rec}$ distribution (left), and the difference between the reconstructed energy of the $\pi^{0}$ $E_{\pi^{0}Rec}$ and the corresponding MC Truth $\pi^{0}$ $E_{\pi^{0}MC}$ (right). Each of the optimal shower generators \optimumXOne (blue), \optimumXNine (green) and \optimumSteps (cyan), as well as the two generative models \CCiii (orange) and \LFlows (violet) are shown, with the \geant reference shown in grey. In both cases, the errors are derived from the standard deviation across three different random seeds used for the detector simulation. The lower panels in each case shows the relative deviation form \geant.}
    \label{fig:pi0_reco_mass_energy}
\end{figure*}

\begin{figure*}[tbhh]
    \centering
    \includegraphics[width=0.4\textwidth]{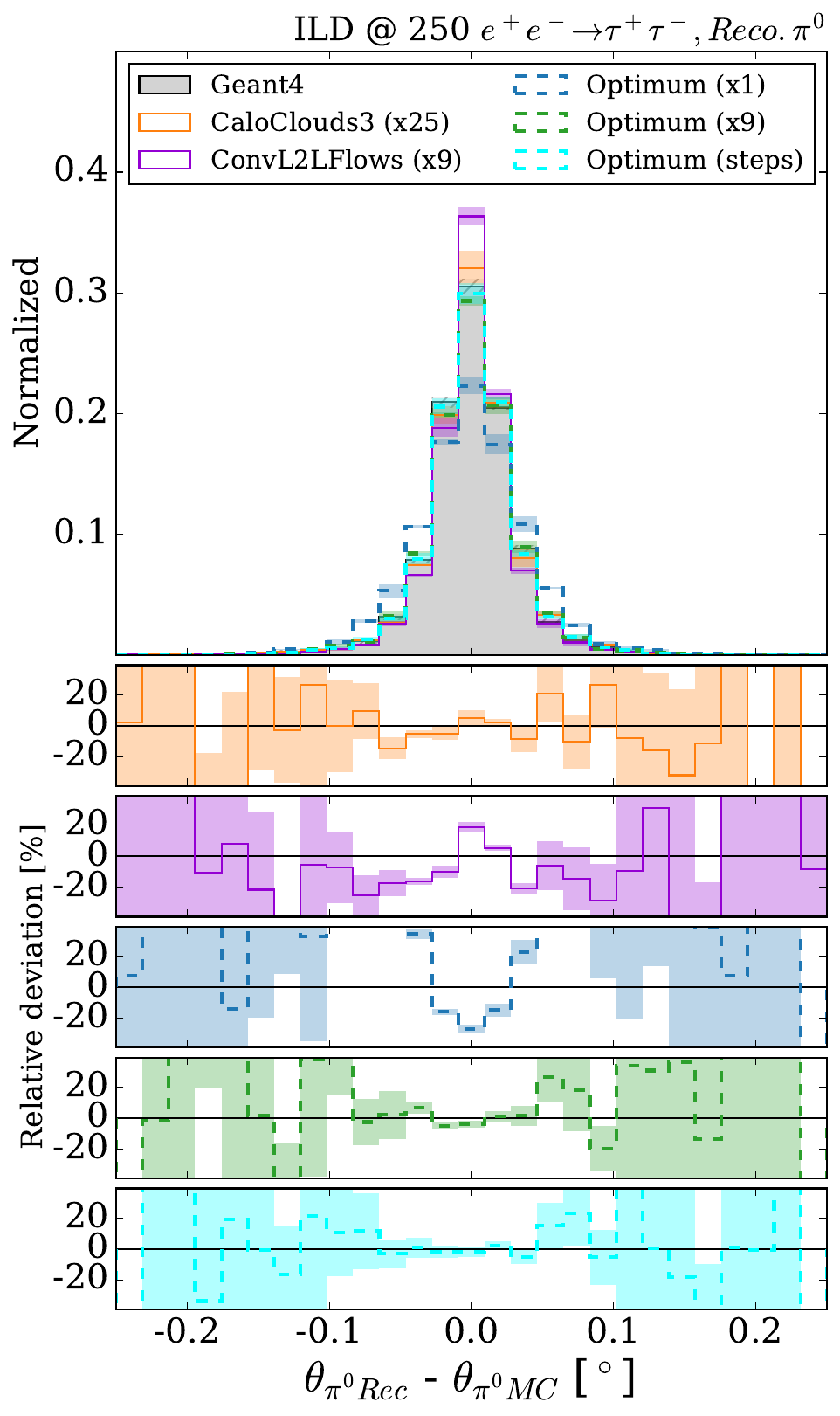}
    \includegraphics[width=0.4\textwidth]{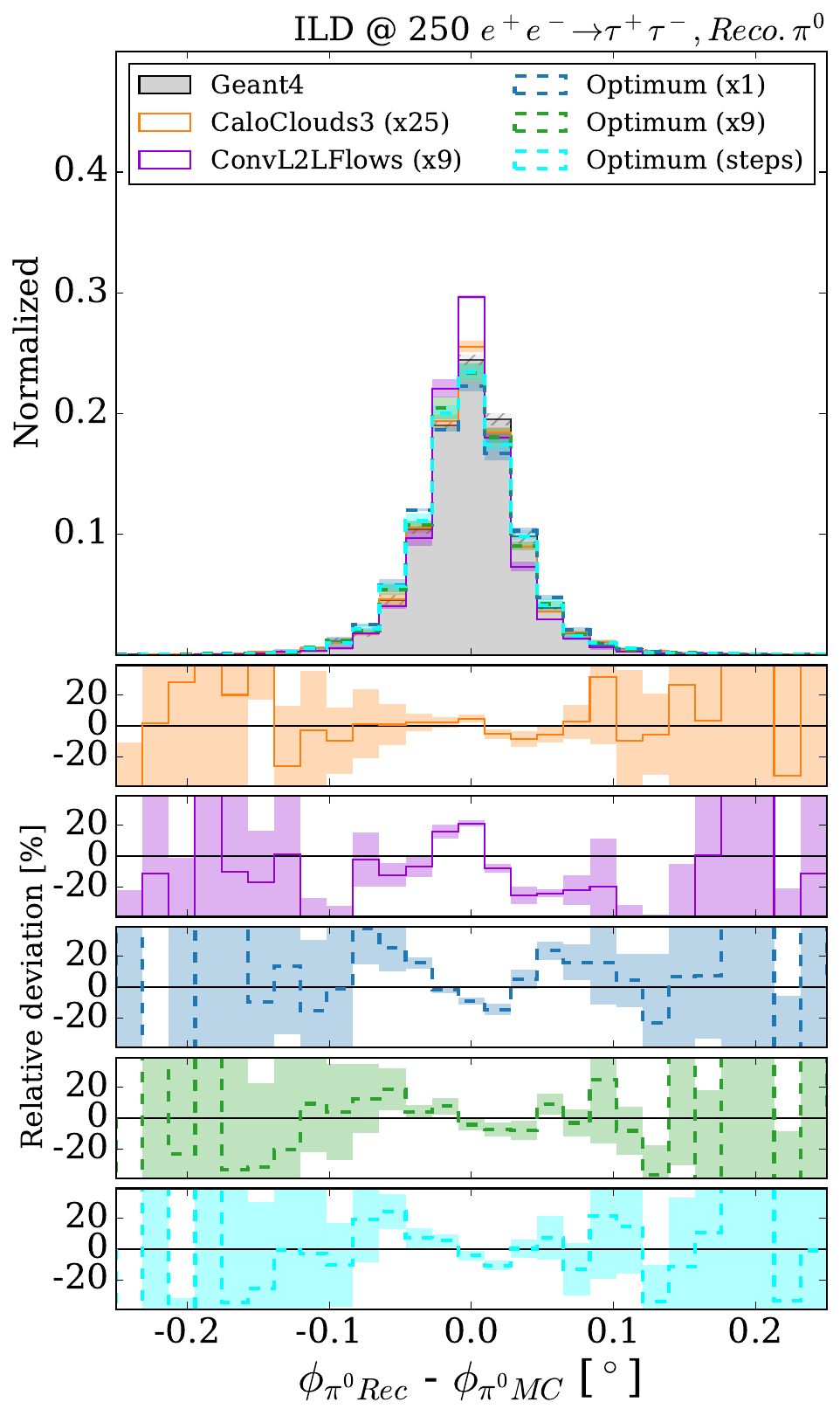}
    \caption{Reconstructed angular deviations for $\pi^{0}s$ produced by tau decays in the process $e^{+}e^{-} \rightarrow \tau^{+}\tau^{-}$: the difference between the reconstructed $\theta_{\pi^{0}Rec}$ and MC Truth $\theta_{\pi^{0}MC}$ direction in $\theta$ (left), and the difference between the reconstructed $\phi_{\pi^{0}Rec}$ and MC Truth $\phi_{\pi^{0}MC}$ direction in $\phi$ (right). Each of the optimal shower generators \optimumXOne (blue), \optimumXNine (green) and \optimumSteps (cyan), as well as the two generative models \CCiii (orange) and \LFlows (violet) are shown, with the \geant reference shown in gray. In both cases, the errors are derived from the standard deviation across three different random seeds used for the detector simulation. The lower panels in each case shows the relative deviation form \geant.}
    \label{fig:pi0_reco_angles}
\end{figure*}

We now study the performance of the individual simulators in terms of key $\pi^{0}$ physics observables. In order to perform a fair comparison, we require that all $\pi^{0}$s selected have been correctly reconstructed (i.e those that fell into the $\text{nGood}$ category described in Section \ref{sec:PhysicsBenchmark:Results:Global}). As before, the results show the average over $3$ different \geant random seeds, with the error calculated from the standard deviation across seeds. It should be noted that some of these distributions are already constrained by the selection of $\pi^{0}$s which fall into the $\text{nGood}$ category. This is due to quality criteria imposed by the $\pi^{0}$ reconstruction procedure, which includes performing a constrained kinematic fit~\cite{List:88030}.

Firstly, we calculate the invariant mass of the reconstructed $\pi^{0}$, $M_{\pi^{0}Rec}$, which is equivalent to the invariant mass of the di-photon system $M_{\gamma \gamma}$ given by 

\begin{equation} \label{eq:PhysicsBenchmark:Inv_Mass}
M_{\gamma\gamma} = \sqrt{2E_{i}E_{j}(1-\cos(\eta))},
\end{equation}

where $E_i$ is the reconstructed energy of photon $i$, $E_j$ is the reconstructed energy of photon $j$, and $\eta$ is the opening angle between their reconstructed flight directions. The invariant mass distributions for \geant, each of the optimum generators, and both the \LFlows and \CCiii models are shown in Figure \ref{fig:pi0_reco_mass_energy}. All models show broad agreement within the stated uncertainties around the bulk of the distribution. Larger relative deviations appear for all optimum shower generators and both models in the tails of the distribution, although the increasing errors on the ratio make the exact discrepancy less clear.

The next observable studied is the difference between the reconstructed energy of the $\pi^{0}$ and the energy of the corresponding MC particle, $E_{\pi^{0}Rec} - E_{\pi^{0}MC}$, which is shown in Figure \ref{fig:pi0_reco_angles}. It can be seen that in comparison to \geant the reconstructed energy tends to be slightly biased towards lower energies than that of the MC particle, although this deviation is heavily suppressed by the large magnitude of the uncertainties. This effect appears to be strongest for \optimumSteps and \LFlows, which correlates with the shifts in linearity for single photons observed for these approaches in Section \ref{sec:SingleParticle}.

Finally, the reconstructed angular differences for the $\pi^{0}$ in both the $\theta$, $\theta_{\pi^{0}Rec} - \theta_{\pi^{0}MC}$, and $\phi$, $\phi_{\pi^{0}Rec} - \phi_{\pi^{0}MC}$, directions are shown in Figure \ref{fig:pi0_reco_angles}. For the distribution in $\theta$, the clearest deviation occurs for the \optimumXOne representation at the level of the detector readout, which produces a noticeably broader distribution with relative deviations quickly exceeding the 30\% level away from the core of the distribution. Clear mismodelings from the \LFlows model are also present in both the $\theta$ and $\phi$ distributions, with relative deviations around the $20\%$ level around the core of the distributions.

\section{\label{sec:Discussion} Discussion}

Developing tools for the fast simulation of showers in highly granular calorimeters is essential to be able to meet the demands of future collider experiments. We have introduced the \DDML library for integrating generative models for fast calorimeter shower simulation into the \ddhep toolkit, enabling model benchmarking in a production-ready software suite and providing access to reconstruction-level physics benchmarks. Two different generative models for fast calorimeter simulation, \LFlows and \CCiii, were integrated into this library. This enabled realistic timing benchmarks of model performance, as well as post reconstruction benchmarks performed using the actual detector geometry. By comparing results using shower representations of different granularities, we were able to disentangle methodological details related to dataset construction from the actual performance of the generative models. Furthermore, we have presented new reconstruction-level benchmarks for the evaluation of generative models designed for electromagnetic shower simulation in highly granular calorimeters, including a first multi-particle benchmark involving di-photon separations, and a full physics benchmark based on hadronic decays of the tau lepton in the process $e^{+}e^{-} \rightarrow \tau^{+} \tau^{-}$. 

We have demonstrated that building a dataset which uses a shower representation directly at the level of the detector readout granularity results in significant distortions in key physics observables when using realistic detector geometries. These distortions are visible in all levels of post reconstruction observables presented -- from single particle observables through to higher level $\pi^{0}$ observables in the process $e^{+}e^{-} \rightarrow \tau^{+} \tau^{-}$ that would propagate directly through to down-stream analysis\footnote{For a summary of all of the observables studied in this paper, see Appendix~\ref{Append:Summary}}. As the quality of shower representation fundamentally limits the maximum achievable performance of a generative model trained on a given dataset, it is essential that datasets use optimized representations that operate on shower information at a lower level than the detector readout. This remains true beyond highly granular calorimeters, as demonstrated by recent work which optimized a voxelized representation for photons in the barrel region of the calorimeter of the ATLAS experiment, showing significant improvement over the current fast simulation tool~\cite{ATL-SOFT-PUB-2025-003}. While the ATLAS calorimeter system has a significantly lower granularity than that studied in this work, combined they emphasize the importance of producing optimized representations across the field of fast calorimeter simulation.

In order to be able to perform these optimizations, it is essential that fast simulation models are integrated back into the software ecosystems used in particle physics experiments- both to study placement into the actual detector geometry and to gain access to reconstruction level observables. The \DDML library introduced as part of this work provides this functionality for detector geometries implemented in \ddhep. Given that this library is designed to be generic, the addition of new ML models and detector geometries should be as straightforward as possible. While this work focused on the ILD detector, other detectors proposed for future colliders such as the FCC-ee are already supported in the library, including both the CLD detector~\cite{Bacchetta:2019fmz} and the ALLEGRO detector~\cite{Mlynarikova:2025skz}.

The generative models included in this study are trained on two distinct data representations. \LFlows is a model designed to operate on a fixed grid structure, while the \CCiii model is point cloud based. Due to the limitations of training directly on the detector readout, as previously discussed, both models are trained on a more granular shower representation. This requirement has greater impact on the \LFlows model, which has to be trained on a representation with a restricted granularity due to the limitations imposed by the use of a regular grid. As a result of using a regular grid of a higher granularity than the detector readout, the \LFlows model shows a poor simulation throughput increase for single particles relative to \geant, as well as showing several deviations in observables, including for higher-level physics quantities. A significant factor in the larger deviations observed for the \LFlows model for single particle observables results from the constrained bounding box necessitated by the used of a regular grid. This is a major reason why this approach to generative modeling exhibited consistently larger deviations than the corresponding \optimumXNine generator, which included a far less restrictive bounding box. By contrast, the \CCiii model is able to operate on a more granular, and therefore more accurate shower representation. Despite this, the model is able to achieve more than two orders of magnitude faster simulation throughput than \geant for single showers with energies between $10-100$~\si{GeV} on a single CPU core. This highlights that a point cloud representation of showers can not only offer a more efficient solution than a regular grid, but also enables a superior speed-accuracy trade-off for highly granular calorimeter simulation in realistic applications. While there are some deviations present in \CCiii observables, in particular for single particle showers, these are suppressed in subsequent reconstruction for the physics case studied. This emphasizes that while studying model performance at the level of single showers provides useful model insight, model performance must ultimately be judged on post-reconstruction level physics observables. This also means that the relative importance of certain shower features depends on the target down-stream analysis. For example, the deviations observed in the intrinsic cluster angles\footnote{The extent to which these deviations appear for \CCiii depend on the approach used to determine the intrinsic angle. See \cite{buss2025caloclouds3ultrafastgeometryindependenthighlygranular} for more details.} in Section~\ref{sec:SingleParticle} may be more relevant for certain searches for physics beyond the standard model~\cite{Lee:2018pag}. 

It should be noted that some deviations remain, even for the \optimumSteps generator (see for example the resolution plot in Figure~\ref{fig:resolution}), fixing an upper limit for \CCiii performance. These deviations are caused by shower placement into isolated regions in the geometry which feature greater irregularities -- more details are presented in Appendix~\ref{Append:Systematics}. While the reconstructed energy here could be corrected by dedicated calibrations, improving the modeling of shower structure in these regions would require the addition of dedicated model trainings.

This work has studied the performance of generative models for highly granular electromagnetic shower simulation for the case of photons. Future work could investigate the case of electron or positron showers. As these particles are charged, they also require track-cluster associations to be performed as part of the reconstruction, and this may add increased importance to particular shower observables. In order to achieve significantly faster simulation throughput at the level of full physics events for detectors with highly granular calorimeters in general, it will be necessary to address the challenge of highly granular hadron shower simulation. Work in \cite{Buss:2025cyw} has recently demonstrated that the use of a diffusion-transformer mechanism in a point cloud model can provide an accurate modeling of hadron showers, including for single particle observables at the post-reconstruction level. However, further work is still required to develop a model which can achieve a significantly faster simulation throughput, as well as to develop benchmarks similar to the ones shown here targeted at hadronic showers.

\begin{acknowledgments}
We thank Dirk Krücker for valuable comments on the manuscript. This research was supported in
part by the Maxwell computational resources operated at Deutsches Elektronen-Synchrotron DESY,
Hamburg, Germany. This project has received funding from the European Union’s Horizon 2020
Research and Innovation programme under Grant Agreement No 101004761. We acknowledge
support by the Deutsche Forschungsgemeinschaft under Germany’s Excellence Strategy – EXC
2121 Quantum Universe – 390833306 and via the KISS consortium (05D23GU4, 13D22CH5)
funded by the German Federal Ministry of Research, Technology and Space (BMFTR) in the ErUM-Data
action plan. A.K. has received support from the Helmholtz Initiative and Networking Fund’s initiative
for refugees as a refugee of the war in Ukraine. P.M. has benefited from support by the CERN Strategic R$\&$D Programme on Technologies for Future Experiments \cite{EPRD}. Some figures and text in this paper have previously appeared in a doctoral thesis \cite{McKeown:2024}.
\end{acknowledgments}

\bibliography{literature}

\newpage
\appendix

\section{\label{sec:Integration} Integration into Standard Software Chains}

In order to study the performance of the generative models described in Section \ref{sec:Models} in a realistic physics simulation including event reconstruction, it is necessary for the models to be combined with the experiment's standard software ecosystems, typically implemented in C++. This appendix provides additional details on this generic library in Section \ref{sec:Integration:DDML}, and on the specific implementation for ILD in Section \ref{sec:Integration:ILD_DDML}.

\subsection{\label{sec:Integration:DDML} The \DDML Library}

\begin{figure*}[tbhh]
    \centering
    \includegraphics[width=0.7\textwidth]{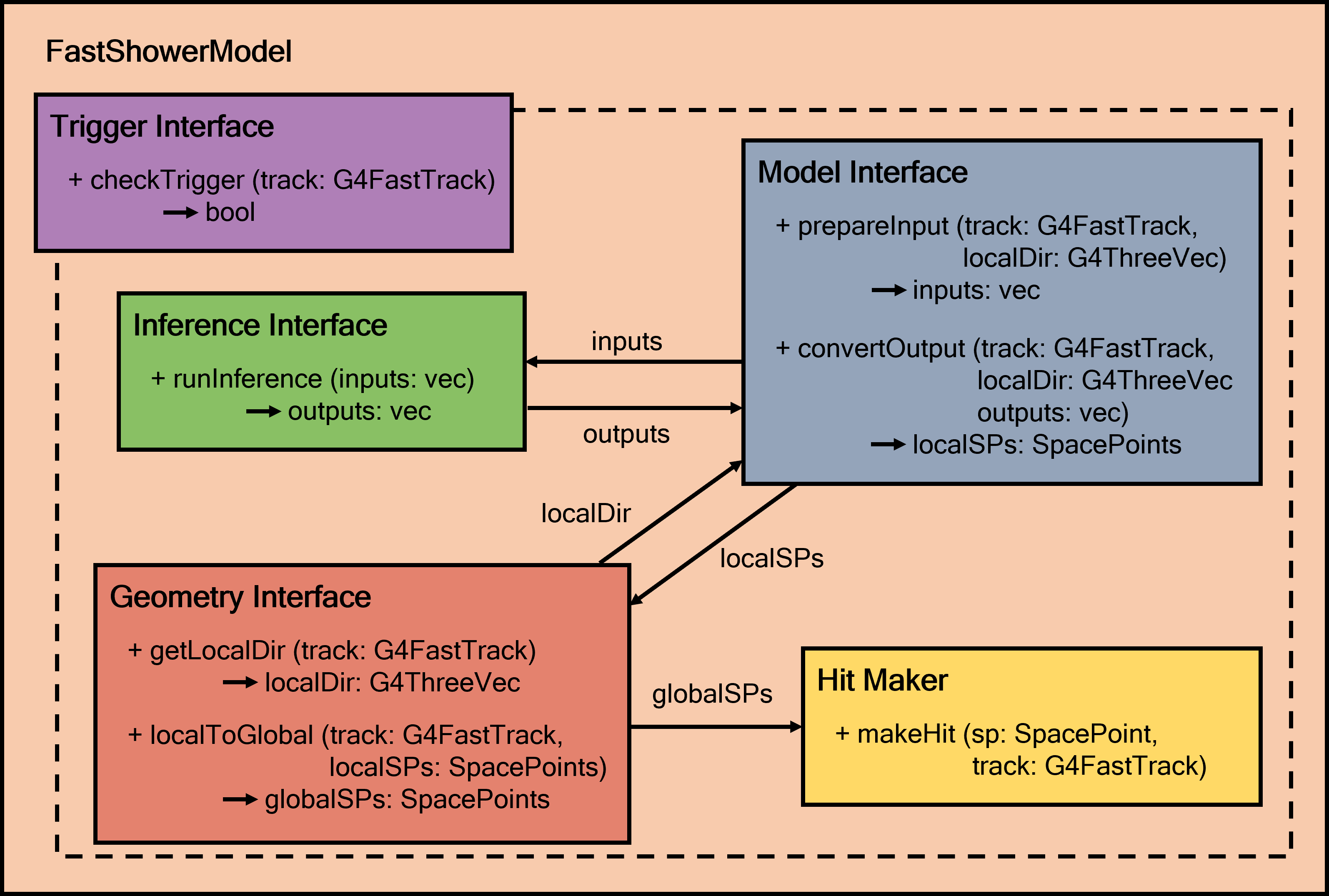}
    \caption{Class diagram illustrating the core components of the \DDML library. See the text for a detailed description of the interfaces. Figure from \cite{McKeown:2024}.}
    \label{fig:DDML_Class_diagram}
\end{figure*}

To integrate generative models for fast calorimeter simulation into full simulation applications, we introduce the \DDML \footnote{\url{https://github.com/key4hep/DDML}} library~\cite{ddml_zenodo} as part of the \textsc{Key4hep} software stack. It follows the \textsc{Geant4 Par04} example for running machine learning inference in fast simulation models \cite{Geant4Par04}. Our library is built on top of the \geant and \ddhep toolkits, providing access to a broad suite of functionality. In particular, this includes an interface to a trigger mechanism present in \geant for terminating physics-based full simulation in favor of an alternative simulation approach. A trigger is associated with a particular geometrical region of the detector, and is activated if an impinging particle satisfies certain criteria (particle type, energy etc.). This provides a seamless way of incorporating a generative model based fast calorimeter simulation tool into a full simulation application.

The \DDML library is designed to support a generic approach to fast simulation with generative models. This can be split into three key requirements:

\begin{enumerate}
    \item Allow the use of different kinds of generative model, in terms of their structure, the inputs they require and the outputs they generate.
    \item Allow different engines for model inference to be employed.
    \item Allow the use of different detector geometries implemented in \ddhep.
\end{enumerate}

This requires two conventions to be chosen for the library. The first is the adoption of a local (right-handed) coordinate system, defined such that the origin is placed at the point of entry into the calorimeter, with the $z'$ axis orthogonal to the calorimeter face, and pointing into the calorimeter system. The $x'$ axis is aligned with the direction of the magnetic field. This allows the model output to be handled in the same manner independent of where in the detector a particular particle is incident. The second is the interpretation of the model outputs as local space points in this coordinate system. This provides a generic means of interpreting the output of a model, independent of its architectural details.

The library is split into a number of different interfaces via a class template. This decouples the three aspects detailed above as far as possible, making it easier to extend and maintain the library. The interfaces are outlined below, with a class diagram of \DDML shown in Figure \ref{fig:DDML_Class_diagram}, and the order of operations being shown in Algorithm \ref{algo:DDML}. \\

\begin{algorithm}
    \caption{Pseudocode illustrating the order of operations for the core components of the \DDML library.}
    \begin{algorithmic}[1]
    \If {\textit{Trigger}.checkTrigger(track) == True}
       \State Kill full simulation of particle
       \State localDir = \textit{Geometry}.getLocalDir(track)
       \State inputs = \textit{Model}.prepareInputs(track, localDir)
       \State outputs = \textit{Inference}.runInference(inputs)
       \State localSPs = \textit{Model}.convertOutput(track, localDir, outputs)
       \State globalSPs = \textit{Geometry}.localToGlobal(track, localSP)

       \For {(sp in globalSPs)}
            \State \textit{HitMaker}.makeHit(sp, track)
        \EndFor
    \Else
       \State Full simulation of particle with \geant
    \EndIf
    \end{algorithmic}
    \label{algo:DDML}
\end{algorithm}

\noindent \textbf{Trigger}\\
The Trigger Interface sits on top of the trigger on particle type and energy that exists in \ddhep and \geant. This interface allows a particular fast simulation model to be excluded from running in certain regions of the detector (\textit{checkTrigger}). This provides a simple means of handling regions of a calorimeter with an irregular structure, where either full simulation or a separate generative model can be run instead.\\

\noindent \textbf{Model}\\
The Model Interface is used to provide a model specific implementation. The role of this interface is two-fold. The first role is the preparation of the input (\textit{prepareInputs}) in the form expected by the model. As part of this, a \textit{localDir} object is available to provide information about the local direction at the calorimeter face. This can then be used as conditioning input for the generative model. The second role is the interpretation of the output of the model such that it can be converted into local space points (\textit{convertOutput}).\\

\noindent \textbf{Inference}\\
The Inference Interface provides a simple means of calling the inference library for a model (\textit{runInference}). Currently both the \textsc{LibTorch} \cite{pytorch} and \textsc{OnnxRuntime} \cite{onnxruntime} inference libraries are supported. Alternatively, functionality is provided for loading a pre-simulated shower library from a \textsc{HDF5} file, which is intended for model prototyping and representation investigations.\\

\noindent \textbf{Geometry}\\
The Geometry Interface performs two separate roles. The first is to compute the local direction (\textit{getLocalDir}), which is provided to the Model Interface as a consistent means of model conditioning. The second is to place the local space points produced as output of the Model Interface into the geometry of the detector (\textit{localToGlobal}). This includes both the conversion from local calorimeter coordinates into global (envelope) coordinates, and the placement of hits onto sensitive detector elements. It must be implemented on a per-geometry basis, with disk endcap, polyhedral and cylindrical barrel calorimeter geometries already supported.\\

\noindent \textbf{HitMaker}\\
A helper class provided by \geant to allow the placement of energy deposits produced by the fast simulation model, given that their position lies within a sensitive element of the detector (\textit{makeHit}).\\

Currently, the \DDML library only supports single shower generation with a generative model (inference with batch size of one) on a CPU, which still represents the dominant hardware available in high energy physics computing infrastructure \cite{Boehnlein:2803119}. Future support for batched shower generation and the addition of GPU support is foreseen. Such a development would then make the most significant simulation speed-ups relative to \geant accessible via parallelization.

\subsection{\label{sec:Integration:ILD_DDML} \DDML Implementation for ILD}

The \DDML implementation for the ILD detector used in this study includes both the \CCiii and \LFlows models described in Section \ref{sec:Models}. The models were converted to a format suitable for use in C++, with the components that composed the architectures being serialized by a combination of tracing and scripting each of the individual operations. This was achieved predominantly by using the utilities provided by \textsc{TorchScript} in \textsc{PyTorch} \cite{pytorch}, with the \textsc{ShowerFlow} component of the \CCiii architecture making use of the \textsc{POUTINE} effect handlers present in the \textsc{PYRO} \cite{bingham2019pyro} deep probabilistic programming library in which the model was implemented.

This study focuses on the barrel region of the ILD detector. In order to leverage the symmetry present in the ILD detector, we generate showers at different positions in the detector using models trained on showers in a single location, as described in Section \ref{sec:Train_Valid_data}. This is possible because the regularized calorimeter used for creating the training dataset has no dead material within an active layer, and no gaps. This means that the model is a valid simulator for the vast majority of the detector, while in regions which are particularly irregular, full simulation is run instead using the trigger interface described in Section \ref{sec:Integration:DDML}.

The first type of region excluded is the transition between the edge of the barrel and the endcap. In this region, there is a gap between the barrel and the endcap, and a change in orientation of the calorimeter layers. For this reason, the edges of the barrel at $\theta < 40$ degrees and $\theta > 140$ degrees in the global ILD coordinate system are excluded from fast simulation. The second type of transition region is the intersection between staves of the octagonal barrel, where an asymmetrical change in orientation of the calorimeter layers occurs. These $8$ transition regions are excluded by $8$ cuts in the global ILD coordinate system, each of which consists of an $8.01$ degree range in $\varphi$.

\section{Observable Comparison Summary}\label{Append:Summary}

Table~\ref{table:all} provides a concise overview of the agreement between the evaluated surrogate models and the \geant reference across all benchmark observables.
The comparison employs the Jensen-Shannon (JS) divergence and the mean absolute error ($L_1$) as quantitative metrics, both measuring the deviation of a model’s reconstructed distributions from those obtained with \geant.
Lower values correspond to better agreement.

Uncertainties are estimated from repeated evaluations using identical event samples simulated with different random seeds, for both \geant and the surrogate models.

Among the reference optimal shower generators, \optimumXOne- trained directly at detector readout granularity- shows systematically larger deviations, indicating that coarse spatial representations limit achievable fidelity. Progressively finer geometrical resolution in \optimumXNine and \optimumSteps significantly improves the match, providing an effective upper bound on achievable performance for surrogates.

\begin{table*}[tbh]
\sisetup{
separate-uncertainty=true,
table-format=4.3(5)
}
\centering
\begin{tabular}{lccccc}
\toprule
Metric & \optimumXOne & \optimumXNine & \optimumSteps & \LFlows (x9) & \CCiii (x25)  \\
\midrule
Single shower observables \\
$\mathrm{JS}^{E_\mathrm{radial(100)}} (\times 10^{-4})$          & 4.71 $\pm$ 0.04 & \textbf{0.04 $\pm$ 0.01} & 0.06 $\pm$ 0.01  & \it{8.37 $\pm$ 0.03} & 1.11 $\pm$ 0.02 \\
$\mathrm{JS}^{E_\mathrm{radial(30)}} (\times 10^{-4})$           & \it{23.33 $\pm$ 0.16} & 0.46 $\pm$ 0.02 & \textbf{0.04 $\pm$ 0.01}  & 1.92 $\pm$ 0.06 & 2.24 $\pm$ 0.06 \\
$\mathrm{JS}^{E_\mathrm{long}} (\times 10^{-4})$                 & \textbf{0.049 $\pm$ 0.013} & 0.050 $\pm$ 0.013 & \textbf{0.049 $\pm$ 0.013}  & \it{0.313 $\pm$ 0.023} & 0.050 $\pm$ 0.011 \\
$\mathrm{JS}^{iPhi_\mathrm{res(100\%)}} (\times 10^{-4})$        & \textbf{13.84 $\pm$ 1.58} & 22.72 $\pm$ 2.07 & 24.68 $\pm$ 2.02 & \it{342.18 $\pm$ 7.96} & 48.64 $\pm$ 3.62 \\
$\mathrm{JS}^{iPhi_\mathrm{res(4\%)}} (\times 10^{-4})$          & \it{299.91 $\pm$ 7.08} & 9.33 $\pm$ 1.20 & \textbf{1.37 $\pm$ 0.53}  & 6.58 $\pm$ 1.20 & 16.85 $\pm$ 1.93 \\
$\mathrm{JS}^{iTheta_\mathrm{res(100\%)}} (\times 10^{-4})$      & 44.40 $\pm$ 3.15 & 37.72 $\pm$ 2.71 & \textbf{37.52 $\pm$ 2.82} & \it{553.91 $\pm$ 10.26} & 374.08 $\pm$ 8.83 \\
$\mathrm{JS}^{iTheta_\mathrm{res(4\%)}} (\times 10^{-4})$        & \it{195.27 $\pm$ 5.54} & 13.23 $\pm$ 1.32 & \textbf{2.02 $\pm$ 0.66}  & 39.12 $\pm$ 2.49 & 32.10 $\pm$ 2.16 \\
$\mathrm{L_1}^{\frac{\sigma_{90}}{\mu_{90}}}$   & \it{114.33 $\pm$ 5.43} & 29.84 $\pm$ 5.13 & \textbf{11.63 $\pm$ 5.08} & 49.31 $\pm$ 5.35 & 17.98 $\pm$ 5.26 \\
$\mathrm{L_1}^{\mu_{90}}$                       & 0.25 $\pm$ 0.04 & 0.29 $\pm$ 0.03 & 0.40 $\pm$ 0.03 & \it{1.06 $\pm$ 0.04} & \textbf{0.09 $\pm$ 0.04} \\
& \\
Multi-particle observables  \\
$\mathrm{JS}^{\gamma\gamma_{rec.}} (\times 10^{-6}) @\mathrm{5GeV}$    & \it{182.69 $\pm$ 27.87} & \textbf{18.09 $\pm$ 10.28} & -- & 28.52 $\pm$ 13.56 & 36.30 $\pm$ 14.21 \\
$\mathrm{JS}^{\gamma\gamma_{rec.}} (\times 10^{-6}) @\mathrm{20GeV}$   & \it{380.74 $\pm$ 32.14} & \textbf{10.54 $\pm$ 6.04} & -- & 19.56 $\pm$ 8.86 & 34.21 $\pm$ 12.79 \\
$\mathrm{JS}^{\gamma\gamma_{rec.}} (\times 10^{-6}) @\mathrm{100GeV}$  & 5.57 $\pm$ 2.94 & \textbf{2.17 $\pm$ 1.24} & -- & \it{7.28 $\pm$ 4.48} & 2.95 $\pm$ 1.77 \\
$\mathrm{JS}^{M_{\pi^0}} (\times 10^{-4})$                       & 23.12 $\pm$ 7.38 & \textbf{13.91 $\pm$ 6.89} & 25.83 $\pm$ 8.58 & \it{32.06 $\pm$ 8.74} & 22.08 $\pm$ 7.40 \\
$\mathrm{JS}^{E_{\pi^0}} (\times 10^{-4})$                       & 31.40 $\pm$ 9.93 & \textbf{19.62 $\pm$ 6.74} & \it{35.88 $\pm$ 8.02} & 27.24 $\pm$ 8.22 & 26.87 $\pm$ 8.82 \\
$\mathrm{JS}^{\theta_{\pi^0res}} (\times 10^{-4})$               & \it{143.62 $\pm$ 16.41} & 10.97 $\pm$ 4.94 & \textbf{7.04 $\pm$ 4.27}  & 34.97 $\pm$ 7.63 & 10.69 $\pm$ 4.90 \\
$\mathrm{JS}^{\phi_{\pi^0res}} (\times 10^{-4})$                 & 25.07 $\pm$ 6.98 & 13.90 $\pm$ 6.05 & 16.11 $\pm$ 6.38 & \it{49.74 $\pm$ 10.14} & \textbf{8.68 $\pm$ 4.91} \\
$\mathrm{JS}^{\pi^0_{rec}} (\times 10^{-4})$ & 9.90 $\pm$ 1.66 & 0.73 $\pm$ 0.28 & \textbf{0.49 $\pm$ 0.33} & \it{12.75 $\pm$ 2.3} & 0.51 $\pm$ 0.28 \\

\bottomrule 
\end{tabular}
\caption{Quantitative comparison of models' performance relative to \geant for single photon, di-photon and tau samples. Metrics are JS divergence or $L_1$ distance as indicated, lower values correspond to better agreement. Bold entries denote the best agreement with \geant per observable, italics denote the worst.
}
\label{table:all}
\end{table*}

\section{Systematics resulting from simulation methodology}\label{Append:Systematics}

\begin{figure*}[tbhh]
    \centering
    \includegraphics[width=0.6\textwidth]{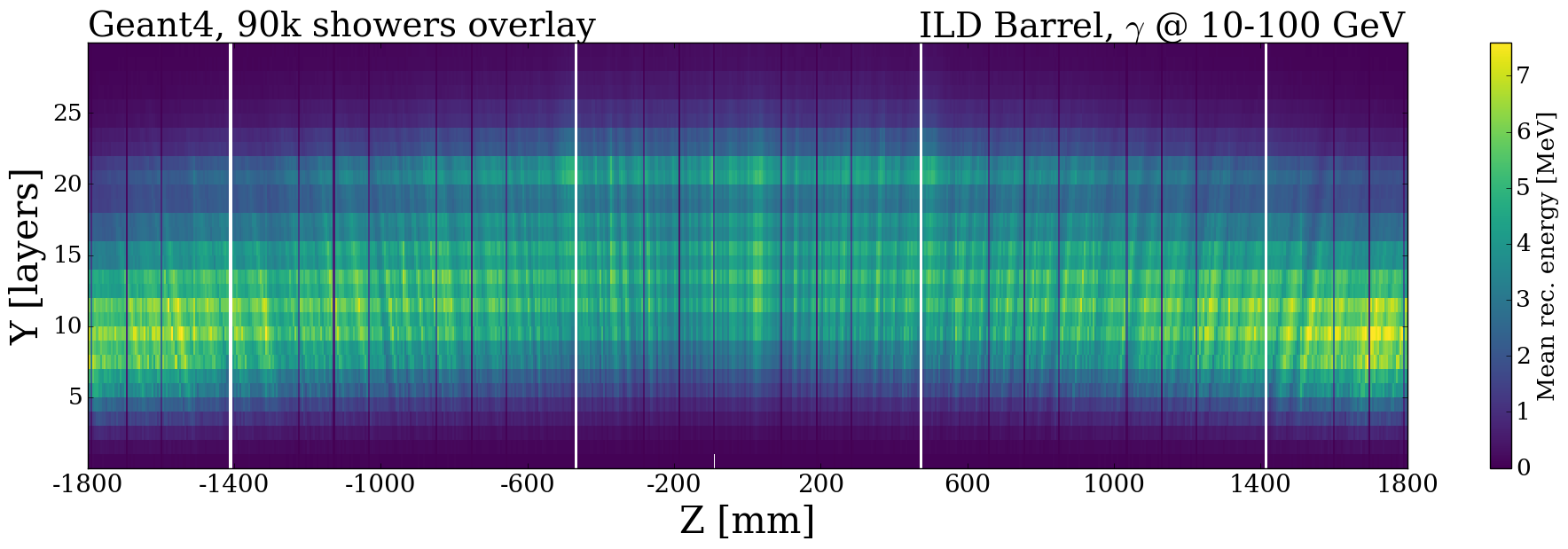}
    \includegraphics[width=0.6\textwidth]{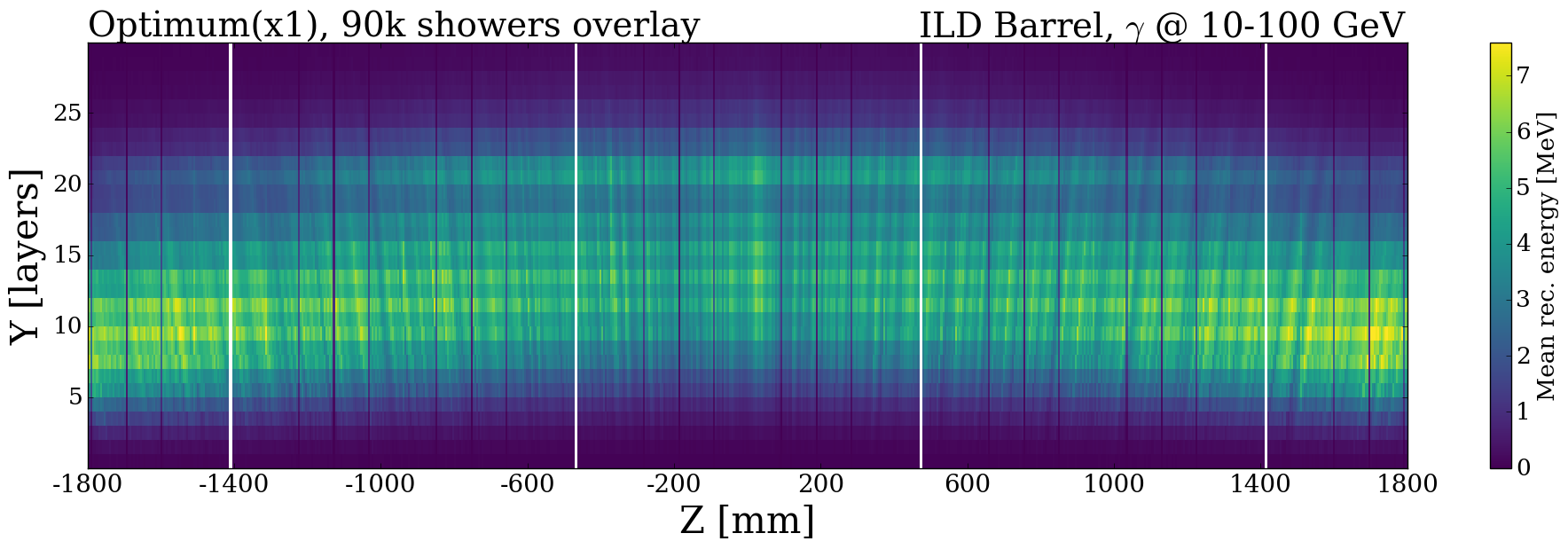}
    \includegraphics[width=0.6\textwidth]{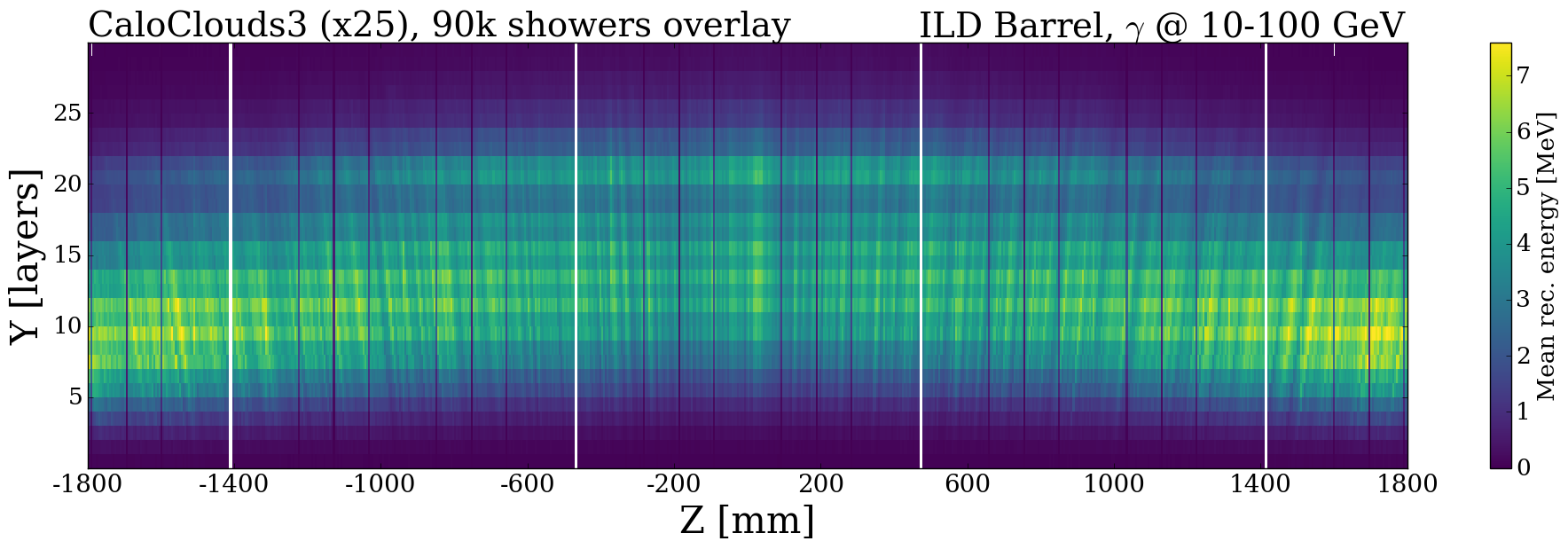}
    \includegraphics[width=0.6\textwidth]{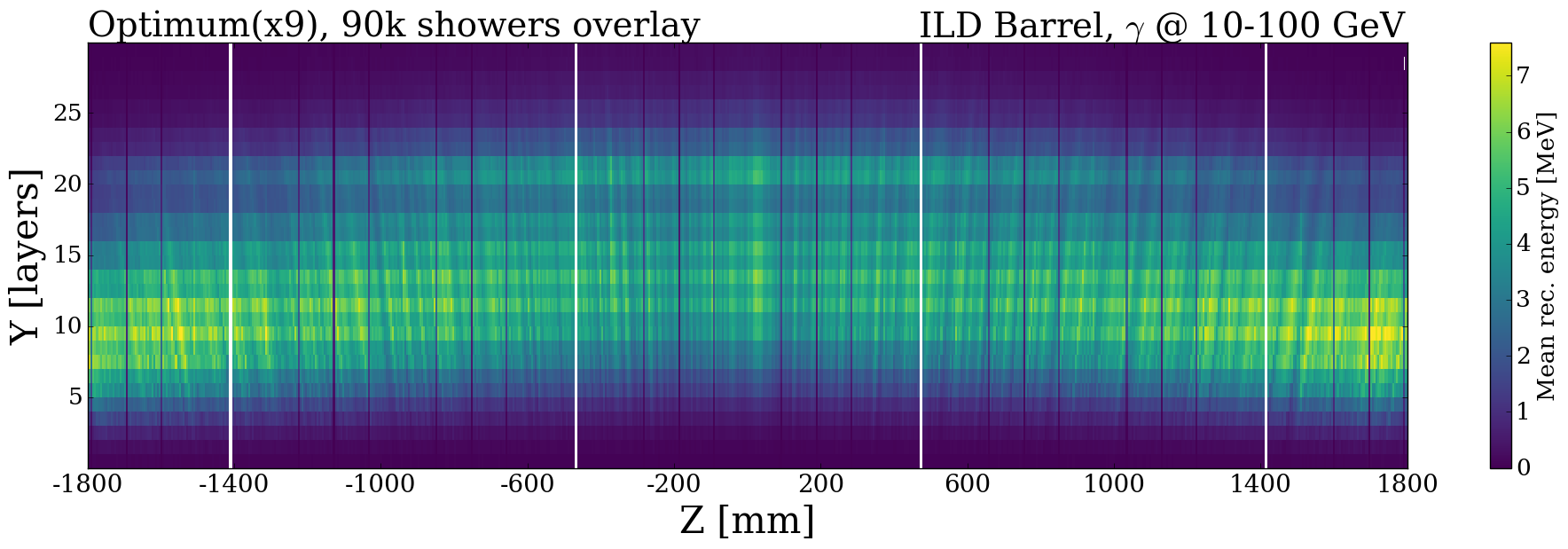}
    \includegraphics[width=0.6\textwidth]{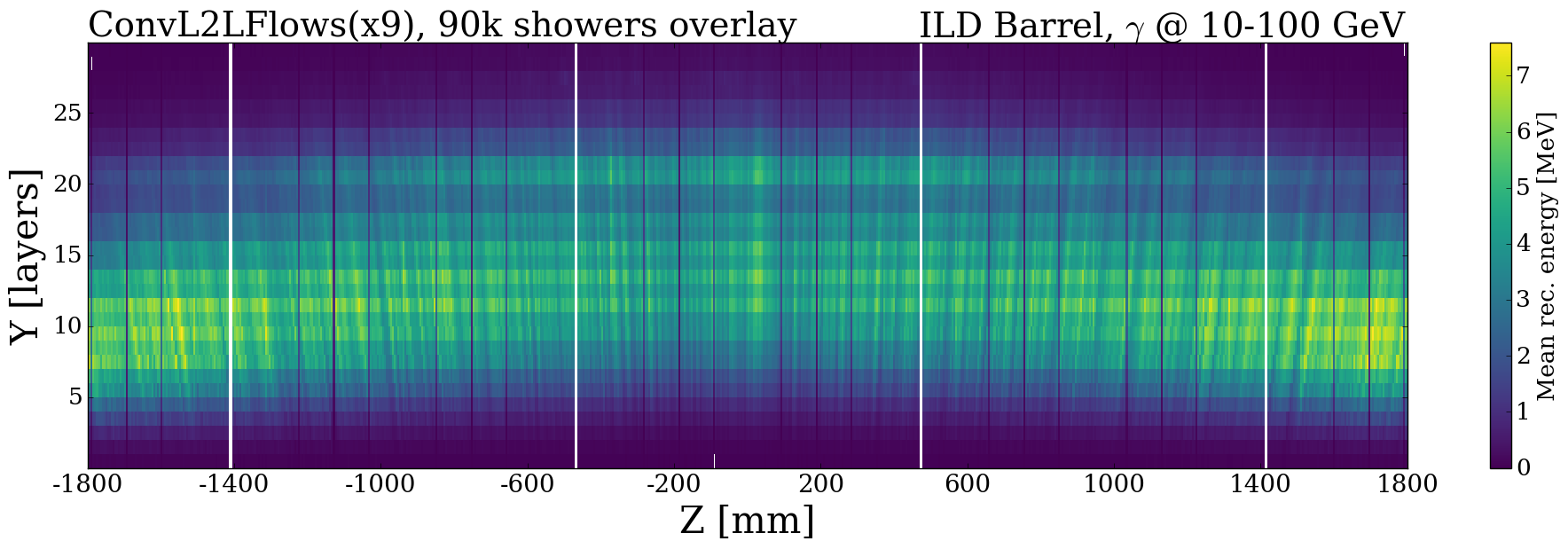}
    \includegraphics[width=0.6\textwidth]{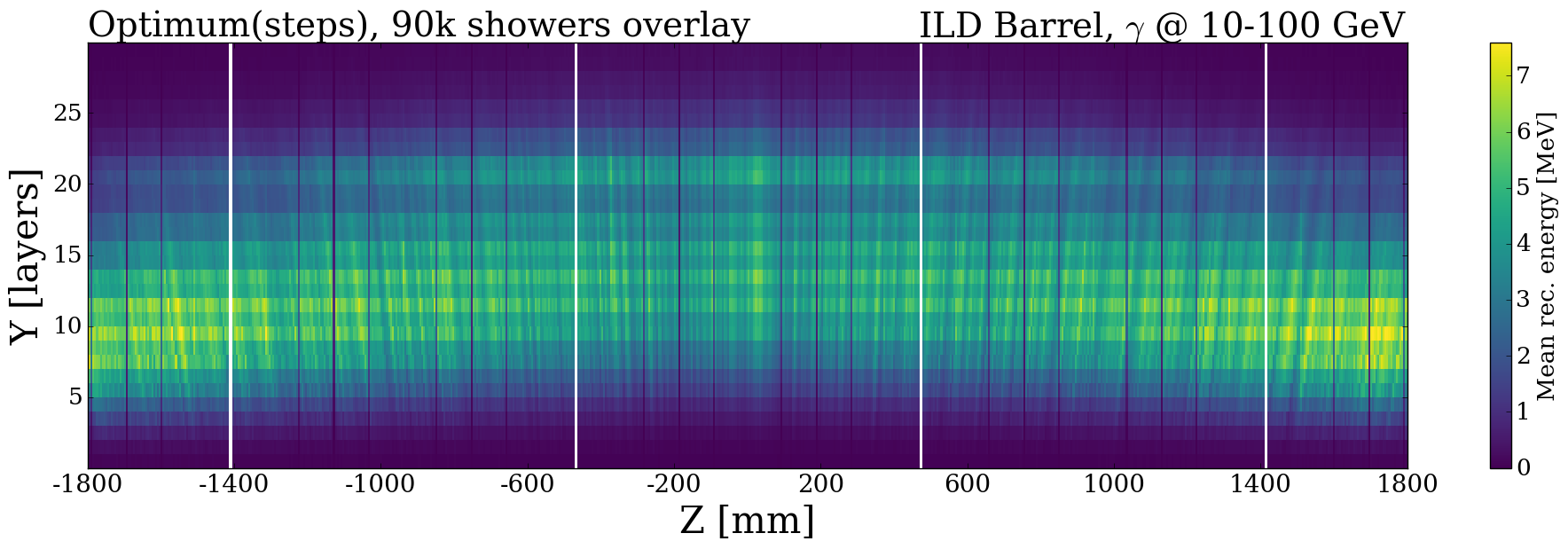}

    \caption{
        Average reconstructed energy of 90,000 photon showers incident on the upper barrel segment of the ILD ECAL, projected onto the YZ plane. Each panel shows one generator or model (name of the model in the title of each panel), with Y denoting the calorimeter depth (layer index) and Z denoting the global detector coordinate in mm. The color scale corresponds to the mean deposited energy per voxel.
    }
    \label{fig:photons_overlay}
\end{figure*}

Figure \ref{fig:photons_overlay} shows the average reconstructed energy pattern for photon showers, visualized as YZ projections of the mean energy per voxel across 90,000 events. The vertical direction Y corresponds to the calorimeter layers, while Z indicates the position within the upper ECAL barrel in mm. All optimal shower generators and generative models reproduce the overall longitudinal and transverse structure observed in the \geant reference, including layer-dependent modulations and geometrical features of the detector segmentation. No significant deviations are visible at this scale, demonstrating that all models are properly integrated into the detector geometry and capture the overall event topology with high fidelity.

\begin{figure*}[tbhh]
    \centering
    \includegraphics[width=0.8\textwidth]{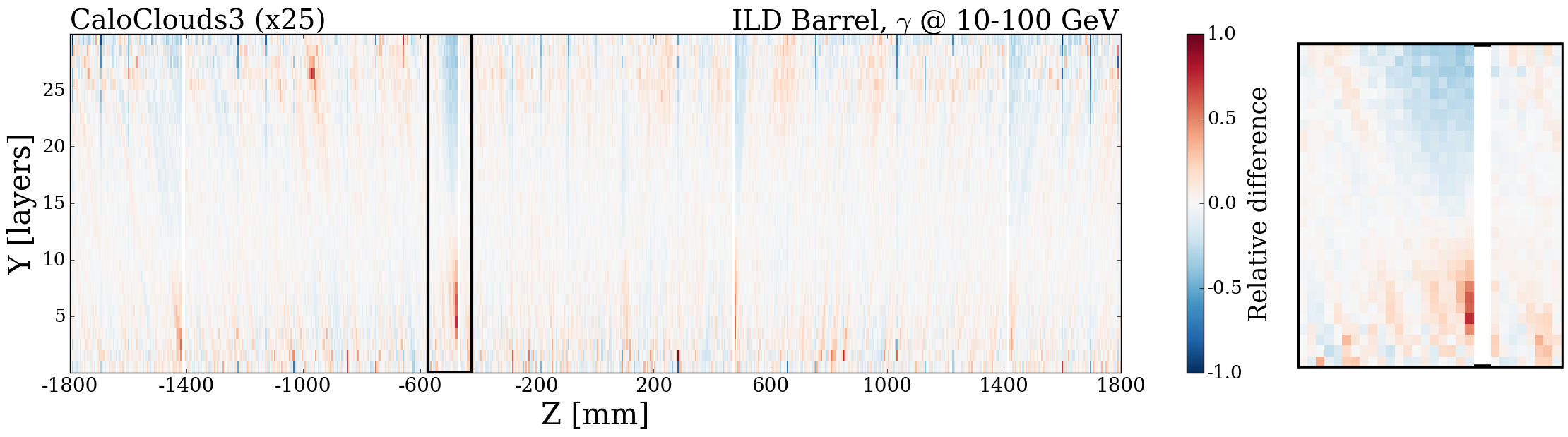}
    \includegraphics[width=0.8\textwidth]{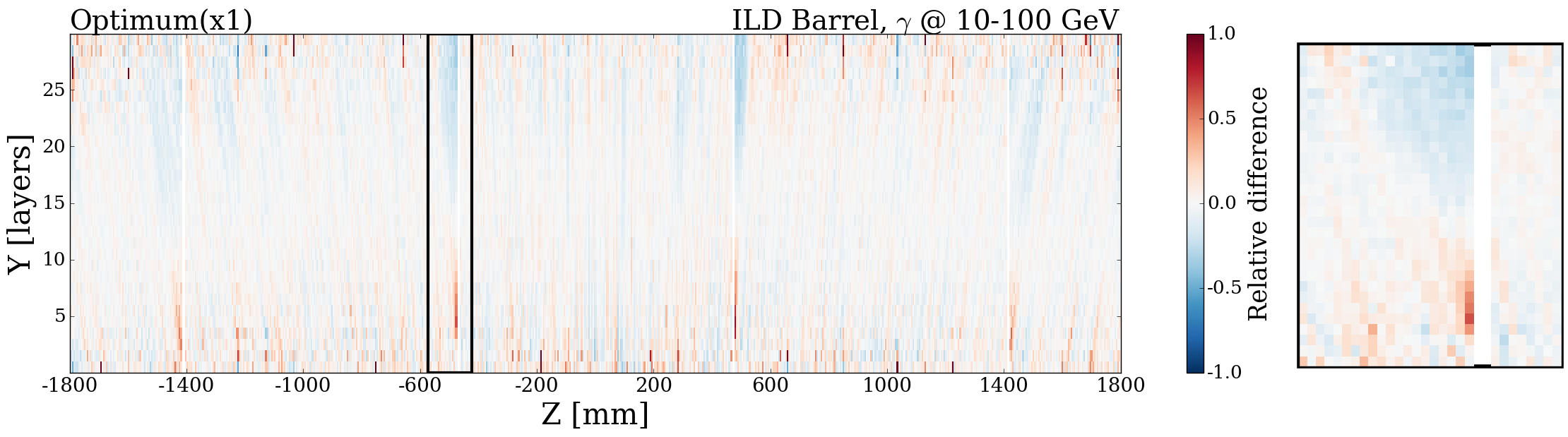}
    \includegraphics[width=0.8\textwidth]{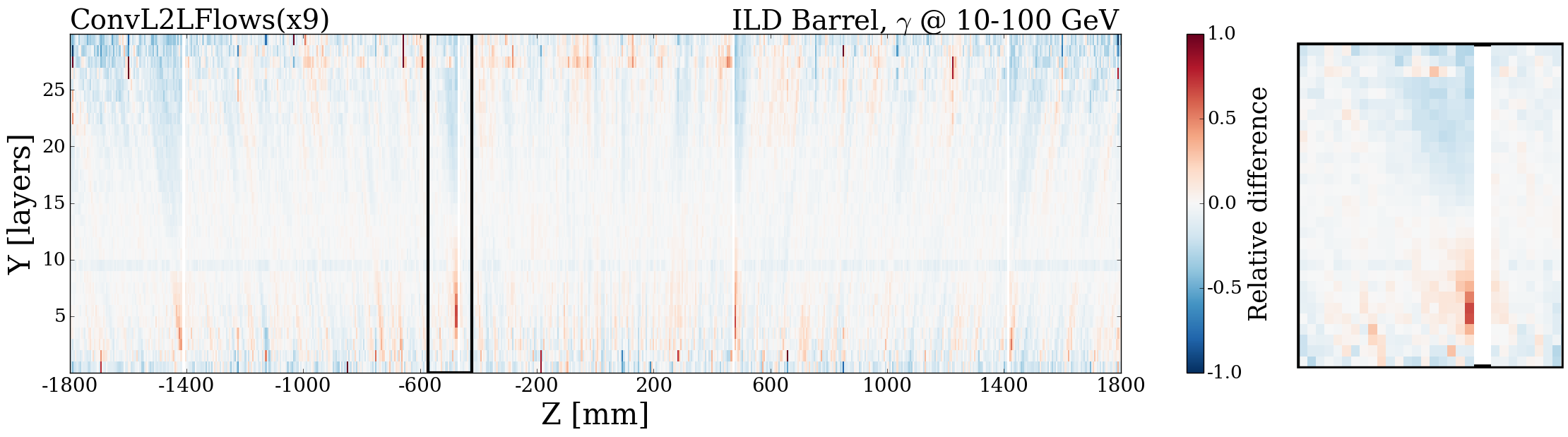}
    \includegraphics[width=0.8\textwidth]{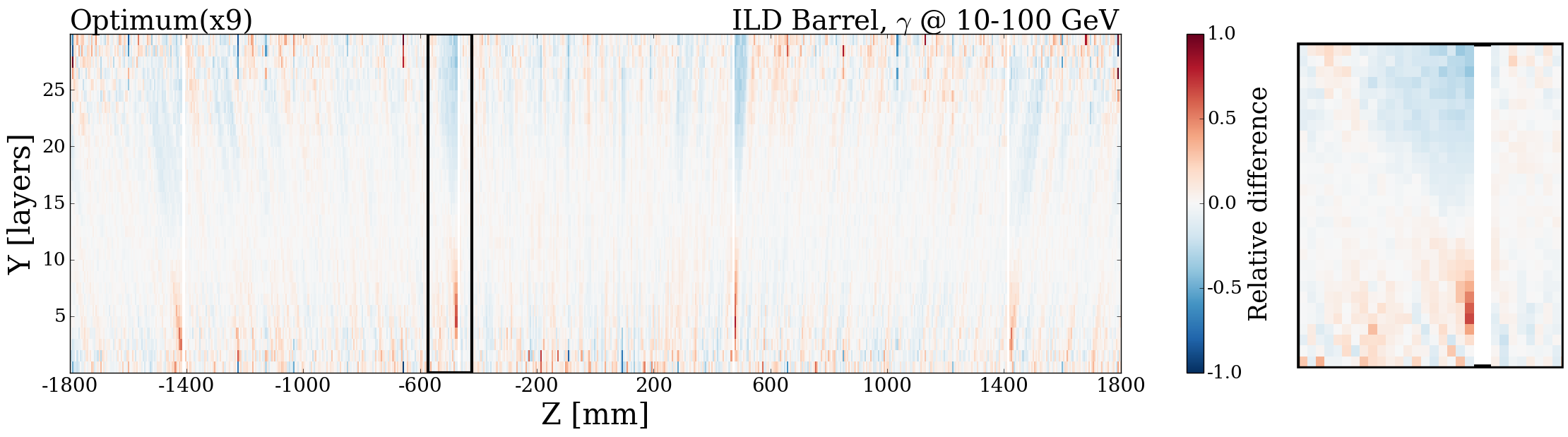}
    \includegraphics[width=0.8\textwidth]{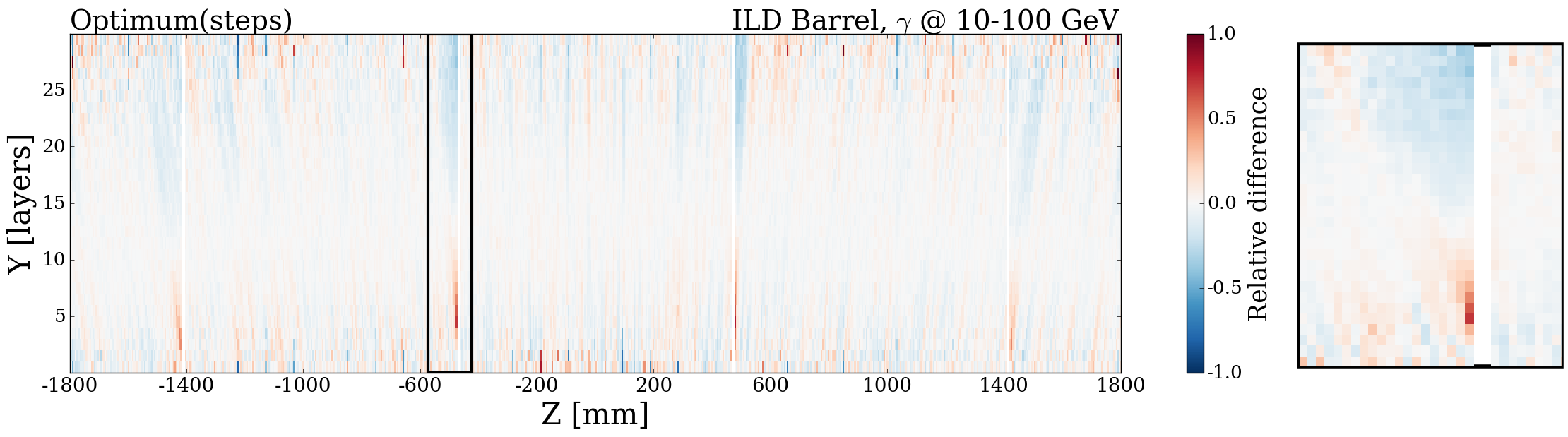}

    \caption{
        Relative per-voxel deviation in mean reconstructed energy in the ILD ECAL with respect to \geant for each surrogate model (name of the model in the title of each panel). The accompanying zoomed-in panels (on the right of each panel) show the regions near calorimeter module gaps. The consistent overestimation at shower onset and underestimation at larger depths arise from the absence of absorber gaps in the idealized training geometry. These localized geometry-dependent effects set a practical limit on surrogate accuracy when transferring models trained at a single reference position to the full ILD geometry.
    }
    \label{fig:photons_overlay_diff}
\end{figure*}

In contrast, the relative per-voxel energy difference maps shown in Fig.~\ref{fig:photons_overlay_diff} reveal small but systematic biases. To the right of each panel, zoomed-in views highlight regions near module boundaries and absorber gaps. Because all surrogate models were trained on idealized geometries without inactive regions, they tend to overestimate the early energy deposition. The showers in the surrogate models start slightly earlier since the absorber material is present in every layer homogeneously in the idealized geometry. In the realistic ILD geometry, however, the presence of structural gaps in the absorber delays the shower onset, allowing the energy to penetrate deeper. This results in a characteristic underestimation of energy in the later layers for all surrogates when compared to the full \geant simulation.

These effects originate from the interplay between the idealized local training setup and the complex global detector structure.  
They represent geometry-dependent systematic biases that persist even for the most accurate surrogate, \optimumSteps, and define a practical upper limit on achievable agreement when a single conditional model is reused across the barrel.  

Because the affected regions are spatially confined, extending the approach with localized trainings or lightweight region-specific calibrations offers a straightforward path to further reduce the remaining discrepancies.

\end{document}